\documentclass[11pt,letter]{article} 
\usepackage{jheppubGreen} %
\usepackage{colortbl} 
\usepackage[section] {placeins}
\definecolor{nodegreen}{rgb}{0,0.4,0.3}
\definecolor{weakred}{rgb}{0.8,0.6,0.6}
\definecolor{darkgreen}{rgb}{0,0.6,0} 
\definecolor{forestgreen}{rgb}{0.133,0.545,0.133}
\definecolor{purple}{rgb}{0.62745098,0.125490196,0.941176471}

\title{Argyres-Douglas Loci, Singularity Structures and Wall-Crossings in Pure ${\cal N}=2$ Gauge Theories with Classical Gauge Groups}
\author{Jihye Seo}
\author{and Keshav Dasgupta}
\affiliation{Ernest Rutherford Physics Department, McGill University,\\
3600 rue University, Montreal, QC H3A 2T8, Canada}
\emailAdd{jihyeseo, keshav@hep.physics.mcgill.ca}
\abstract{${\cal N} = 2$ Seiberg-Witten theories allow an interesting interplay between the Argyres-Douglas loci, singularity structures and wall-crossing formulae. In this 
paper we investigate this connection by first studying the 
singularity structures of hyper-elliptic Seiberg-Witten curves for pure ${\cal{N}} = 2$ gauge theories with $SU(r + 1)$ and $Sp(2r)$ gauge groups, and propose new methods to locate the Argyres-Douglas loci in the moduli space, where multiple mutually non-local BPS states become massless. In a region of the moduli space, we compute dyon charges for all $2r + 2$ and $2r + 1$ massless dyons
for $SU(r + 1)$ and $Sp(2r)$ gauge groups respectively for rank $r > 1$. From here
we elucidate the connection to the wall-crossing phenomena for pure $Sp(4)$ Seiberg-Witten theory near the Argyres-Douglas 
loci, despite our emphasis being only at the 
massless sector of the BPS spectra.  
We also present $2r-1$ candidates for the maximal Argyres-Douglas points for pure $SO(2r+1)$ Seiberg-Witten theory.
}
\arxivnumber{1203.6357}
\begin{document}
\maketitle
\flushbottom
 \toccontinuoustrue

\section{Introduction}

Ever since Seiberg and Witten proposed the exact solution to the low energy limit of ${\cal N} = 2$ SYM theory with at most two derivatives and four fermions 
\cite{SeibergWittenNoMatter, SeibergWittenWithMatter}, there have been tremendous activity in this field. The list of results we now have are impressive, and yet every time 
with every new investigations novel and surprising results seem to come up, for example \cite{Gaiotto, AGT}, to name a few. Even oft-investigated directions, for example 
the study of Argyres-Douglas loci \cite{ArgyresDouglas}, hold surprises with sufficient number of interesting new physics, rendering a detailed study 
not only necessary but also inevitable.  One of 
the primary aim of this paper is to analyze the physics of Argyres-Douglas loci using pure ${\cal N} = 2$ theories with classical gauge groups. In particular 
we study singularity structures of hyper-elliptic Seiberg-Witten (SW) curves for ${\cal{N}} = 2$ pure gauge theories with $SU(r + 1)$ and $Sp(2r)$ gauge groups, and propose new methods to locate Argyres-Douglas loci in the moduli space, where multiple mutually non-local BPS states become massless. As one might have expected, 
from the behaviour of SW 1-form, we observe that singularity structure of moduli space of the SW curve survives as singularity of SW-theory. Indeed,
in discriminant loci of the curve, which is a ${\rm codim}_{\mathbb{C}}$-$1$ loci in the moduli space satisfying a complex relation $\Delta f=0$, a BPS state become massless. 
Here by $f$ we will mean the (hyper-)elliptic equation of the curve (to be explained in more details a bit later), and $\Delta$ to be the discriminant. As an added bonus, 
we will compute, in a region of moduli space, dyon charges for all $2r + 2$ and $2r + 1$ massless dyons
for $SU(r + 1)$ and $Sp(2r)$ gauge groups respectively with rank $r > 1$.

Another interesting surprise that came out of more recent investigations on the subject, which is not just restricted to 
${\cal N} = 2$ theories, is the idea of wall-crossing \cite{GMNwall} leading to numerous developments in both physics and mathematics. 
The wall-crossing phenomena start from the observation that the BPS spectra are not globally invariant in the full moduli space. Instead, as we vary moduli values, the BPS spectra will jump in discrete manners, as if the moduli spaces were broken into multiple chambers, whose walls are real codimension 1 curve, which is wall of marginal stability. 
What happens at the wall is that the central charges of the BPS states line up the phases of their charges, so that the bound state of BPS states becomes marginally stable. In simple words, for a given amount of charges, the system is trying to reduce the energy by {\it minimizing the masses} of the BPS states. The basic idea is that on one side of the wall there can be a bound state, and on the other side it can break down into multiple BPS states, therefore rendering BPS spectra different on either sides. The readers are encouraged to read \cite{GMNwall} for more complete explanation of the wall-crossing formula. In this paper, we will see some remnant of wall-crossing phenomena, even though we are only looking at the {\it massless sector} of the BPS spectra. In subsection \ref{Sp4}, we will see an example of wall-crossing phenomena in the massless sector of pure $Sp(4)$ SW theory, near Argyres-Douglas loci of the theory.

The connection between Argyres-Douglas loci and certain aspects of wall-crossing is not new. 
 Argyres-Douglas points of $SU(3)$ were studied in detail in \cite{AF, GiveonRocek, CH}; and 
 \cite{Marino:Moore, RSVV} also discuss wall-crossing.
 In particular, the works of \cite{ShapereVafa}
and \cite{GMNwall} specified the marginal stability walls near Argyres-Douglas points where the massive bound states may decompose into multiple BPS states. In this paper,    
using the $Sp(4)$ curve, we observe that BPS charges for some vanishing 1-cycles 
jump as we go across Argyres-Douglas loci, giving a concrete example of wall-crossing phenomena for the massless BPS states. 

Before moving ahead let us present few concepts that will prove useful while navigating the paper. Two BPS states become massless where two
different solutions of vanishing discriminant intersect. This happens in ${\rm codim}_{\mathbb{C}}$-$2$ loci in the moduli space, which can be located by demanding discriminant and its exterior
derivative to vanish, so that we go to the region in the moduli space where the discriminant locus itself becomes singular. 
In these regions, double discriminant, (or as we like to call them: discriminant of discriminant) vanishes as well. 
Two massless BPS states can be mutually local or non-local, giving degeneration of the curves into node or cusp singularity. This can be distinguished by the order of vanishing of the double discriminant. It can also be seen from the shape of the intersection locus inside the moduli space: it is node-like
and cusp-like for mutually local and non-local pairs of BPS dyons
respectively. 
Alternatively, we can also demand a Seiberg-Witten curve to take a certain form, and it will restrict us to certain regions in the moduli space of the curve. Using that method we present $2r-1$ candidates for maximal Argyres-Douglas points for pure $SO(2r+1)$ Seiberg-Witten theory.

Our plan in the next few subsections is as follows: we will  provide some
extra motivation to look at the $Sp$ gauge group, coming from the F-theory picture followed by some preliminary materials for studying the geometry of hyperelliptic curve. 
We will then be required to brush upon two standard tools for singularity search, namely exterior derivative and discriminant. This will help us to 
understand how cusp-like singularity is related to Argyres-Douglas theory. At this point it will be important to know how to 
read off massless dyon charges by observing the branch points movement for vanishing 1-cycles. This will help us to 
show the dyon charges of vanishing cycles for Seiberg-Witten curves in two different forms. We will demonstrate this explicitly and also
provide a familiar example of rank 1 SW curve, where the moduli spaces for geometry and brane dynamics have a clear 1-1 mapping. Interestingly, our 
 study of vanishing BPS dyon charges will also teach us about the wall-crossing phenomena: a result that will have important implications later on. 

In the following we will start by reviewing some aspects of Seiberg-Witten theory that can be immediately used in later sections.

\subsection{Review of ${\cal N}=2$ Seiberg-Witten theory curves, and 1-forms: BPS states and 1-cycles \label{ReviewN2}}

We study a class of ${\mathcal{N}}=2$ $d=4$ gauge theories.  ${\mathcal{N}}=2$ is a good middle-ground to learn about field theory, which is more realistic than ${\mathcal{N}}\ge 3 $ and more controllable than ${\mathcal{N}}=0,1$. The reason for this is well explained in other papers, for example in \cite{LercheReview}, which the readers may 
look up for details. 
When amount of supersymmetry ${\mathcal{N}} \ge 2$, we say supersymmetry is extended, and it allows non-trivial anti-commutators between supercharges, 
\begin{equation}
\{Q^A, Q^B\} =Z^{AB} , \qquad A, B = 1, 2, \cdots , {\mathcal N},
\end{equation}
which we call central charges of superalgebra. Mass of a state is given by $M \ge \| Z\| $, where equality holds for BPS states.

If a ${\mathcal{N}}=2$ $d=4$ gauge theory has a low energy effective action, then it enjoys holomorphicity and can be solved exactly thanks to Seiberg-Witten theory. Seiberg-Witten geometry comes in package with Seiberg-Witten curve and SW differential 1 form.
 The SW curve is a complex curve (or a real 2 dimensional Riemann surface) some of whose 1-cycles correspond to a BPS states. Writing down the 1-cycle in terms of symplectic basis with $\mathbb{Z}$ coefficients, we can read off quantized electro-magnetic (dyonic) charges of the corresponding BPS state.

As reviewed in \cite{LercheReview}, central charge of a BPS state is given by integrating 1-form $\lambda_{\rm SW}$ over 1-cycle $\nu$, 
\begin{equation} Z= \oint_\nu \lambda_{\rm SW}. \label{Zlambda} \end{equation}
Assuming $\lambda_{\rm SW}$ is free of delta-function behaviour, 
vanishing 1-cycle gives massless BPS states. For pure $SU(r+1)$ and $Sp(2r)$ cases, we will see that $\lambda_{\rm SW}$ does not blow up near vanishing cycles. Therefore, study of vanishing 1-cycles can teach us about massless BPS states in the system. For $SU(r+1)$ theories with flavors,  \cite{GST} shows that this is true only up to a subtlety related to scaling dimensions. See \cite{EHIY} and \cite{EH} for earlier works on scaling behaviour at Argyres-Douglas loci. In this paper we won't discuss the scaling dimensions 
(this will be relegated to future work \cite{Seo}), instead we will only check that 1-form does not blow up.
We therefore assume that:

\vskip.1in

\framebox[1.1\width]{Singularity loci of SW curve $\subset$  Singularity loci of SW theory.}    
 
\vskip.1in

On the other hand, it is not forbidden to integrate 1-form over non-vanishing 1-cycle only to get zero value for the integration.  In that case, Seiberg-Witten theory will contain massless BPS state, which is not captured by vanishing 1-cycle of Seiberg-Witten curve. This will be highly non-generic, but currently we do not know whether this is prohibited either. We will however not discuss this possibility in this paper, if it exists it will make the former a proper subset of the latter, in the box above. 

 Our aim would then be to focus on the singularities of this theory (i.e dyon charges of massless BPS states), which are 
reflected in the singularities of the geometry. Recall that 
 Seiberg Witten curves were written and studied, and especially massless dyon charges were explicitly computed, for some low rank cases. For example the  
curves, massless dyons (monodromies), and some aspects of singularity aspects were studied for 
$SU(2)$ theory with and without matter in \cite{SeibergWittenNoMatter, SeibergWittenWithMatter}, for $SU(n)$ with and without matter 
in \cite{KLYTsimpleADE, ArgyresFaraggiSU, HananyOzSU}, for $SO(2r)$ and $SO(2r+1)$ without matter in  \cite{BLSo} and \cite{SO5DS} respectively, 
and $SO(r)$ with matter in \cite{HananySO}.
   
Here we will study the singularity structure of the moduli space of the Seiberg-Witten curve for pure $SU(r+1)$ and $Sp(2r)$ theory.  
For $SU(n)$ Seiberg-Witten theories, in certain region of moduli space, new superconformal exotic theories were uncovered and studied in 
\cite{ArgyresDouglas, ArgyresFaraggiSU}, where we have {\emph {mutually non-local}} massless dyons. Recall from \eqref{Zlambda} that massless dyons are associated with vanishing 1-cycles of the SW curve. If two 1-cycles have non-zero intersection number, we call them mutually non-local. In physics terms, no symplectic transformation will make them electronic simultaneously. If the SW curve degenerates into a cusp form, i.e when more than three branch points collide on the $x$-plane, then multiple {\emph {mutually non-local}} (i.e. having non-zero intersection) 1-cycles vanish at the same time. Interestingly these points are where the connection between the Argyres-Douglas loci and wall-crossing 
becomes more transparent. 
Argyres-Douglas points of $SU(3)$ were studied in detail in \cite{AF, GiveonRocek, CH}, and 
 \cite{Marino:Moore, RSVV} also discuss wall-crossing.
 In particular, Argyres-Douglas points are on the marginal stability walls \cite{ShapereVafa,GMNwall} and we will give concrete examples of how BPS dyon charges of vanishing cycles change discretely across Argyres-Douglas points for the $Sp(4)$ case.

For pure $Sp(2r)$ theory, \cite{DSW} proposed $r+1$ candidates for maximal Argyres-Douglas theories. Here similarly, we propose $2r-1$ candidates for maximal Argyres-Douglas points of pure $SO(2r+1)$ theory. Note that these numbers also match dual Coxeter number of the gauge group. Scaling behaviour at maximal Argyres-Douglas points for pure ABCD SW theory were studied in \cite{EHIY} and \cite{EH} (recently \cite{GST} pointed out some subtleties), and there are two such points in moduli space for A and D groups ($SU(r+1) $ and $SO(2r)$, whose maximal Argyres-Douglas points we discuss again in \eqref{SUMaxAD} and \eqref{SOevenMaxAD}). For B and C ($SO(2r+1)$ and $Sp(2r)$), scaling behaviour is also being studied in \cite{Seo} in the spirit of \cite{GST}.

\subsubsection{Motivation to look at $Sp(2r)$ gauge group from F-theory\label{SpMot}}

\begin{figure}[htb]
        \begin{center}
\includegraphics[width=.45\textwidth]{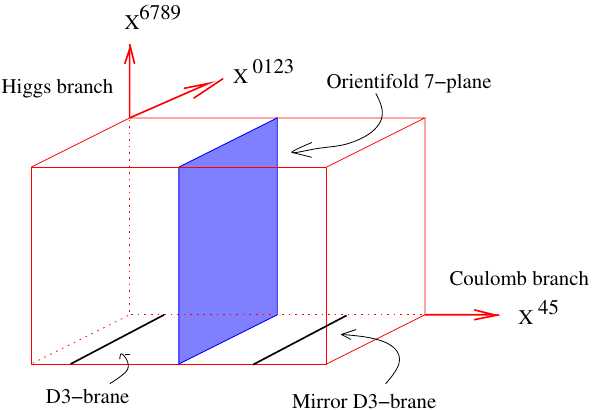} \includegraphics[width=.45\textwidth]{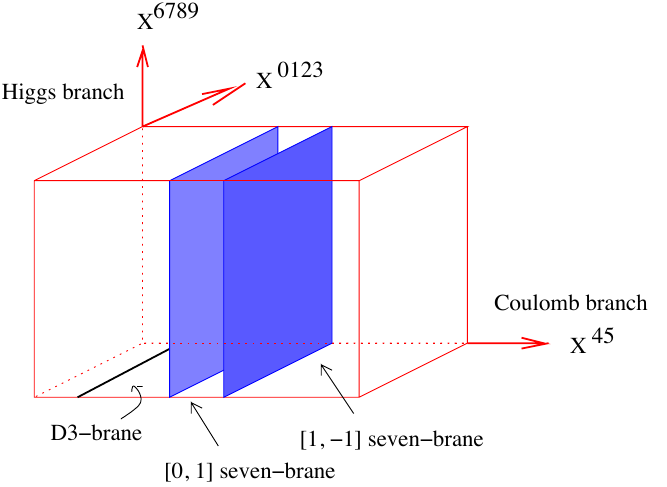}
        \caption{{In terms of D3-branes probing 7-branes, F-theory describes the $Sp(2r)$ SW theory with an antisymmetric matter.}}
        \label{o7pq7}
        \end{center}
        \end{figure}

Another physical motivation to study $Sp(2r)$ curve comes from the F-theory lift of SW theories and D3/O7 brane construction.   
Original $SU(2)$ SW theory has an F-theory interpretation in terms of a D3 brane probing quantum corrected O7-plane geometry \cite{Sen1O7split, BDS1D3}\footnote{Dirichlet branes and Orientifold planes are boundary conditions of open string theory. The number next to D and O denotes the spatial dimensions of the object. By $SL(2,{\mathbb{Z}})$ transformation of D7-brane, we can obtain $[p,q]$-7-branes.}. F-theory breaks the O7-plane
into two $[p,q]$-7-branes as in {\bf figure \ref{o7pq7}}. 

When we have $r$ D3 branes
probing the same geometry, we obtain $Sp(2r)$ gauge theory \cite{DLSmultiD3, ShapereTachikawa}, 
which is captured by $Sp(2r)$ SW theory with an antisymmetric matter \cite{AMP}. 
 In the rank 1 case the anti-symmetric traceless hypermultiplet matter is null, and it reduces to original Seiberg-Witten theory of $SU(2)=Sp(2)$ gauge group (with some flavors given by D7 branes). 
 For rank 2 or higher, one would need to use the curves given in \cite{AMP} which take into account the anti-symmetric matter with arbitrary mass.
 
 It was asked in \cite{DSW} what the effects of adding extra probe D3-branes are on 7-brane dynamics, and more specifically whether O7-plane still splits into the same two 7-branes. It also motivates one to see explicit duality and 1-1 mapping between the F-theory picture and the SW theory. 
 As a first step toward this goal, we will study the pure SW theory for simplicity.

     \subsection{Geometry review\label{GeomReview}}

In \eqref{Zlambda}, we learned that various (vanishing) 1-cycles of SW curve are related to (massless) BPS states of SW theory. 
Here we will review some geometric tools. 
 
\begin{figure}[htb]
   \begin{center}
        \includegraphics[width=.7\textwidth]{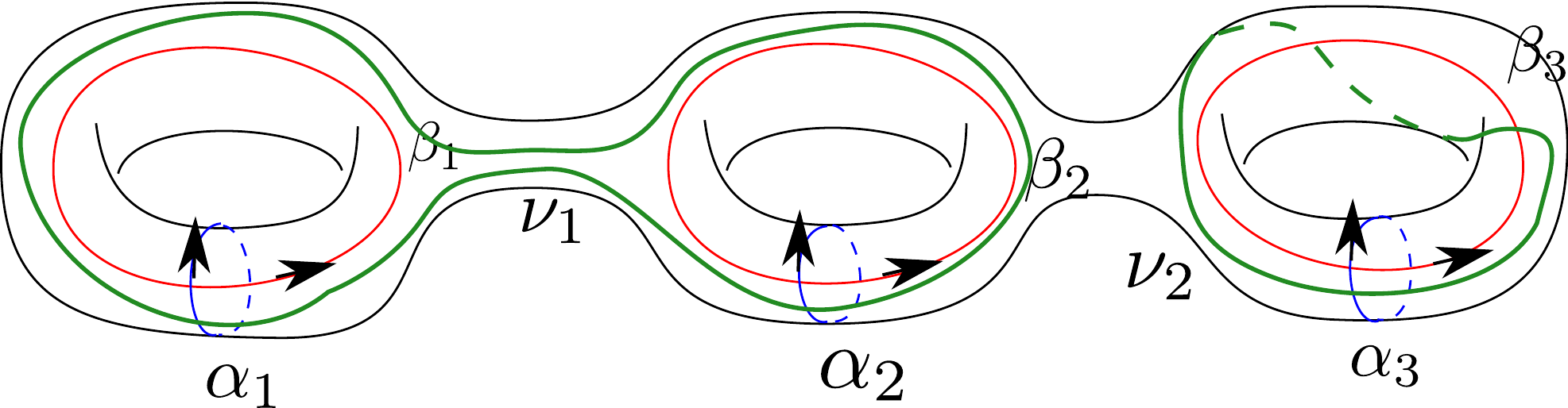}
   \caption{Various 1-cycles and their symplectic basis for a Riemann surface of genus 3}
      \label{nu}
    \end{center}
   \end{figure}

 On the surface of a Riemann surface (e.g. SW curve), we can draw various 1-cycles as in {\bf figure \ref{nu}}.
Symplectic basis for 1-cycles are given by ${\color{blue}\alpha_i}$ and ${\color{red}\beta_i}$
whose non-vanishing intersection numbers are ${\color{blue}\alpha_i } \cap {\color{red}\beta_j} =\delta_{ij}$.

   Any 1-cycle can be written in terms of symplectic base 1-cycles ${\color{blue}\alpha_{i}}, {\color{red}\beta_i}$'s, with integer $\mathbb{Z}$ coefficients, (which are interpreted as quantized magnetic and electronic charges for dyon states). For example, in {\bf figure \ref{nu}}, we draw two 1-cycles $\nu_{1,2}$ which can be written in terms of basis 1-cycles as follows:
    \begin{equation}
  {\color{forestgreen}\nu_1} = {\color{red}\beta_{1}} +{\color{red}\beta_{2}} , \qquad
    {\color{forestgreen}\nu_2} = -{\color{blue}\alpha_{3}} +{\color{red}\beta_{3}} .
    \end{equation}
 A BPS state is associated with a 1-cycle, with $\mathbb{Z}$ coefficients giving electric (${\color{blue}\alpha}$) and magnetic (${\color{red}\beta} $) charge for each $U(1)$.  Its crucial to note that not all 1-cycles are paired up with BPS states. They need to pass at least the test of wall-crossing formula. Thus not all states are BPS states, 
and the BPS property is determined by the marginal stability, namely by comparing the masses $-$ which is then just the energy argument. 
   
Among 1-cycles, there exists an anti-symmetric and bilinear operation of taking intersection numbers. Since each cycle has an orientation (as seen by the arrow in figure), the contribution comes with a sign. When the intersection number is (non) zero, we say two cycles are mutually (non) local. For example,  
\begin{itemize}
\item mutually local pairs are: $   {\color{blue}\alpha_{1,2}} \cap    {\color{forestgreen}\nu_2}=0, \quad    {\color{red}\beta_{1,2}} \cap    {\color{forestgreen}\nu_2}=0$, $   {\color{forestgreen}\nu_1} \cap    {\color{forestgreen}\nu_2} =0, \quad    {\color{forestgreen}\nu_1} \cap    {\color{red}\beta_{1,2,3}}=0, \quad     {\color{forestgreen}\nu_1} \cap    {\color{blue}\alpha_3} =0 $,
\item mutually non-local pairs are: $   {\color{forestgreen}\nu_1} \cap  {\color{blue}\alpha_{1,2}}= -1, \quad    {\color{forestgreen}\nu_2} \cap {\color{blue} \alpha_3} =-1, \quad    {\color{forestgreen}\nu_2} \cap {\color{red}\beta_{3}}=1$.
\end{itemize}

\begin{figure}[t]
   \begin{center}
        \includegraphics[width=.7\textwidth]{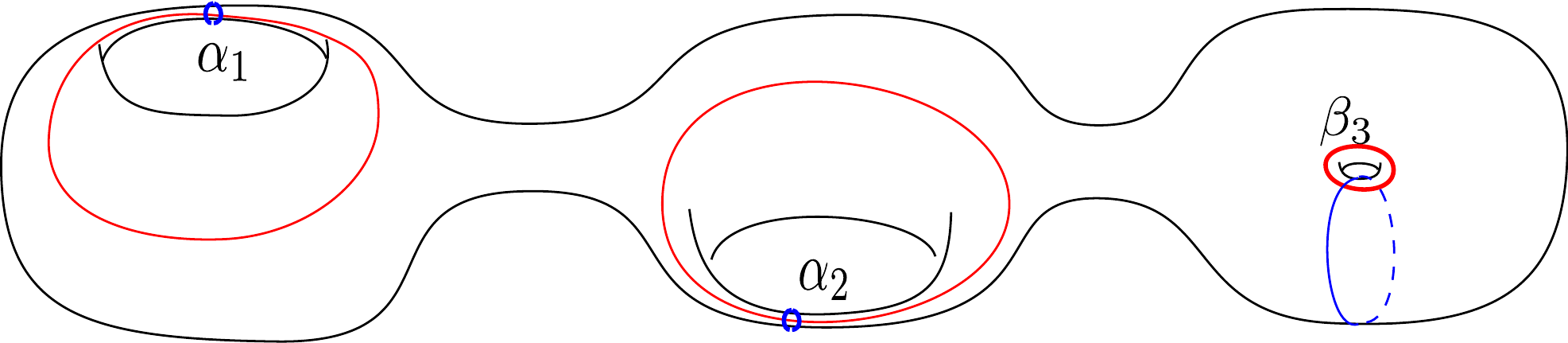}
    \end{center}
    \caption{Vanishing cycles of genus-3 Riemann surface. All these 3 cycles are mutually local, since intersection numbers all vanish.}
    \label{local}
   \end{figure}

\begin{figure}[t]
   \begin{center}
   \includegraphics[width=.7\textwidth]{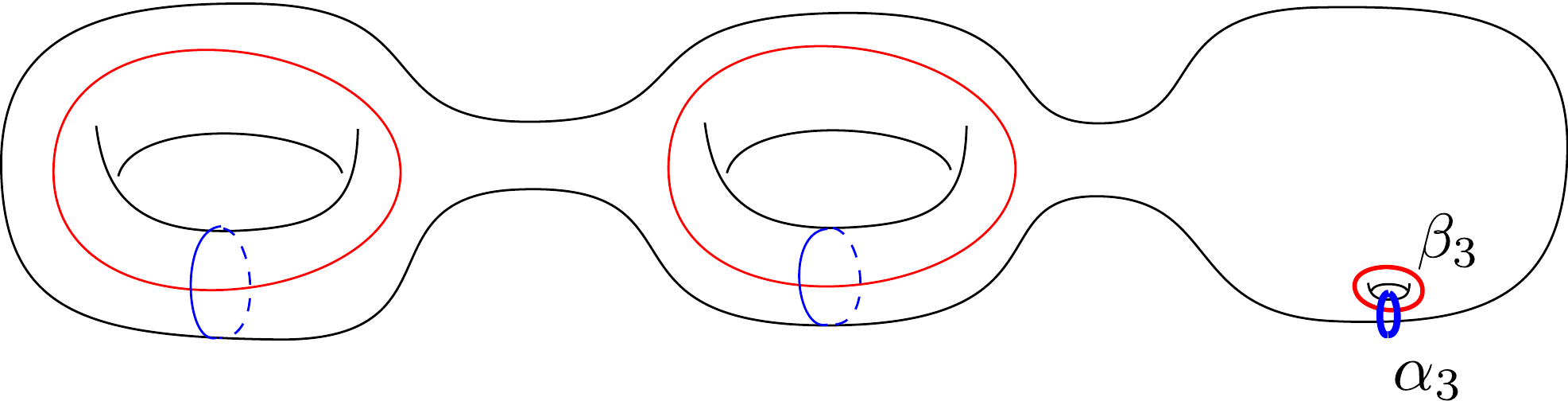}
       \caption{Mutually non-local vanishing cycles of genus-3 Riemann surface. Their intersection number is non-zero.}
    \label{nonlocal}
     \end{center}
   \end{figure}

In pure Seiberg-Witten theory the number of moduli is related to the genus which, in turn, is related to the rank of the gauge group. These moduli are allowed to vary.
At a generic point in the moduli space, the SW curve is smooth and all the 1-cycles are non-vanishing. However, we could tune to less generic location in the moduli space where we have vanishing 1-cycles. 
In {\bf figure \ref{local}}, we show examples of three vanishing 1-cycles. Even if 1-cycles vanish, we still can think of intersection numbers among them. In {\bf figure \ref{local}},
 all the vanishing cycles $\alpha_1, \alpha_2, \beta_3$ are mutually local. 
  However, in {\bf figure \ref{nonlocal}},    
   two vanishing cycles are mutually non-local since $   {\color{blue}\alpha_3} \cap {\color{red}\beta_{3}}=1\ne 0$.

Recall that vanishing 1-cycles correspond to massless BPS states in physics.  
Now we can interpret vanishing cycles in {\bf figures \ref{local}} and {\bf \ref{nonlocal}} in terms of massless BPS states in physics\footnote{Of course, here we are assuming that all vanishing 1-cycles correspond to BPS states. Our belief is that it will pass the wall-crossing test, because it costs no energy. It will be interesting and important to test this assumption.}.

In {\bf figure \ref{local}},  
   three vanishing cycles correspond to mutually local massless BPS states.
  All the massless BPS states can be made electronic by proper electric-magnetic (e-m) duality, and one can write down Lagrangian by adding each local pieces.

On the other hand in {\bf figure \ref{nonlocal}}, two vanishing cycles correspond to two massless BPS dyons which are mutually non-local. In other words, they correspond to massless electron and magnetic monopole for the same $U(1)$. 
Lagrangian description does not exist, but this nontrivial theory has been studied by \cite{ArgyresDouglas} in $SU(r+1)$ SW theories. When maximal number of mutually non-local BPS dyons become massless, it is called maximal Argyres-Douglas theory, and for pure $Sp(2r)$ it was solved in \cite{DSW}.

\subsubsection{Tools for singularity search: exterior derivative and discriminant \label{EXandDisc}} 
 Recall a standard fact that an algebraic variety given by $F=0$ has singularity where exterior derivative vanishes i.e $F=dF=0$ (and the tangent space becomes sick). 
To apply this to the hyperelliptic curve, let us  write down Riemann surface as the following complex curve: 
\begin{equation}\label{ccurve}
y^2 = f(x; u_1, u_2, \cdots, u_r),
\end{equation}
where $u_i$'s are coordinates along the moduli space. The function       
 $ F(x,y)  \equiv y^2 - f(x)=0$ becomes singular where $F=dF=0$. The exterior derivative $d$ can be written in terms of the partial derivatives with respect to all the coordinates. In this case, the Riemann surface is embedded in a space spanned by $x$ and $y$. Therefore the exterior derivative is given as $d= dx \frac{\partial  }{\partial x} + dy \frac{\partial  }{\partial y}$, and $dF=0$ is equivalent to having $\frac{\partial F}{\partial x}=\frac{\partial F}{\partial y}=0$. (Later, we will take exterior derivative inside the moduli space spanned by coordinates $u_i$'s, then $d= \sum_i du_i \frac{\partial  }{\partial u_i} $.) 
 The relations $F=\frac{\partial F}{\partial x}=\frac{\partial F}{\partial y}=0$ boil down to $f(x)$ having repeated roots, namely vanishing discriminant $\Delta_x f =0$.
    
So far we have considered $f(x; u_1, u_2, \cdots, u_r)$ to be a generic function expressed in terms of polynomials of certain degrees. We can make this a bit more 
precise. Consider 
$f_n(x)$ to be a degree-$n$ polynomial in $x$. It can be factorized in terms of its $n$ roots $e_i$'s as below:
\begin{equation}
f_n(x)=\sum_{i=0}^n a_i x^i = a_n \prod_{i=1}^n (x-e_i). \label{polynomialF}
\end{equation}
Its discriminant $\Delta_x f_n$, expressed in terms of its roots and denoted as:
\begin{equation}
\Delta_x \left( f_n(x) \right) =  a_n^{2n-2} \prod_{i<j} (e_i-e_j)^2 \label{DiscDef}
\end{equation} 
 vanishes if and only if the polynomial has a multiple (i.e repeating or degenerate) root. 
Because of the squaring, the discriminant is symmetric in the roots. Therefore, the discriminant can also be expressed in terms of the coefficients of the polynomial $a_i$'s.
With no loss of generality we can set the coefficient of the highest degree term of $f_n(x)$ to be $a_n=1$. We observe that $\Delta_x \left( f_n(x) \right)$ is a polynomial in terms of $a_i$'s, with a power of $a_0$ to be $n-1$, and the power of $a_i$ to be $n$ for $i\ne0, n$. 

As examples, we display discriminants of polynomials with degrees in $x$ to be 2, 3, and 4 below:
\begin{eqnarray} 
\Delta_x (ax ^2+bx+c) &=& {\color{blue} b^2} -4 a{\color{red}  c}, \label{DiscDegExamples}\\
\Delta_x (ax ^3+bx^2+cx+d) &=& b^2  c^2 -4 a {\color{blue} c^3}-4{\color{blue} b^3 d} +18abcd - 27 a^2 {\color{red} d^2 }, \nonumber \\
\Delta_x (ax ^4+bx^3+cx^2+dx+e) &=& b^2 c^2 d^2 -4ac^3d^2 -4b^3 d^3 +18abcd^3- 27 a^2 {\color{blue} d^4}\nonumber \\
&&-4b^2 c^3 e +16 a{\color{blue} c^4}e+18b^3cde\nonumber \\ 
&& -80 abc^2 de -6ab^2 d^2 e+ 144 a^2 c d^2 e- 27{\color{blue} b^4 }e^2 \nonumber \\
&& +144 ab^2 ce^2   -128 a^2 c^2 e^2   -192 a^2 bd e^2
 + 256 a^3 {\color{red}  e^3}. \nonumber
\end{eqnarray}
Note that for each coefficient, the highest degree term in the discriminant has the power which is same as the degree of the polynomial. The only exception is the coefficient of the constant piece, whose highest degree term in the discriminant has the power one less than the degree of the polynomial in $x$. (Top degree coefficient $a$ has a slightly different behaviour, but it is fixed to be $1$ in our cases, so we can ignore that.) To summarize, we can write this as 
 \begin{eqnarray}   
\Delta_x f_{n}(x)  &=& \#  \left( a_{i}^{n} + \cdots \right) ,\quad i\ne 0, n    \nonumber \\
   &=&  \# \left( a_0^{n-1} + \cdots \right),   \label{DiscDeg}
\end{eqnarray}  
where $f_n(x)$ was given in \eqref{polynomialF}. This denotes that if we write $\Delta_x f_{n}(x) $ as a polynomial in $ a_{i}$, then the top degree will be $n$ ($n-1$ resp.) for $i \ne 0, n$ ($i=0$ resp.).

If we are solving $\Delta_x  \left( f_n(x) \right)=0$ in terms of $a_0$ (the constant piece in the polynomial), we will have $n-1$ solutions. If we solve for $a_i$ ($i\ne0$), then we will have $n$ solutions. On the vanishing discriminant loci, $a_0$ will be not as good as other moduli as a coordinate. We will give a simple example for $f_2(x)$ in the next paragraph.  

For an algebraic curve given as $y^2=f(x)$, discriminant of the right hand side $\Delta_x f$ of the the relation is also called in short as a discriminant of the curve. Singularity of the curve is captured by colliding roots on the $x$-plane, at vanishing discriminant $\Delta_x$.  
As a simple example, an algebraic curve given as 
\begin{equation}
y^2 = f_2(x) = x^2 +a_1 x +a_0 
\label{simplecurve}
\end{equation}
has a discriminant 
\begin{equation}
\Delta_x \left( x^2 +a_1 x +a_0  \right) = {a_1}^2 -4 {a_0}. \label{degree2}
\end{equation}
In the complex $2$ dimensional moduli space of the curve \eqref{simplecurve} spanned by $a_1$ and $a_0$, complex codimension 1 locus given by $\Delta_x f_2 =a_1^2 -4 a_0=0$ specifies where the curve becomes singular. Note that $a_1$ makes a good coordinate of $\Delta_x f_2 = 0$ singular locus, however $a_0$ cannot fully act as one. Specifying the value of $a_0$ is not enough to pin down the value of $a_1$, the sign ambiguity still remains for $a_1$. This is related to the fact that under the $\mathbb{Z}_2$ transformation of the curve given as $x \rightarrow -x$, moduli transform as $a_1\rightarrow -a_1, a_0\rightarrow a_0$. In other words, $a_0$ is blind to this $\mathbb{Z}_2$ transformation. Later in subsection \ref{blind}, we will explain similar phenomenon for more complicated curves: One of the moduli (again the constant piece) is less useful as a coordinate on vanishing discriminant loci. 
Among $r$ moduli $u_i$'s of rank $r$ curve of $SU(r+1)$ ($Sp(2r)$ resp.) gauge group, $u_r$ is blind to $\mathbb{Z}_{r+1}$ ($\mathbb{Z}_{r}$ resp.) phase rotation on the $x$-plane, making it less useful as a coordinate on $\Delta_x f_n =0$ loci. (Of course this is due to the property explained in \eqref{DiscDeg} earlier.)

Note that we keep a subscript for the discriminant symbol, as a reminder of which variable we take discriminant with respect to. This will be useful when we have a polynomial in multiple variables.
For example, the following polynomial in $x$ and $y$
\begin{equation}
g(x,y) = (x-3y)(x-y) = 3(y-x)(y-x/3)
\end{equation}
has two possible discriminant operators $\Delta_x$ and $\Delta_y$, whose actions are given as from \eqref{DiscDef}
\begin{equation}
\Delta_x g = 1^{2 \cdot 2-2} (3y -y)^2 = 4y^2, \qquad \Delta_y g = 3^{2 \cdot 2-2} (x -x/3)^2 = 9 \cdot \frac{4}{9}x^2 = 4x^2.
\end{equation}
If we have repeated roots, it is reflected by vanishing of \eqref{DiscDef}.
The number of roots repeated is called the degeneracy, multiplicity of zero, or
order of vanishing. 
Later we will not only ask whether a discriminant vanishes, but also how fast it vanishes. The order of vanishing of double discriminant will play an important role in distinguishing Argyres-Douglas loci.

\subsubsection{Cusp-like singularity and Argyres-Douglas theory \label{CuspAD}}
   It's nothing new that a repeated root gives singularity. For example, $y^2=x^n$ ($n\ge 2$) is singular at the origin, because, $x=0$ is a repeated root with various degeneracy. 
For example, $y^2=x^2$ has a node like singularity, while 
 $y^2=x^3$ (and higher, $n\ge 3$) has a cusp like singularity.

Similarly,  
 the hyper-elliptic curve $y^2 = f(x)$ degenerates into a cusp form
\begin{equation}
y^2 = (x-a)^m \times \cdots,
\end{equation} 
 when $m\ge 3$ branch points collide on $x$-plane. We will argue in this paper, that this corresponds to an Argyres-Douglas theory, with {\emph {mutually non-local}} massless dyons. Each time, we demand a branch point to collide with another, we are using up a degree of freedom. Therefore, by demanding 3 branch points to collide all together, we use up two degree of freedom. This means that the Argyres-Douglas theory occurs in ${\rm codim}_{\mathbb{C}}$-$2$ loci, and so one would expect it to occur in SW theory 
with ${\rm dim}~ {\cal M} \ge 2$.  This explains why Argyres-Douglas theory was found for $SU(n)$ SW theory for rank 2 or higher originally
 in \cite{ArgyresDouglas}.
 When $m$ is maximized, this is the called maximal Argyres-Douglas point. 
  For pure case, the degree of freedom is $r$, and we can bring $r+1$ branch points together by using up all the degrees of freedom. Therefore there will be maximal Argyres-Douglas points in the moduli space where $r+1$ branch points collide, forming a strong cusp.

Scaling behaviour for maximal Argyres-Douglas points of pure SW curves with ABCDE gauge groups were studied in \cite{EHIY}, and it is easy to find 
2 points in moduli space which are maximal Argyres-Douglas for ADE groups.   
Recently, the $r+1$ candidates for the maximal Argyres-Douglas points were given for $C_r=Sp(2r)$ SW theory in \cite{DSW}. In subsection \ref{SOoddADmax}, we will show 
the $2r-1$ maximal Argyres-Douglas points for $B_r=SO(2r+1)$ SW theory. 

Our aim 
in this paper therefore is to provide a recipe for locating Argyres-Douglas loci in the moduli space of pure SW theory. The procedure that we will follow can be 
expressed in few easy steps.  
First we start with hyper-elliptic Seiberg-Witten curve $y^2=f(x;u,v,\cdots)$. Then demanding $\Delta_x f=0$ and $d (\Delta_x f)=0$ gives two massless BPS dyons.
Note that here $\Delta_u \Delta_x f=0$ too. Once we have this, our next step would be to
check the {\bf o}rder {\bf o}f {\bf v}anishing (o.o.v.) of each solution from $\Delta_u \Delta_x f=0$.
If o.o.v. $\ge$ 3, we have our Argyres-Douglas loci, namely: the hyperelliptic curve degenerates into a cusp-like singularity $y^2=(x-a)^3 \times \cdots$
and two mutually non-local dyons become massless. We will apply this technique empirically by checking it up to rank 5 for the pure $Sp$ and $SU$ SW theories.
 
 \subsection{How to read off monodromies of the Seiberg-Witten curves \label{monodromySec} }

     \begin{figure}[htb]
\begin{center}
\includegraphics[
width=4.3618in 
]{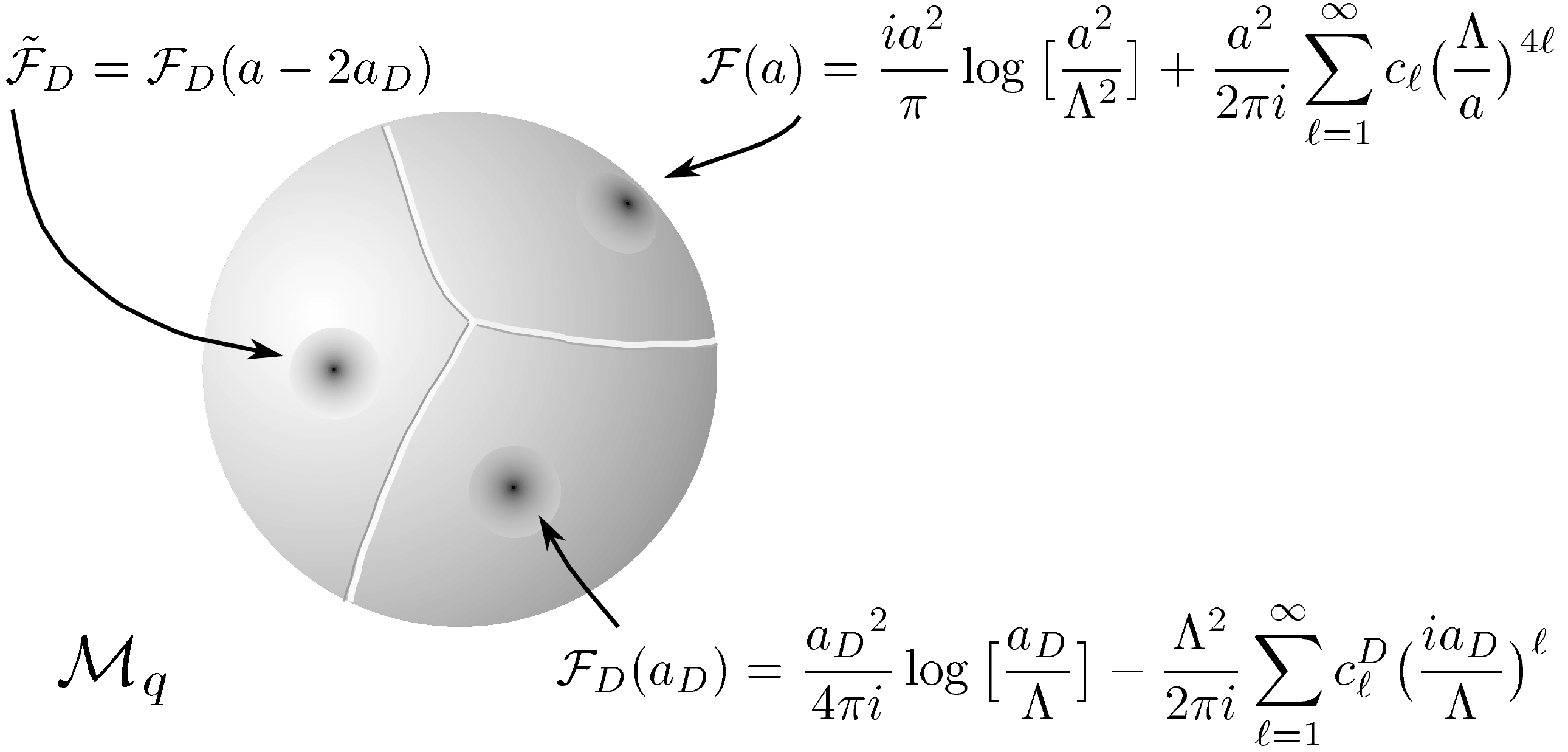}
\end{center}
\caption{Three special points on the moduli space of pure $SU(2)$ Seiberg-Witten theory. The top right point corresponds to singularity at $u \rightarrow \infty$, and the given expression for prepotential ${\cal F}$ is valid for large $u$. The bottom right point corresponds to a magnetic monopole point at $u=\Lambda^2$, and the left point corresponds to a dyon point at $u=-\Lambda^2$.
This figure is again taken from Lerche's review article \cite{LercheReview}.}
\label{rank1M}
\end{figure}

In the literature, Seiberg Witten curves for $SU(2)$ are written in multiple
forms. In \cite{SeibergWittenNoMatter}, pure $SU(2)$ theory is discussed and the Seiberg-Witten curve 
\begin{equation}
y^2 = (x-\Lambda^2)(x+\Lambda^2)(x-u) \label{origSW}
\end{equation}
was introduced. In \cite{SeibergWittenWithMatter}, $SU(2)$ theories with flavors are discussed, whose low energy limit is the pure $SU(2)$ theory with Seiberg-Witten curve  
\begin{equation}
y^{2}=x\left( x(x-u)+\frac{1}{4}\Lambda ^{4}\right).   \label{sp2SW}
\end{equation}
$Sp(2r)$ curve in \cite{ArgyresShapere} reduces to this form \eqref{sp2SW} at rank 1, therefore let us denote this curve as $Sp(2)$ curve. 

 \begin{figure}[htb]
\begin{center}
\includegraphics[
width=4.3618in 
]{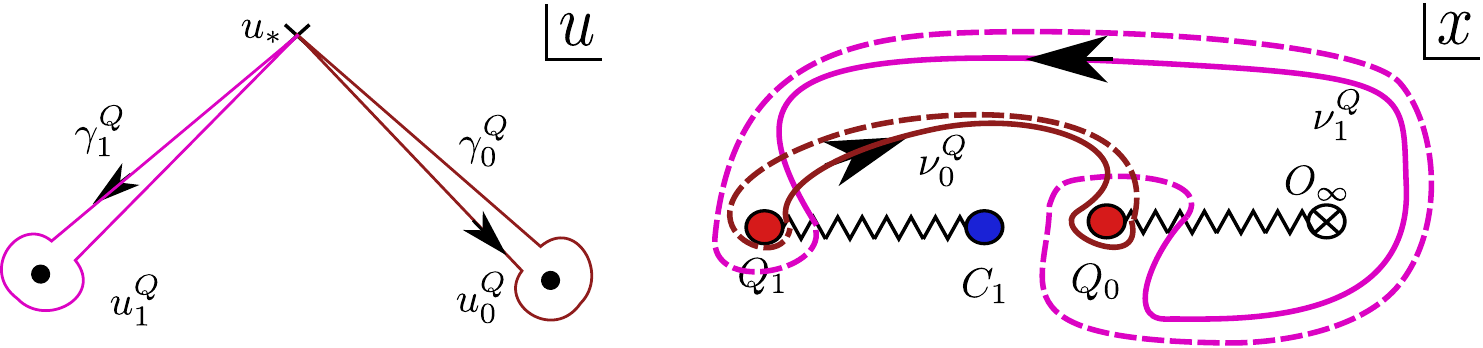}
\end{center}
\caption{Vanishing cycles for pure $Sp(2)$: For Seiberg Witten curve $y^{2}=x\left( x(x-u)+\frac{1}{4}\Lambda ^{4}\right)$, discriminant vanishes at $u=\pm \Lambda^2$. As $u$ varies on $u$-plane on red and purple paths given, the branch points will move along the path given in respective colors on $x$-plane. Each purple and red cycles on the right correspond to massless monopole and dyon of $Sp(2)$ theory.}
\label{sp1}
\end{figure}

On the other hand, \cite{KLYTsimpleADE} proposed another form (so called $SU(2)$
curve)
\begin{equation}
y^{2}=(x^{2}-u)^{2}-\Lambda ^{4}=(x^{2}-{u+\Lambda ^{2}})(x^{2}-{u-\Lambda
^{2}}),  \label{su2Lerche}
\end{equation}
which has an advantage of immediate generalization into ADE series \cite{KLYTsimpleADE}.

\begin{figure}[htb]
\begin{center}
\includegraphics[
width=4.1517in 
]{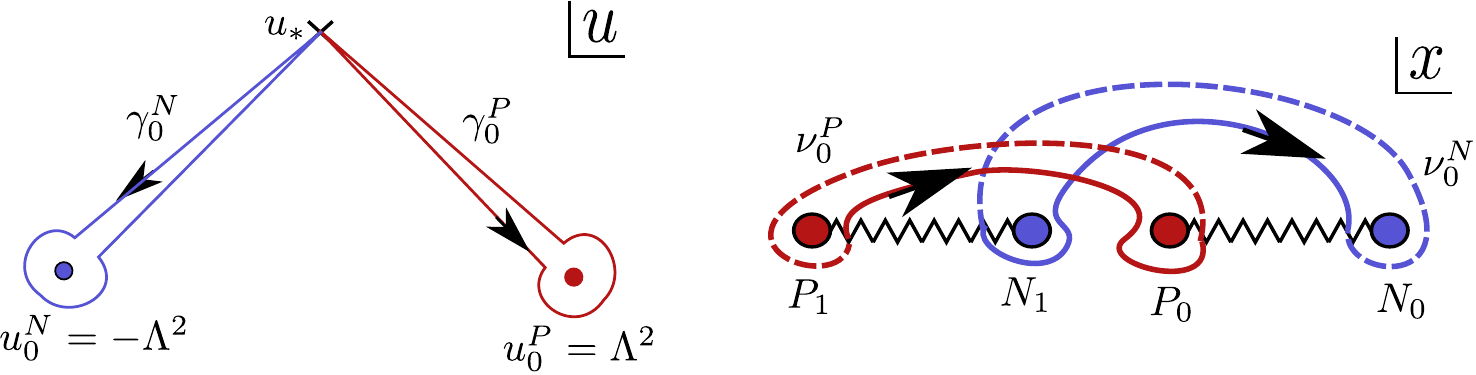}
\end{center}
\caption{Vanishing cycles for pure $SU(2)$: For Seiberg Witten curve $y^{2}=(x^{2}-u)^{2}-\Lambda ^{4}=(x^{2}-{u+\Lambda ^{2}})(x^{2}-{u-\Lambda^{2}})$, discriminant vanishes at $u=\pm \Lambda^2$. As $u$ varies on $u$-plane on red and blue paths given, the branch points will move along the path given in respective colors on $x$-plane. Each blue and red cycles on the right correspond to massless monopole and dyon of $SU(2)$ theory.}
\label{su2}
\end{figure}

Discriminant vanishes at $u=\pm \Lambda^2$ for all of three curves for pure $SU(2)$ theory given by \eqref{origSW}, \eqref{sp2SW}, \eqref{su2Lerche}.
In this paper, we discuss $SU(r+1)$ and $Sp(2r)$ curves, which reduce to \eqref{su2Lerche} and \eqref{sp2SW} respectively at rank 1. Therefore curves \eqref{su2Lerche} and \eqref{sp2SW} are particularly interesting to us, and their monodromy is explained briefly in {\bf figures \ref{su2}} and {\bf \ref{sp1}}. 

We will consider Seiberg-Witten curves for pure $Sp(2)=SU(2)$ theory written in two different forms given in \eqref{sp2SW} and \eqref{su2Lerche}. Let us name those curves $Sp(2)$ and $SU(2)$ curves respectively, given their generalizations into $Sp(2r)$ and $SU(r+1)$ curves. We will discuss their vanishing cycles in detail to demonstrate that
these two curves indeed capture the same physics. 

Soon we will demonstrate how to capture 
massless dyon and monopole at $\Delta=0$ locus of each curve.

\subsubsection{Review of $SU(2)$ monodromy \label{su2monodromy}}

\begin{figure}[htb]
\begin{center}
\includegraphics[
width=\textwidth
]{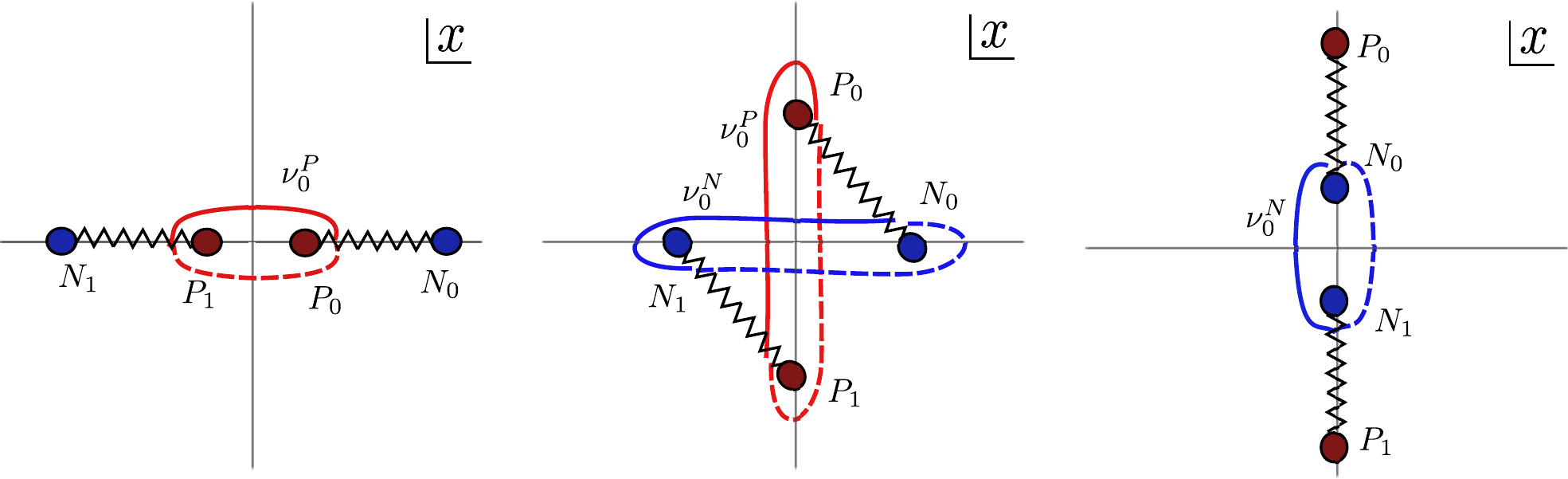}
\end{center}
\caption{Branch points move on $x$-plane, as we change the moduli of $u$ along the real axis. From the left, each figure corresponds to
different ranges of u value as  ${u\sim \Lambda ^{2},}$ $-{\Lambda ^{2}<}$ ${
u<\Lambda ^{2}~}$, and  ${u\sim \Lambda ^{2}}$. In the middle figure, all
the branch points are separated, and two cycles are drawn which correspond
to a monopole and a dyon. As we vary the value of $u$, each cycle can vanish
at a point in the moduli space (all drawn in consistent colors),
corresponding to either a monopole or a dyon becoming massless.}
\label{rank1su}
\end{figure}

As explained in \cite{KLYTsimpleADE}, the monodromy of $SU(2)$ curve captures
both massless dyon and monopole. The curve \eqref{su2Lerche} has four branch
points
\begin{equation}
 N_{0,1} = \pm \sqrt{u+\Lambda^2}, \qquad P_{0,1}=\pm \sqrt{u-\Lambda^2}, 
\end{equation} which are all distinct at a generic value of modulus $u$. 
 As we vary $u$, a different pair of branch points will collide at different
singular point in the moduli space: $N_0$ and $N_1$ collide as $u\rightarrow -\Lambda^2$ and $P_0$ and $P_1$ collide as $u\rightarrow  \Lambda^2$  as denoted with blue and red in {\bf figure
\ref{su2}}. Explicitly how this happens is described in {\bf figure
\ref{rank1su}}. As we change the moduli of $u$ along the real axis, we see
how branch points move on $x$-plane. From the left, each figure happens at
different values of $u$ as:
\begin{equation}
u\sim -\Lambda ^{2}, \qquad -\Lambda^{2}< u<  \Lambda^{2}, \qquad u\sim  \Lambda ^{2}
\end{equation} respectively.
In the middle of {\bf figure
\ref{rank1su}}, all
the branch points are separated, and two cycles are drawn which correspond
to a monopole and a dyon. As we vary the value of $u$ (moving to left and right figures), each cycle vanishes
at a point in the moduli space (all drawn in consistent colors),
corresponding to either a monopole or a dyon becoming massless as in {\bf figure \ref{rank1M}} \footnote{Note that what trajectory each cycle takes does matter. It is important
not only which two branch points are connected, but also through what
trajectory they are connected. Without this piece of information, we could
mis-identify vanishing cycles. For example, If in the middle picture, if
blue and red cycles were drawn so that they don't intersect, then they
together correspond to a dyon or a monopole. See subsection \ref{whytrajectory} more details along this line.}.

\subsubsection{$Sp(2)$ monodromy \label{sp1monodromy}}
\begin{figure}[htb]
\begin{center}
\includegraphics[
width=\textwidth
]{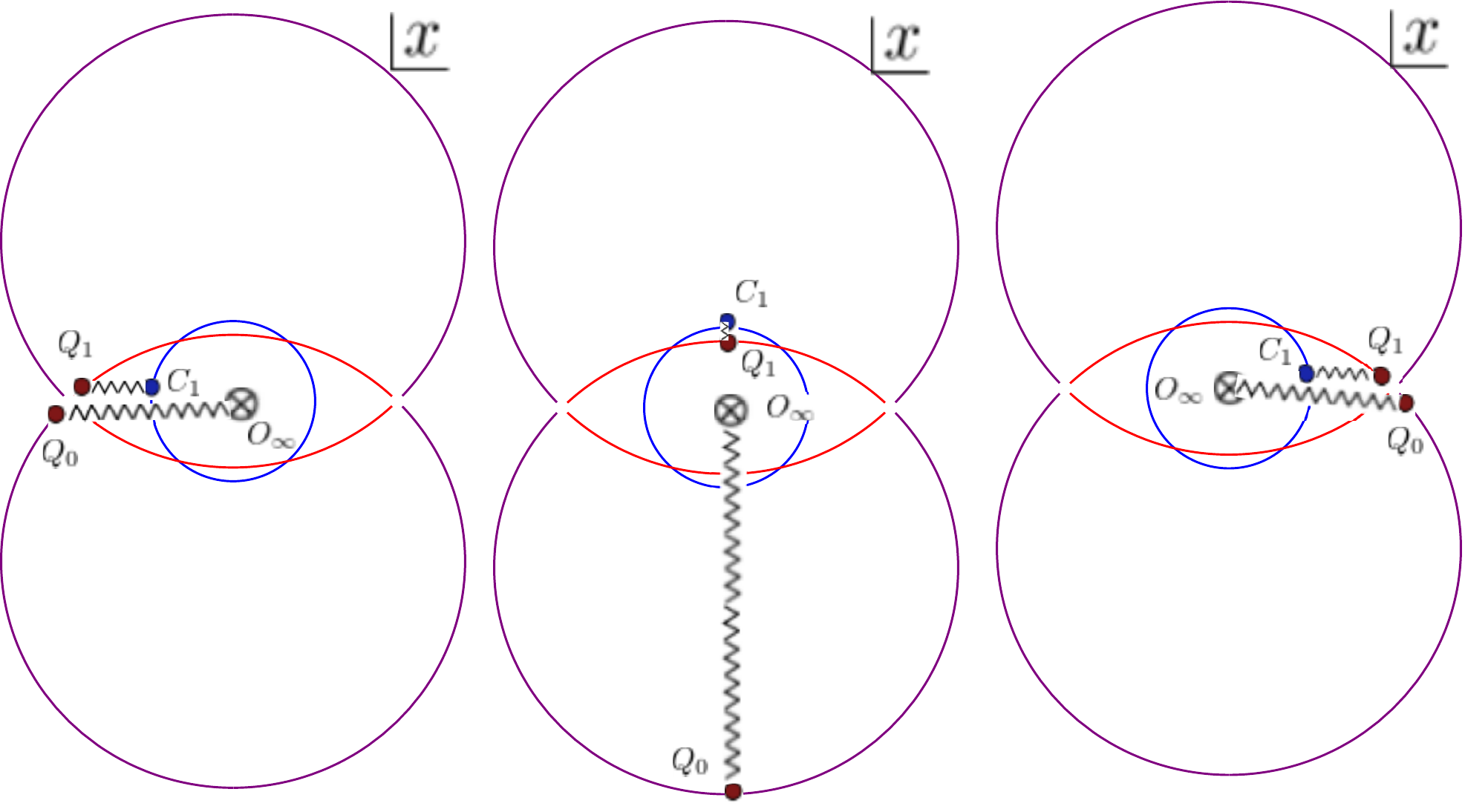}
\end{center}
\caption{For the $Sp(2)$ case the branch points moves around once we
change the phase of the moduli $u$. One may see that the behavior is slightly
different from the $SU(2)$ cases plotted earlier. Blue, purple, and red tracks are trajectories of branch points $C_1, \ Q_0$, and $Q_1$ respectively of \eqref{sp1fourpoints}, as we vary phase of $u$ while fixing its magnitude $|u|={\Lambda ^{2}}$. From the left, the values of $u$ are $\Lambda^{2}, \  i\Lambda^{2},  \ -\Lambda^{2}$ respectively. In other words, the phase of $u$ are $0,\ \pi/2, \ \pi$.}
\label{rank1spnoaxes}
\end{figure}

Now, let us examine the monodromy of the $Sp(2)$ curve of \eqref{sp2SW}. We
have total four branch points,
which are three at finite values of $x$, and
one at infinity. 
As in \cite{SeibergWittenWithMatter}, we can shift $x$ by $x\rightarrow x+u$ to obtain
\begin{equation}
y^{2}=(x+u)\left( x(x+u)+\frac{1}{4}\Lambda ^{4}\right).   \label{sp2SWshift}
\end{equation}
In order to bring a branch
point at infinity to the origin (so that it is easier to keep track of how
cycles change), perform $x\rightarrow 1/2x, y\rightarrow x^2 y$ transformation to obtain
\begin{equation}
y^{2}=x(1+2 u x)\left( 1 +2 ux+ \Lambda ^{4} x^2 \right),   \label{sp2SWshift2}
\end{equation}
whose four branch points are
\begin{equation}
O_{\infty}=0, \qquad  C_1=-\frac{1}{2u}, \qquad Q_{0,1}=-\frac{1}{\Lambda^2}\left(u\pm\sqrt{-\Lambda^2+u^2}\right).\label{sp1fourpoints}
\end{equation} 
 At generic value of $u$, all four branch points are separated, but as we vary $u$, they can collide with each other.  {\bf Figure \ref{rank1spnoaxes}} shows how branch points move on $x$-plane under
changing the phase of moduli $u$ while its magnitude is fixed at $|u|={\Lambda ^{2}}$.
Each of three non-zero branch points follows the track with the
corresponding color. Blue, purple, and red tracks are trajectories of branch points $C_1, \ Q_0$, and $Q_1$.

\begin{figure}[htb]
\begin{center}
\includegraphics[
width=\textwidth
]{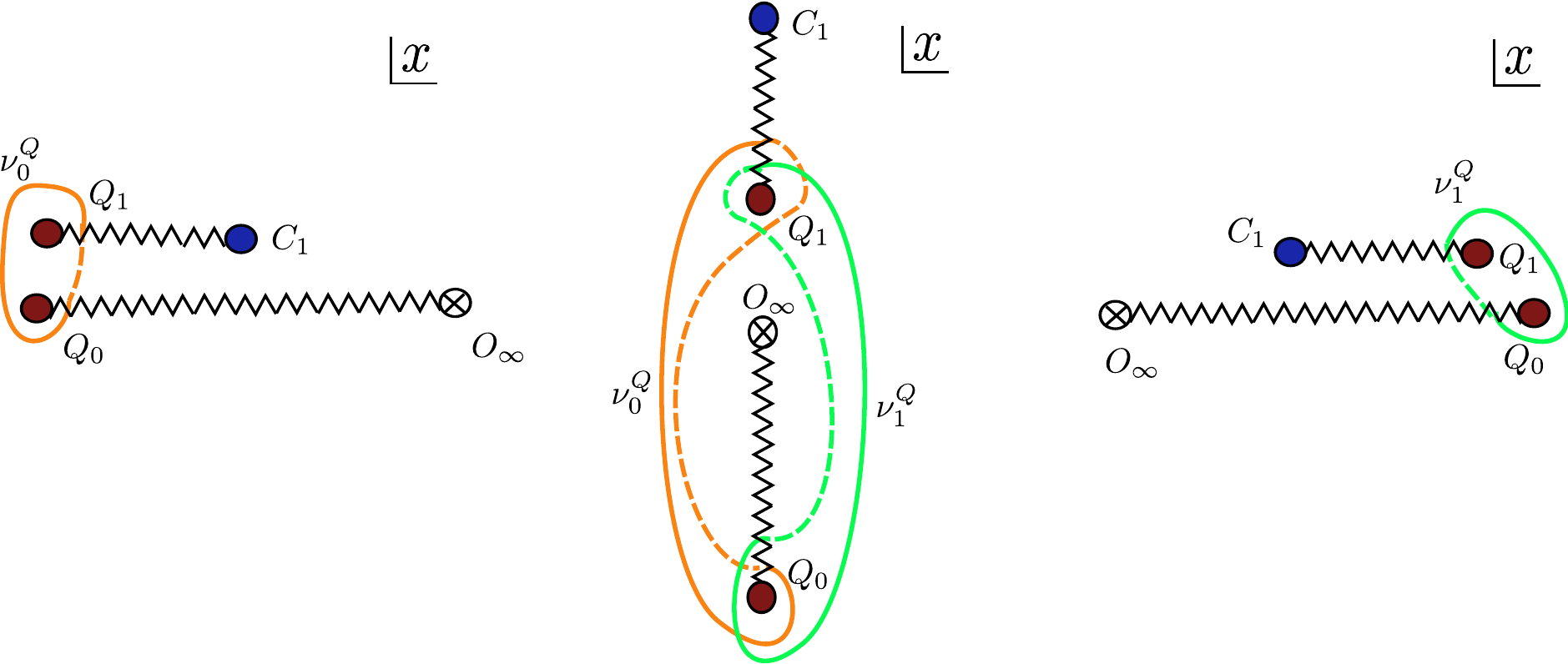}
\end{center}
\caption{This is similar to the earlier figure for the $Sp(2)$ case except that we magnify the
various branch points and the branch cuts. One may now easily compare the $Sp(2)$ story with the corresponding
$SU(2)$ case.}
\label{rank1sp}
\end{figure}

{\bf Figure \ref{rank1sp}} shows the magnified view of the
branch points and branch cuts. In the middle figure, we have two 1-cycles in
orange and green, which vanish in left and right figures respectively. As in
SU(2) case in {\bf figure \ref{rank1su}}, they again correspond to massless dyon and monopole at appropriate locations in moduli space depicted in {\bf figure \ref{rank1M}}.  Note that
each cycle connects the same pair of branch points, but through different
trajectories.

Note that the middle pictures of {\bf figures \ref{rank1su}} and {\bf \ref{rank1sp}} capture the vanishing cycles in the configuration. Instead of showing all the animations, we will only draw the middle picture for the rest of the paper. 

Furthermore, starting at a generic point in moduli space, the branch points on $x$-plane are separated. 
 As we move around a singular locus in $u$-plane, a pair of branch points approach. We 
can then easily read off BPS charge of vanishing cycles from the relative trajectory of branch points.

\subsection{Why we need information about trajectory of branch points \label{whytrajectory}}
\begin{figure}[htb]
\begin{center}
\includegraphics[
width=2.0375in 
]{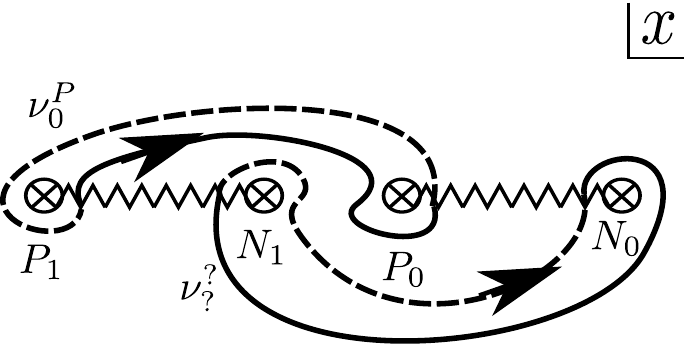}
\end{center}
\caption{If we connect the branch points with wrong trajectory, we get wrong
answers for vanishing cycles for $SU(2)$. Note that these two cycles shown here do not intersect, while monopole and dyon of $SU(2)$ SW theory have nonzero intersection number for their charges.}
\label{su2wrong}
\end{figure}

We argue here why it is essential to know the trajectory of the branch points on the $x$-plane in order to read off the vanishing cycles. In particular, it is not enough to know which two branch points are colliding with each other. Even if we choose different pair, it might still be the same cycle. Near {\bf figure \ref{su2}}, we explained how to obtain vanishing cycles for $SU(2)$ theory. Each colliding pair gave a different vanishing cycle. However, if we did not know what trajectory each root is taking, we might have gotten wrong cycles, as depicted in {\bf figure \ref{su2wrong}}.

\begin{figure}[htb]
\begin{center}
\includegraphics[
width=2.5in 
]{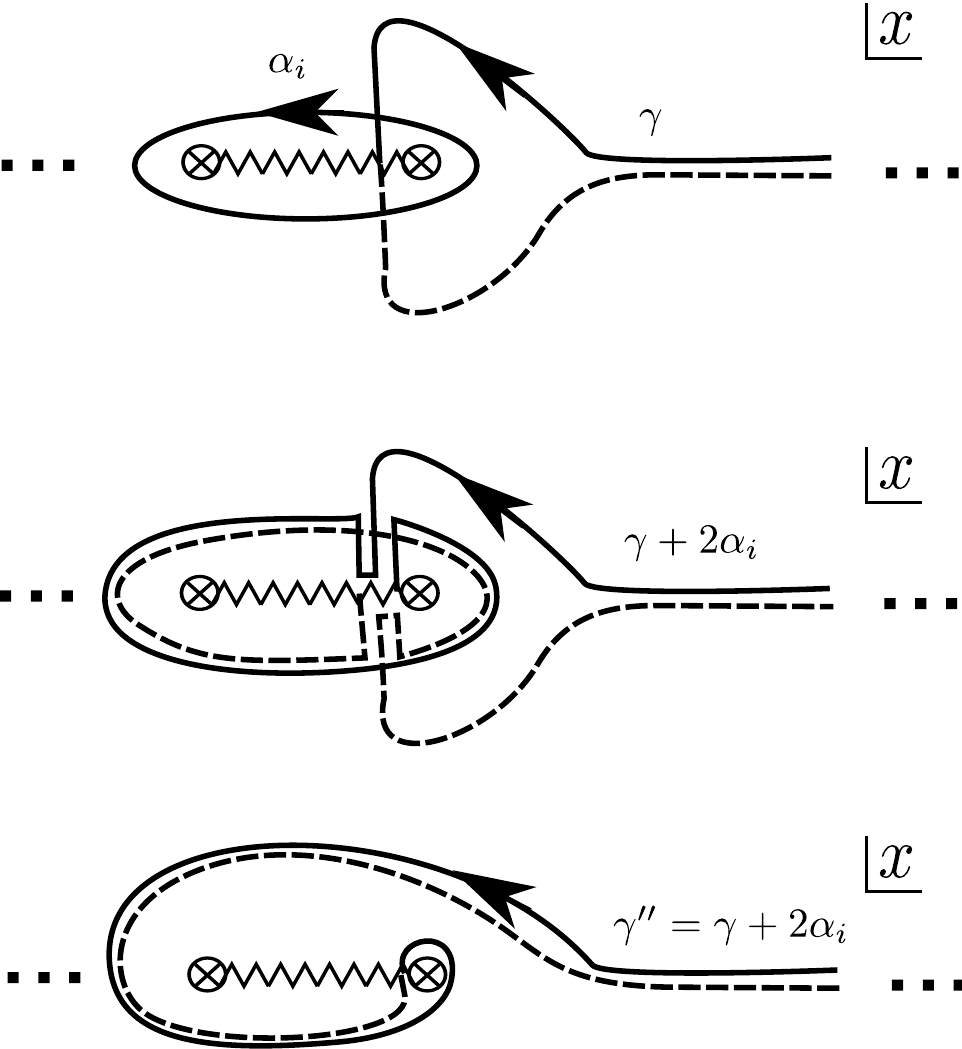}
\end{center}
\caption{Depending on which direction a cycle originates from a branch cut, we have an
ambiguity of $2n\alpha$'s.}
\label{rootaround}
\end{figure}

\begin{figure}[htb]
\begin{center}
\includegraphics[
width=3.3in
]{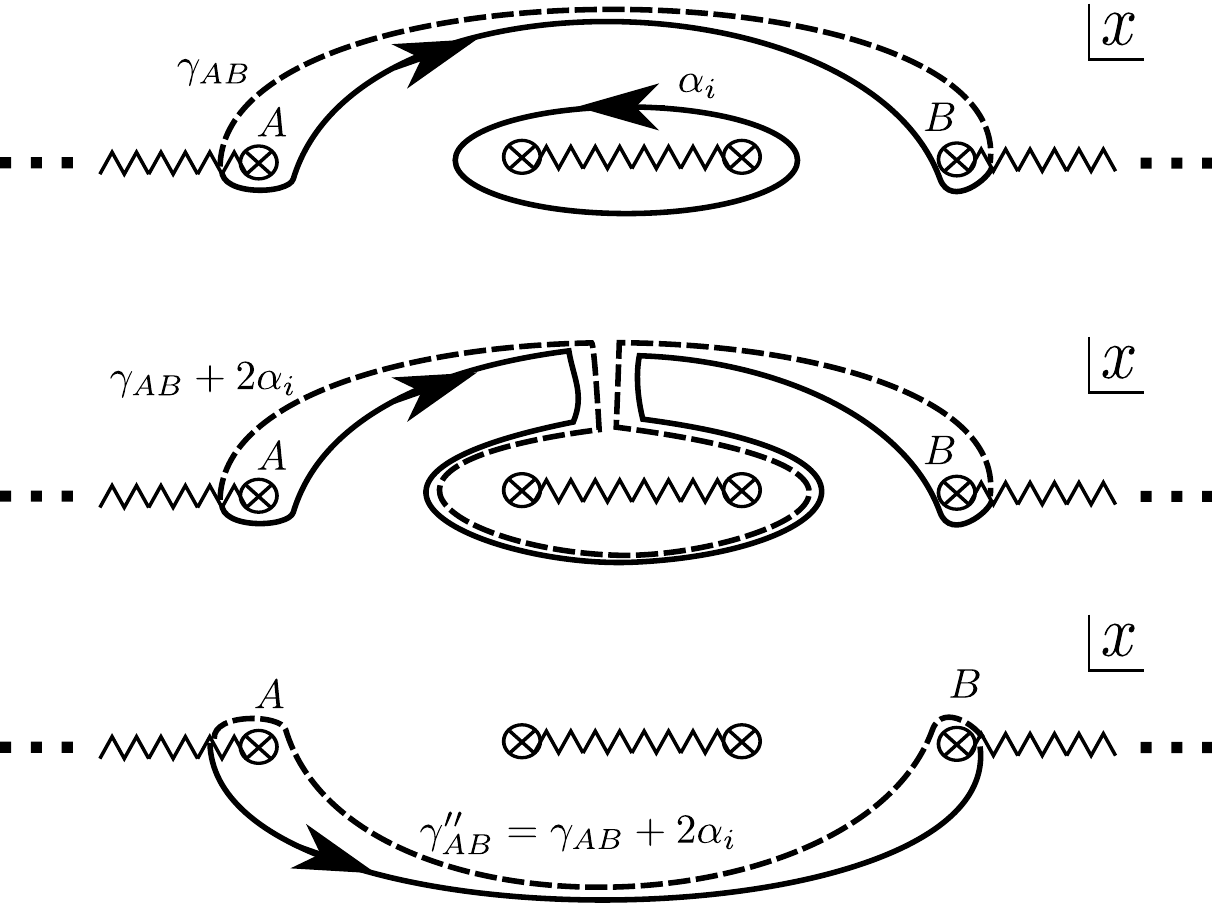}
\end{center}
\caption{Depending on how a cycle wraps around a branch cut, we have an
ambiguity of $2n\alpha$'s.}
\label{goaroundacut}
\end{figure}

However if we only know which two branch points are colliding, then there is an ambiguity in reading off vanishing cycles, up to $2\alpha_{i}$'s. {\bf Figure \ref{rootaround}} gives an example of wrapping a trajectory around a branch cut, which gives extra $2\alpha_i$ to the vanishing cycle. 
Similarly, the location of the trajectory relative to other branch cuts also changes the vanishing cycle as in {\bf figure \ref{goaroundacut}}.
The above examples were about ambiguity up to even number of $\alpha$ cycles. However, 
if we allow a trajectory to surround a branch cut, we can also get any integer number ambiguity of $\alpha$ cycles.
\begin{figure}[htb]
\begin{center}
\includegraphics[
width=3.3in 
]{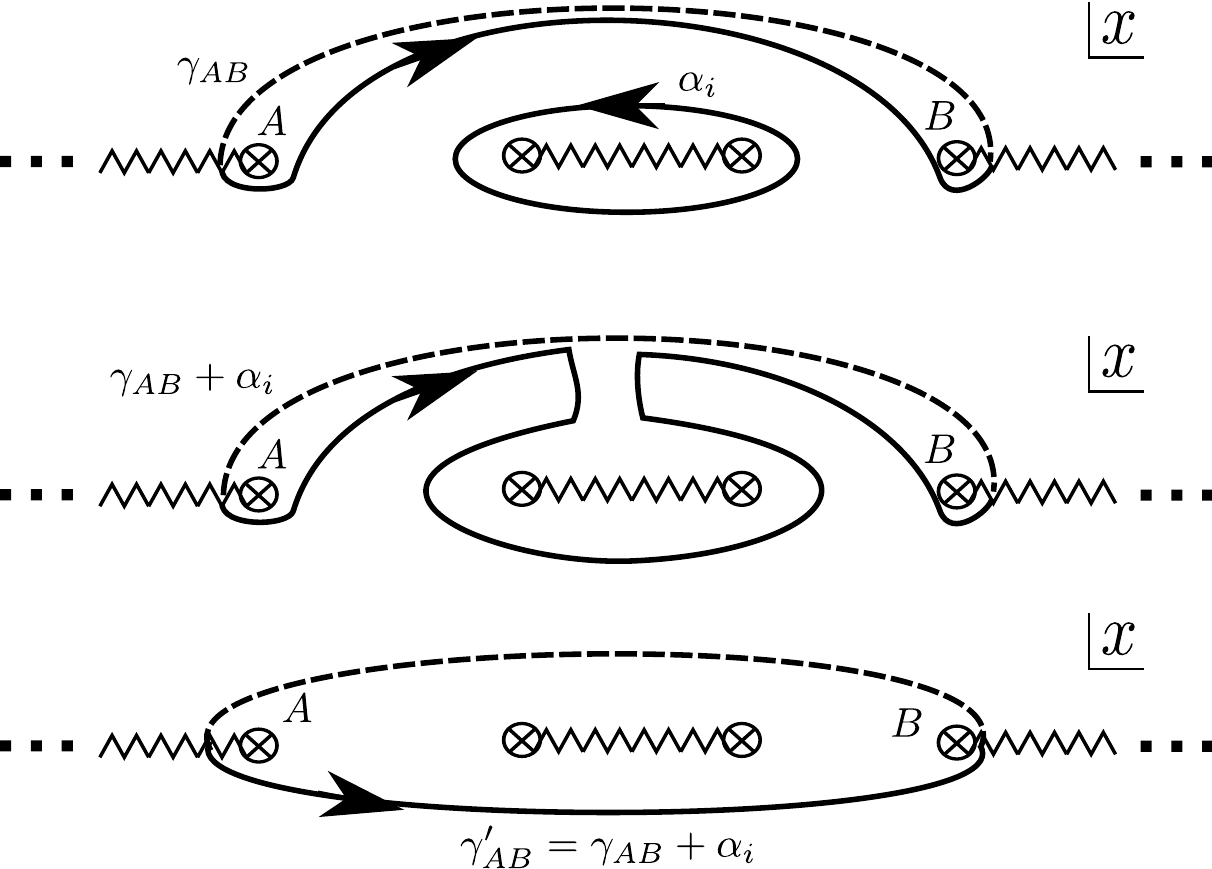}
\end{center}
\caption{If a cycle surrounds a branch cut, we have a further 
ambiguity of $n\alpha$'s. Compare with {\bf figure \ref{goaroundacut}} where only $2n \alpha$ ambiguity was allowed.}
\label{goaroundacutodd}
\end{figure}

\subsubsection{Relative trajectories of branch points - time information \label{RelativeChoreo}}  

To study the relative trajectories of branch points one would need more information than just the collection of singular loci. For example to see how the various points 
are connected, one may need temporal information for the evolution of the trajectories\footnote{An example from classical mechanics can readily explain this. Imagine 
a set of two particles connected by springs. Each of these pair of particles are allowed to perform random motions. Imagine also that these random motions are periodically 
captured by their {\it footprints}.  
If one wants to learn the full dynamics of these pair of particles, one can make certain guesses based on the footprints, however not the full dynamics. 
With time information lost, we cannot learn the relative location of each particle in a pair. We need a few key time informations. 
This is thus exactly related to the problem at hand:
to read off the dynamics of the branch cuts on $x$-plane we will require certain {\it temporal} information over and above the knowledge of the 
singularity loci. \label{dancers}}. 

\begin{figure}[htb]
\begin{center}
\includegraphics[
width=4.7in 
]{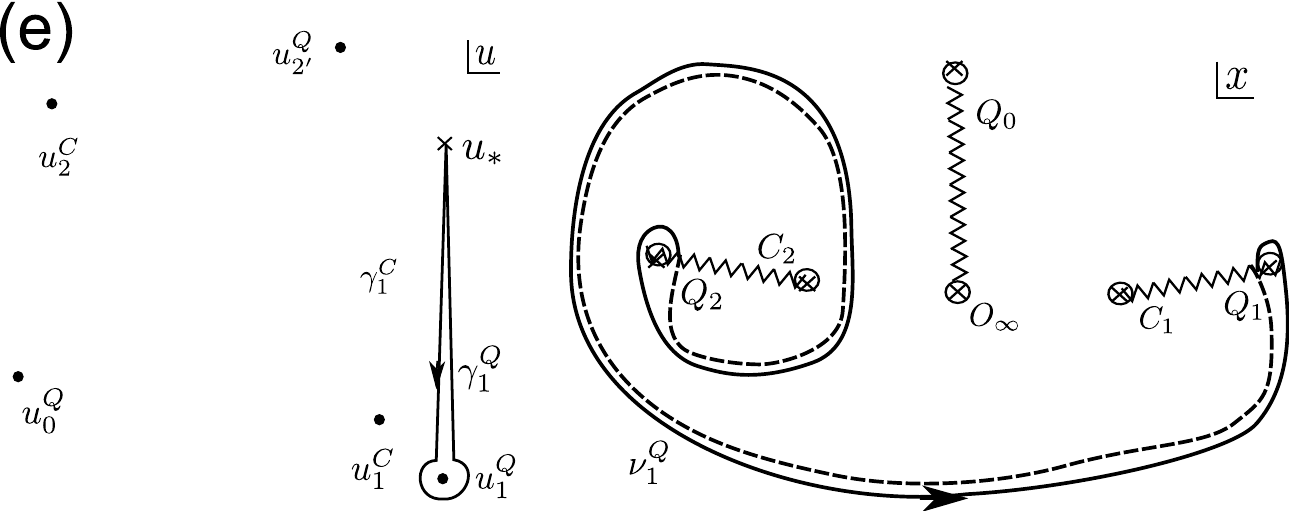}
\end{center}
\caption{Take a part of {\bf figure \ref{sp4jihye2}}. In the current subsection, we will discuss how we obtain such result for the vanishing cycle.}
\label{concise}
\end{figure}

As an example, let us give a preview of what we will see in subsection \ref{Sp4}, and explain how we got the result.
In subsection \ref{Sp4}, we will look at vanishing cycles of $Sp(4)$ pure SW theory. The result is summarized in {\bf figure \ref{sp4jihye2}}. Let us explain how 
to get these results. For simplicity, we will only sketch how to compute dyon charge for a vanishing cycle $\nu^Q_1$. In {\bf Figure \ref{concise}} we depict
the result from subsection \ref{Sp4}. Is there a simple way to understand the {\it dynamics} of the figure?  

\begin{figure}[htb]
\begin{center}
\includegraphics[
width=3in 
]{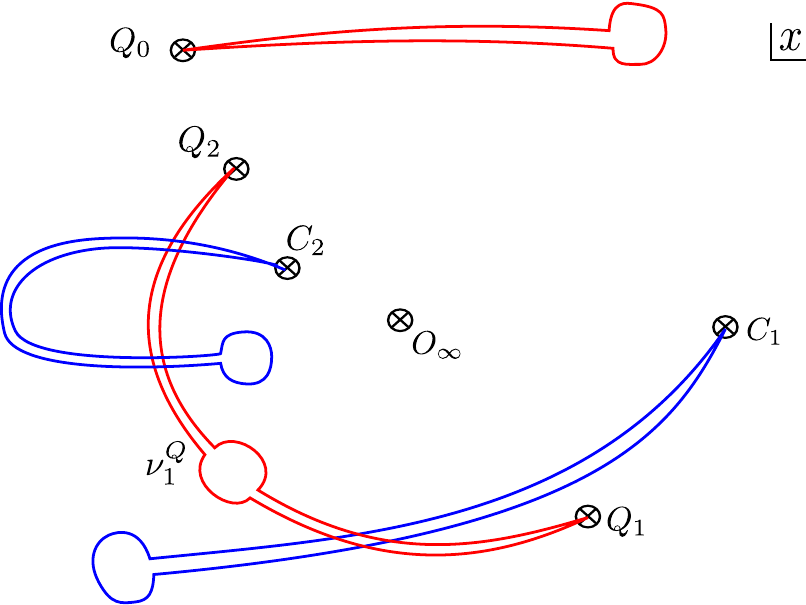}
\end{center}
\caption{Here we draw trajectories of branch points on $x$-plane, with time information flattened out. This does not contain enough information for the charge of vanishing cycle. We need a relative trajectory as in {\bf figure \ref{dance}.}}
\label{snow}
\end{figure}

The technique that we will use to analyze such dynamics of the curves will be a 
semi-numerical method with Mathematica. The way it goes is as follows: we start by simultaneously looking at both $u$ and the $x$ planes.  
As we vary {\it time} or the moduli (on the left side of the {\bf figure \ref{concise}}) along a non-contractible loop in the moduli space, we observe that on the $x$-plane, the branch points move around. {\bf Figure \ref{snow}} is the trajectories of branch points, with all the time information flattened out. This is 
then exactly equivalent to the {\it footprints} of the pair of classical particles at the end of their random motions (see footnote \ref{dancers}). 

\begin{figure}[htb]
\begin{center}
\includegraphics[
width=6in 
]{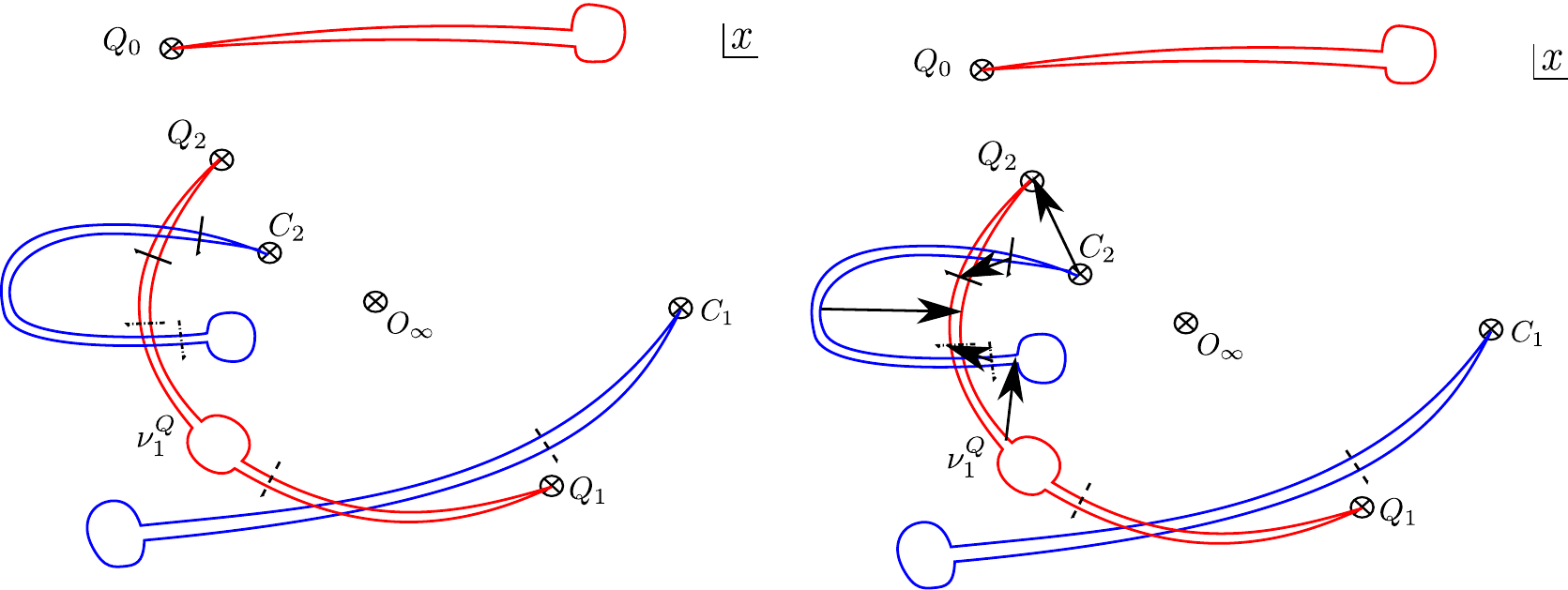}
\end{center}
\caption{Now with certain key time information explicitly given, we can read off dyon charges for massless BPS state. Note the counter-clockwise rotation of the branch cut, which is drawn as an arrow connecting two branch points pointing from $C_2$ to $Q_2$. This is why the vanishing cycle wraps around a branch cut in {\bf figure \ref{concise}}.}
\label{dance}
\end{figure}

In fact things are even better now. 
By adding a few more information, we can read of vanishing cycles correctly. The important point is when the (flattened) trajectories intersect, which branch point crossed the crossing first? For example, in left of {\bf figure \ref{dance}}, we put a few time markers. A pair of markers with each distinctive line patterns are placed, to denote the time information. The branch cut connecting $Q_2$ and $C_2$ (denoted as an arrow to manifest the direction of rotation) is drawn on right of {\bf figure \ref{dance}}. Thanks to the time markers, we see that the branch cut rotates counter-clockwise.

\begin{figure}[htb]
\begin{center}
\includegraphics[
width=6in 
]{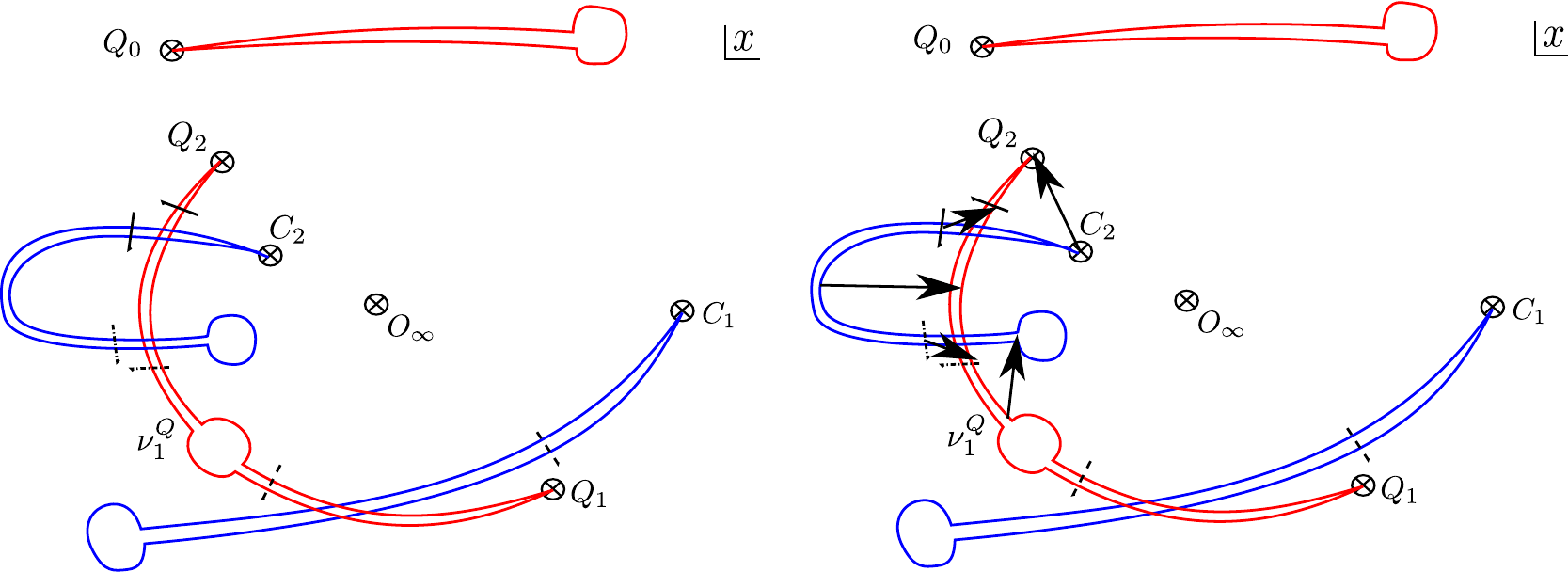}
\end{center}
\caption{In order to emphasize the importance of the time information, imagine a hypothetical situation where we had a different time information. Now the branch cut can rotate the opposite direction, and this adds ambiguity of multiple of $\alpha_2$ on the vanishing cycle, as explained in {\bf figure \ref{rootaround}}. In {\bf figure \ref{concise}}, the vanishing cycle will wrap around the branch cut in the opposite direction.}
\label{wrongdance}
\end{figure}

The relative usefulness of our approach is immediately obvious when we compare {\bf figure \ref{dance}} with {\bf figure \ref{wrongdance}}. 
{\bf Figure \ref{wrongdance}} is a hypothetical (wrong) situation where we have markers placed at wrong places, leading us into a different dyon charge for vanishing cycles. 

\subsection{Plan of the paper \label{Plan}}
 
%

The plan of the paper is as follows. 
In section \ref{suSec}, we compute the monodromy of pure $SU$ curves. In a certain region of the moduli space, we can compute all the vanishing cycles exactly, as shown in subsection \ref{exactSUmonodromy}. 
We will discuss certain singularity loci for the $SU$ case with two massless BPS dyons which are mutually local in subsection \ref{doublepointsing}.
In subsection \ref{reduceSUrank}, we will discuss how to reduce the rank, and how the dyon charges change.

In section \ref{spSec}, we study pure $Sp$ curves to compute the maximal Argyres-Douglas points and monodromies.
First we will study the behaviour of the branch points of pure $Sp$ curve in subsection \ref{sprootstructure}. 
Specifically, in subsection \ref{inverseX}, we take $x \rightarrow 1/x $ transformation, in order to bring a branch point at infinity to origin. In subsection \ref{infAlone}, we will see that this branch point at infinity does not participate in any of the vanishing cycles related to the stable singularities.
In subsection \ref{monodromySp}, we discuss the vanishing cycles of $Sp$ pure curves.
Subsection \ref{conjectSp} computes in semi-numerical way, all the vanishing cycles for the pure 
$Sp$ curves up to rank 6. This computation will help us to suggest a generalization to arbitrary ranks.
In subsection \ref{MinftyHighrank} we discuss the monodromies with respect to the point at infinity, and explain why arguments from codimensions etc show 
the non-existence of these for rank 2 or higher.
In subsection \ref{Sp4}, we take a detailed look at the singularity structures of pure $Sp(4)$ case, and observe wall-crossing phenomena for 
massless BPS states as we go through an Argyres-Douglas point.
In subsection \ref{reduceSp}, we will discuss how to reduce the rank, and how dyon charges change in a somewhat similar vein as we did for the $SU$ case. 

The next section, i.e section \ref{doublediscsection}, discusses methods to capture Argyres-Douglas loci in both $SU$ and $Sp$ theories. 
In subsection \ref{exteriordVSdd}, we see that exterior derivative will capture the loci of two vanishing cycles. In subsections \ref{suexample} and 
\ref{spexample}, we compare the exterior derivative and double discriminant for $SU$ and $Sp$ curves.  
In subsection \ref{oov3}, we argue how the double discriminant will distinguish the case where the two vanishing cycles are mutually non-local.
We elaborate the combinatoric meaning for factorization of double discriminant in subsection \ref{FactorDD}, and in subsection \ref{curvedegen} we elaborate this
with explicit examples and explanation for the $Sp$ cases. We also show therein how this could be generalized in a straightforward way for the $SU$ cases. 
In subsection \ref{blind}, we will see that certain moduli are more useful than others for taking the double discriminant. 

So far our discussions have mostly been for the $SU$ and the $Sp$ cases. In section \ref{demandCurveForm} we turn to study the $SO$ cases concentrating mostly on the 
$SO$-odd cases. The technique that we will use here will be slightly different from what we had used in the earlier sections. We elaborate this procedure and show  
how we go to the singular regions by explicitly demanding the curves to take certain forms. In subsection
\ref{SOoddADmax}, we compute candidates for maximal Argyres-Douglas points for the $SO(2r+1)$ cases.
 
Finally we conclude with open questions in section \ref{conclusion}. The two appendices are arranged in the following way.
In {\bf appendix \ref{PUREfromAS}} 
we show how one may take the Seiberg-Witten curve and one-form with flavors for the $Sp(2r)$ cases, and from there get the pure $Sp(2r)$ curves.
This appendix has appeared earlier in \cite{DSW}, but we keep it here for completeness. In {\bf appendix \ref{binoid}} 
we prove some binomial identities that will prove very useful 
for deriving certain relations in subsection \ref{SOoddADmax}.

\newpage

\section{Monodromies of pure ${\mathcal N}=2$ $SU(r+1)$ theories \label{suSec}} 

To see how the techniques that we developed in the earlier subsections can be readily used to determine the singularity structures and the Argyres-Douglas loci, 
we will start-off with $SU(r+1)$ 
SW theory. The 
SW curve and SW 1-form for pure $SU(r+1)$ \cite{KLYTsimpleADE} are rewritten as
  \begin{equation}  
 y^{2}=f_{SU(r+1)} = f_{+} f_{ -} , \quad   \lambda_{\mathrm SW} = -dx \log \left( - \frac{1}{2}(f_+ + f_-) - \sqrt{f_+ f_-}\right), \label{su1curve1form}
 \end{equation}
where $f_{\pm}$ are given in terms of $r$ gauge invariant moduli $u_i \in \mathbb{C}$ ($i=1,\cdots, r$) as:
  \begin{equation}  
  f_{\pm} \equiv  x^{r+1} + \sum_{i=1}^r u_i x^{r-i}   \pm \Lambda
^{2r+2},  \label{fpm}
\end{equation}   
whose roots can be denoted by\footnote{Note that both $f_\pm$ depends on $\Lambda$. For the $Sp$ case it will be slightly different.} ($P, N$ stands for positive, negative in front of $\Lambda$ term.)
 \begin{equation}f_{+} \equiv \prod_{i=0}^{r} (x-P_i), \qquad f_{-} \equiv \prod_{i=0}^{r} (x-N_i).\end{equation}
Few comments are in order. First note that the curve equation \eqref{su1curve1form}  has an underlying $\mathbb{Z}_h$ symmetry that will be elucidated later. 
Secondly, 
note that $f_+=f_-=0$ may happen only when $\Lambda=0$. For quantum theory, $\Lambda\ne 0$ and $f_{\pm}$ can never vanish at the same time. Therefore $f_+$ and $f_-$ can never share a root, and there is no vanishing cycle mixing these two groups of roots.  This justifies binary color coding in figures in the current section for branch points and vanishing cycles:
When we draw vanishing cycles and collision of branch points, we can use binary coloring. The branch points and vanishing cycles are all grouped into two mutually exclusive groups (for $P$ and $N$ respectively.). 
On the $x$-plane, only $P_i$'s (or $N_i$'s) can collide among themselves. The discriminant of the curve will effectively factorize as \cite{KLYTsimpleADE}
 \begin{equation} 
 \Delta_x f_{SU(r+1)} = (2 \Lambda^{2r+2})^{2r+2} \Delta_x f_{+} \Delta_x f_{-}. \label{SUfacD}
 \end{equation}
 When $ \Delta_x f_+=0$ ($  \Delta_x f_-=0$), a cycle connecting two $P_i$ ($N_i$) branch points are vanishing.

From the degree in each moduli, we can read off the number of solutions to the vanishing discriminant. 
As explained earlier near \eqref{DiscDeg}, the discriminant will appear as a polynomial in terms of its coefficients, in our case the moduli $u_i$'s, with a specific pattern for their powers. All the moduli $u_i$ will have the top degree to be $r+1$, except for the $u_r$ which has the top degree to be $r$ in the discriminant of $f_\pm$ as below 
\begin{eqnarray}   
\Delta_x f_{\pm}  &=& \#  \left( u_{i}^{r+1} + \cdots \right) ,\quad i\ne r,    \nonumber \\
   &=&  \# \left( u_r^{r} + \cdots \right),   \label{urbad}
\end{eqnarray}  
where we borrowed the notation from \eqref{DiscDeg} to denote the top degree in terms of each modulus.

If we are to solve $\Delta _x f_{\pm}=0$ in terms of one of the modulus, there are $r+1$ solutions for any $u=u_i$ ($i\ne r$) which is not a constant piece. 
However, for $u=u_{r}$, the degree is lower by one as in second line of \eqref{urbad}. 
See \eqref{degree2} for the explanation. 

Therefore for $r>1$, we expect $2(r+1)$ solutions to the vanishing discriminant from the first line of \eqref{urbad},  corresponding to $2(r+1)$ massless dyons. For $r=1$, we cannot take the first line of \eqref{urbad} any more, because there is only one modulus, and we go with the second line. Therefore for $r=1$, we have only 2 solutions. Since rank 1 case is studied much, we focus on $r>1$ cases here.

Each solution to the vanishing discriminant corresponds to a vanishing cycle. Number of the solutions of vanishing discriminant of the curve equals the number of vanishing cycles. We will study the vanishing cycles soon in subsection \ref{exactSUmonodromy} and show how the $( {\color{purple}r+1})+( {\color{forestgreen}r+1})$ vanishing cycles for 
$SU(r+1)$, for bringing two ${\color{purple}P}$ (or ${\color{forestgreen}N}$) points together on the $x$-plane, appear from our analysis.

We can also find maximally singular case, where maximum number (rank +1) of branch points collide on the $x$-plane.  
{Argyres-Douglas loci} occur by bringing $n \ge 3$ branch points together on the $x$-plane, such that the SW curve degenerates into a cusp form $y^2 = (x-a)^n \times \cdots $.
 In particular, the maximal Argyres-Douglas points occur at two points in the moduli space given by 
 \begin{equation}
 u_i=0,~~~ u_r =   \Lambda^{r+1} \label{SUMaxAD}
 \end{equation}
 ($u_r =  - \Lambda^{r+1}$ resp.), where all $N_i=N_j$ (all $P_i=P_j$ resp.) collide together \cite{ArgyresDouglas, LercheReview}.

 At discriminant loci $ \Delta_x f_{SU(r+1)}=0$ and near the corresponding vanishing 1-cycle, 
the 1-form of \eqref{su1curve1form} is regular
\begin{equation}
\lambda_{\mathrm SW} = -dx \log \left(       \pm   \Lambda^{2r+2}   \right), \qquad {\mathrm{near}}  \ f_\pm = \Delta_x f_{\pm}=0,
\end{equation} confirming that the singularity of the SW curve is indeed the singularity of the SW theory.

\subsection{Exact BPS spectrum in a $\mathbb{Z}_{r+1}$-symmetric region of moduli space for arbitrary ranks \label{exactSUmonodromy}} 

\begin{figure}[htb]
\begin{center}
\includegraphics[
width=5.5737in,
height=2.5278in
]{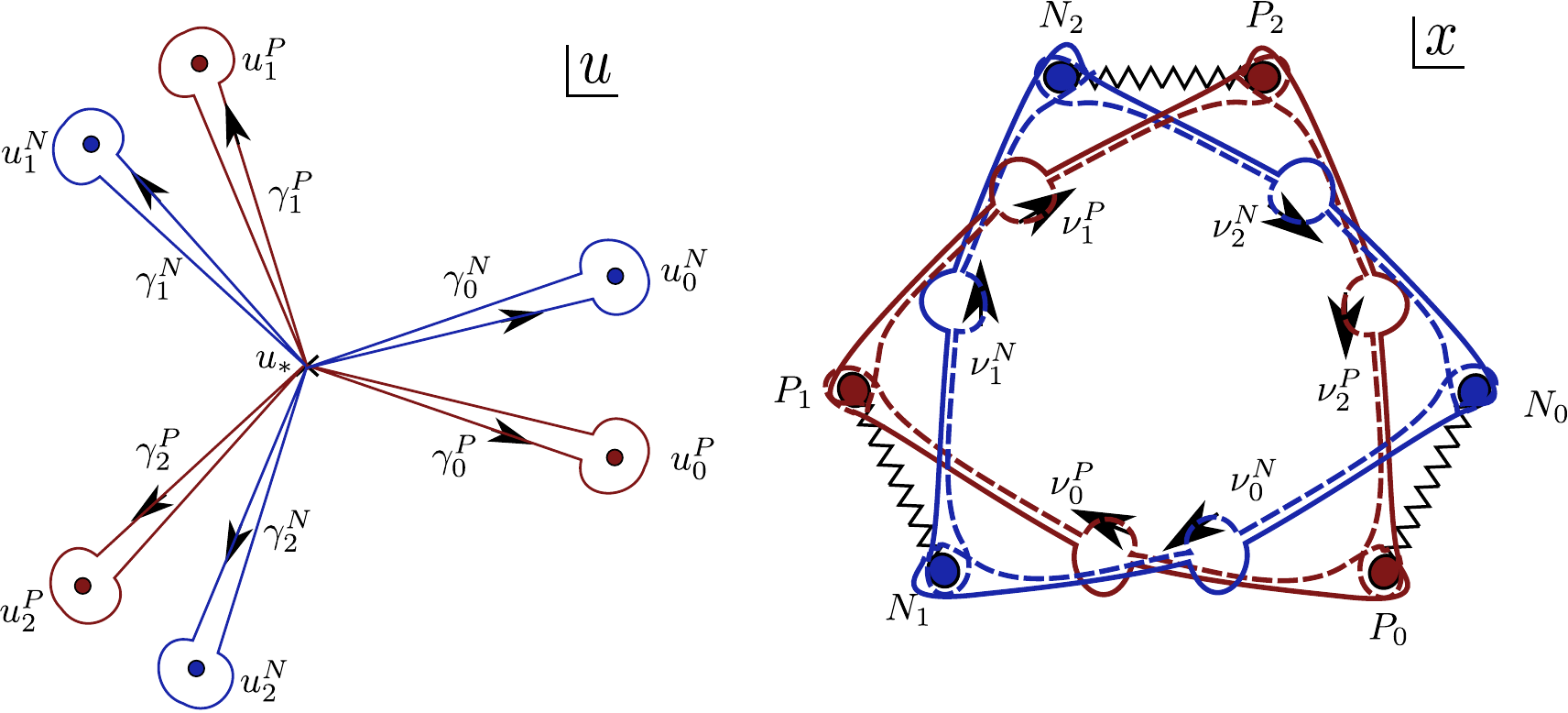}
\end{center}
\caption{Vanishing cycles of $SU(3)$ curve are related to massless BPS dyons. Restricting the moduli space by setting $v\equiv u_2 \in \mathbb{I}$ to a constant, we have six singular points on the $u \equiv u_1$-plane. As we go around each singular point, we have a vanishing cycle on the $x$-plane as shown.}
\label{su3cycles}
\end{figure}

Earlier studies have discussed 
 six vanishing cycles of the SW curve for pure $SU(3)$ for $v \in \mathbb{R}$ \cite{KLYTsimpleADE, KLYTmndrmSU}. For these cases the Argyres-Douglas points occur at ${\rm codim}_{\mathbb{C}}$-$2$ loci in the moduli space. (Similarly for $G_2$ as well \cite{LPGg2}.). Motivated by these references, we can compute the  
$SU(3)$ spectra in $v\equiv u_2 \in \mathbb{I}$ region, where $\mathbb{Z}_2 \times \mathbb{Z}_{r+1}$ symmetry is manifest\footnote{$\mathbb{Z}_2$ is for flopping $P$ and $N$; and 
 $\mathbb{Z}_{r+1}$ is for phase rotation by $\frac{2 \pi}{r+1}$ on the $x$-plane.}. Non-vanishing intersection numbers come only from 
\begin{equation}
\nu_{i}^P \cap \nu_{i+1}^P=\nu_{i}^N \cap \nu_{i+1}^N= 1, \qquad \nu_{i}^P \cap \nu_{i}^N= -2,\qquad \nu_{i}^P \cap \nu_{i+1}^N= 2 \label{SUinter}
\end{equation} as described in {\bf figure \ref{su3cycles}}. 
In the following we will extend this to higher ranks. Our starting point would be to     
choose a moduli slice (for example, in \eqref{1plane} and \eqref{r1plane} with $\frac{ u_r}{\Lambda^{r+1} }\in \mathbb{I}$) with the intersection number 
given in  \eqref{SUinter}. Examples of these are depicted in {\bf figures \ref{su7flower}} and {\bf \ref{rank9SU10xplane}} for higher ranks. 

\subsubsection{Vanishing cycles of $SU(r+1)$ curve in a hypersurface of the moduli space}

\begin{figure}[htb]
\begin{center}
\includegraphics[
width=3.5in
]{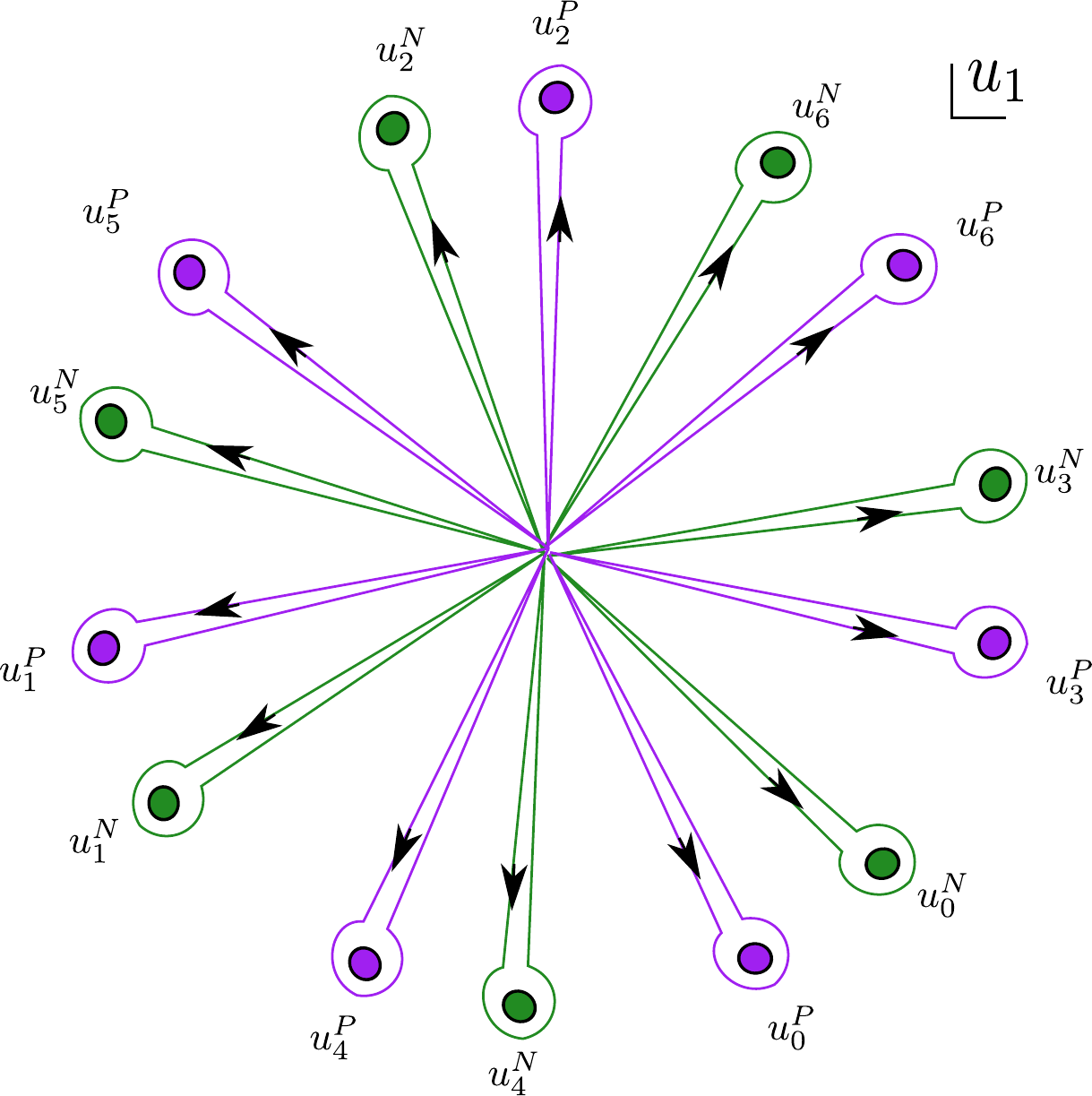}
\caption{Vanishing cycles of $SU(7)$ curve are related to massless BPS dyons. On a moduli slice $u_1$-plane given by $u_2= \cdots = u_{r-1}=0, u_r \ne 0$ and $\frac{ u_r}{\Lambda^{r+1} }\in \mathbb{I}$, we see $2(r+1)$ singular points, for even $r$. Here $r=6$ case is drawn. As we go around each singular point on the $u_1$-plane, we have a vanishing cycle on the $x$-plane as shown in {\bf figure \ref{su7flower}}.}
\label{rank6SU7u1Iur}
\end{center}  
\end{figure}

\begin{figure}[htb]
\begin{center}
\includegraphics[
width=4in
]{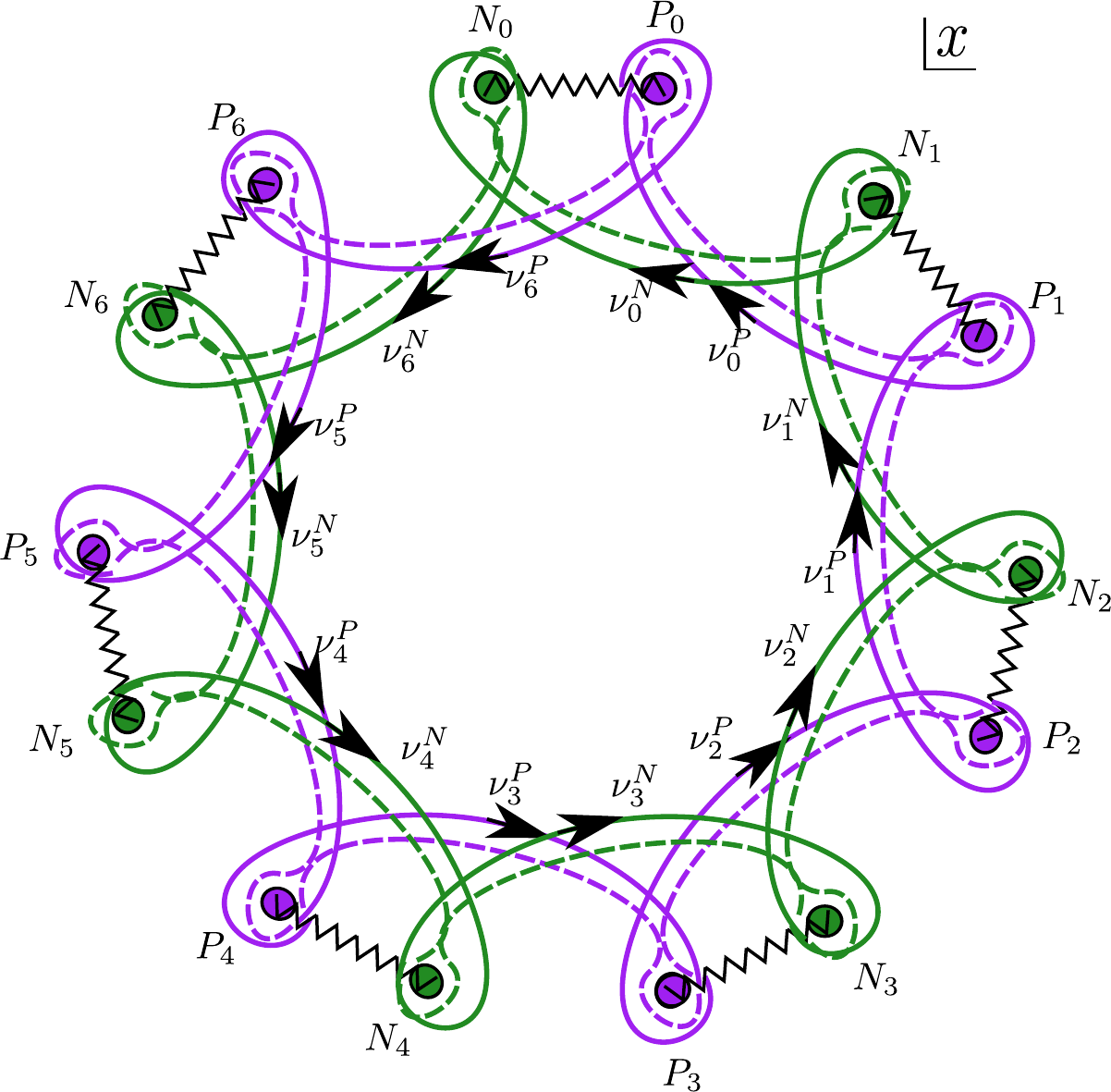}
\caption{Vanishing cycles of $SU(r+1)$ curve are related to massless BPS dyons. Drawn here for rank 6, at a moduli slice (a $u_{5}$-plane) given by $ u_1=u_2= \cdots = u_{4}=0$ and $\frac{ u_6}{\Lambda^{7} }\in {\mathbb{I}} \cup \{ 0\}$. (also the same figure for)
Drawn here for rank 6, at a moduli slice (a $u_{1}$-plane) given by $ u_2=u_3= \cdots = u_{5}=0$ and $\frac{ u_6}{\Lambda^{7} }\in {\mathbb{I}} \cup \{ 0\}$}
\label{su7flower}
\end{center}  
\end{figure}

Starting from the SW curve of \eqref{su1curve1form}, let us choose values of moduli which make $\mathbb{Z}_{r+1}$ symmetry manifest. 
For example, we can choose   
a hypersurface in a moduli space given as 
\begin{equation}
u_2= \cdots = u_{r-1}=0, \qquad u_r = {\rm const.} \label{1plane}
\end{equation} 
For even $r$, the following factors\footnote{Gauss' symbol is used: $[x]$ is the greatest integer that is less than or equal to $x$.} 
  \begin{equation} \Delta_x f_{\pm} = (-1)^{\left[ \frac{r+1}{2} \right]} (u_r  \pm \Lambda^{r+1})^{r-2} \left( 2^2 (r-1)^{r-1}  u_1^{r+1} +(r+1)^{r+1} (u_r  \pm \Lambda^{r+1})^2 \right) \label{evenRsu} \end{equation}
 of discriminant gives us $2(r+1)$ solutions on the 
$u_1$-plane. Let us also demand $\frac{ u_r}{\Lambda^{r+1} }\in \mathbb{I}$ for $\mathbb{Z}_2 \times \mathbb{Z}_{r+1}$ symmetry.
 {\bf Figure \ref{rank6SU7u1Iur}} shows $2(r+1)$ points on such the $u_1$-plane where discriminant of the curve vanishes. Each singular point corresponds to a vanishing cycle on the $x$-plane. 
We obtain configuration of $2(r+1)$ vanishing cycles as in {\bf figure \ref{su7flower}}. At a singular point on the $u_1$-plane given by $u^P_i$ ($u^N_i$ resp.) of {\bf figure \ref{rank6SU7u1Iur}}, the corresponding 1-cycle $\nu^P_i$ ($\nu^N_i$ resp.)  will vanish in {\bf figure \ref{su7flower}}. 

\begin{figure}[htb]
\begin{center}
\includegraphics[
width=3.5in
]{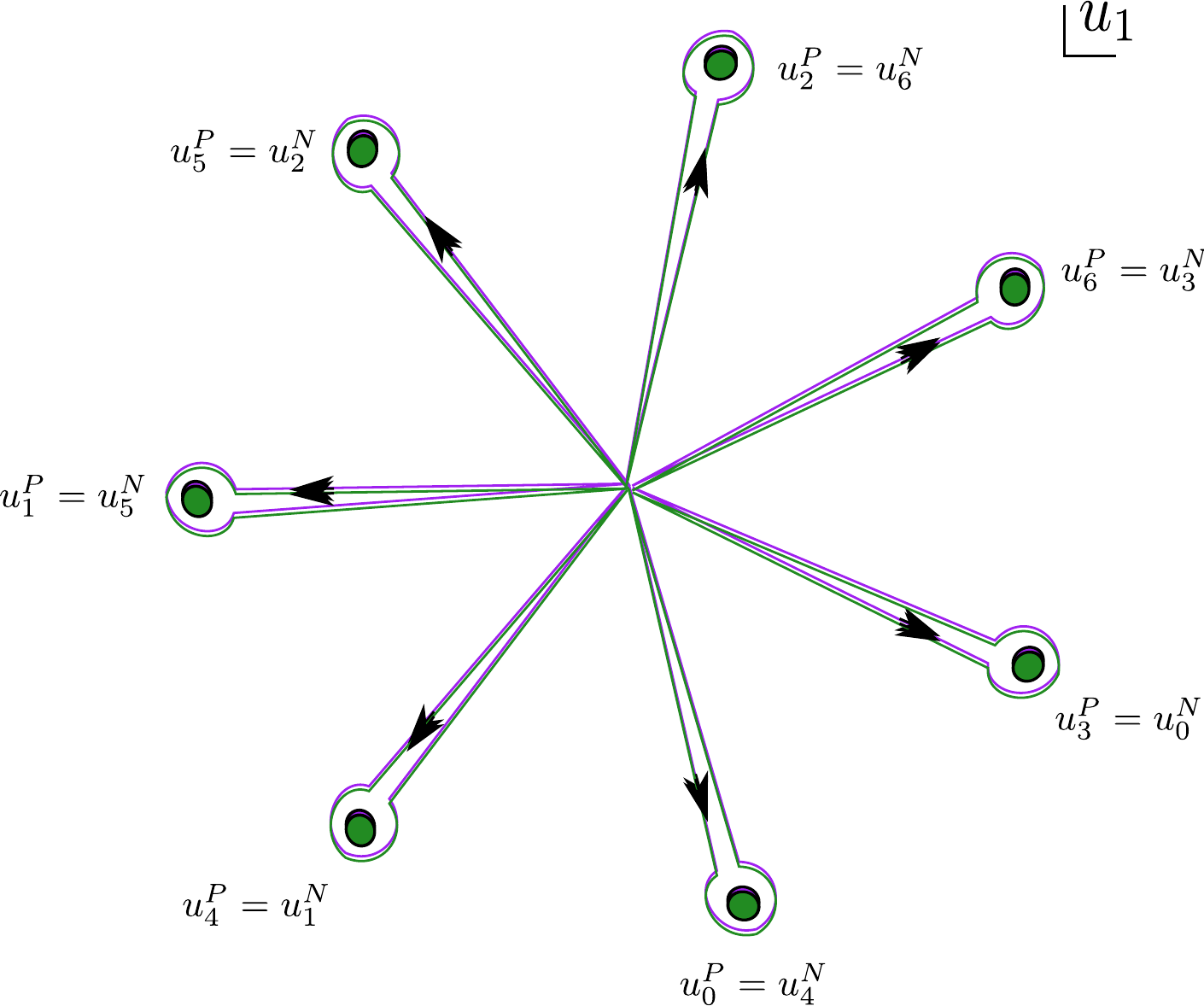}
\caption{
Singularity structure of $SU(7)$ curve on a moduli slice of the $u_1$-plane given by $u_2= \cdots = u_{r-1}=0, u_r = 0$. Compare with {\bf figure \ref{rank6SU7u1Iur}}. Original $2(r+1)$ singular points all collide pairwise to form $r+1$ double-points, for even $r$. Here $r=6$ case is drawn. As we go around each singular double-point on the $u_1$-plane, we have two vanishing cycle on the $x$-plane as shown in {\bf figure \ref{su7flower}}. As denoted by subscripts here, $\nu^P_i $ and $\nu^N_{i+4}$ vanish at the same time on the $x$-plane.}
\label{rank6SU7u1zerour}
\end{center}  
\end{figure}

If we vary $u_r$ such that $u_r=0$, singularity structure shown on the $u_1$-plane will change such that all singular points will collide pairwise, as shown in {\bf figure \ref{rank6SU7u1zerour}}. Each pair of $u^N_i$ and $u^P_{r/2}$ collide on the $u_1$-plane. Therefore each pair of 1-cycles $\nu^N_i$ and $\nu^P_{r/2}$ vanish at the same time. Note that they are mutually local (cycles do not intersect each other), therefore this does not lead to Argyres-Douglas theories.

This is easily seen from \eqref{fpm}. For $
u_2= \cdots = u_{r-1}= { u_r}=0$ and even rank,  \eqref{fpm} reduces to
\begin{equation}  
  f_{\pm} \equiv  x^{r+1} +  u_1 x^{r-i}   \pm \Lambda
^{2r+2},   
\end{equation}   
and as $x\rightarrow -x$, $f_{\pm} \rightarrow f_{\mp}$. In other words, When $u_1$ takes a value where $f_+$ takes a double root, $f_-$ will also take a double root (at a value opposite in sign on the $x$-plane). It is no accident that each pair of 1-cycles $\nu^N_i$ and $\nu^P_{r/2}$ vanish at the same time, and that they are just 180 degrees phase rotation of each other on the $x$-plane of {\bf figure \ref{su7flower}}.

\begin{figure}[htb]
\begin{center}
\includegraphics[
width=4in
]{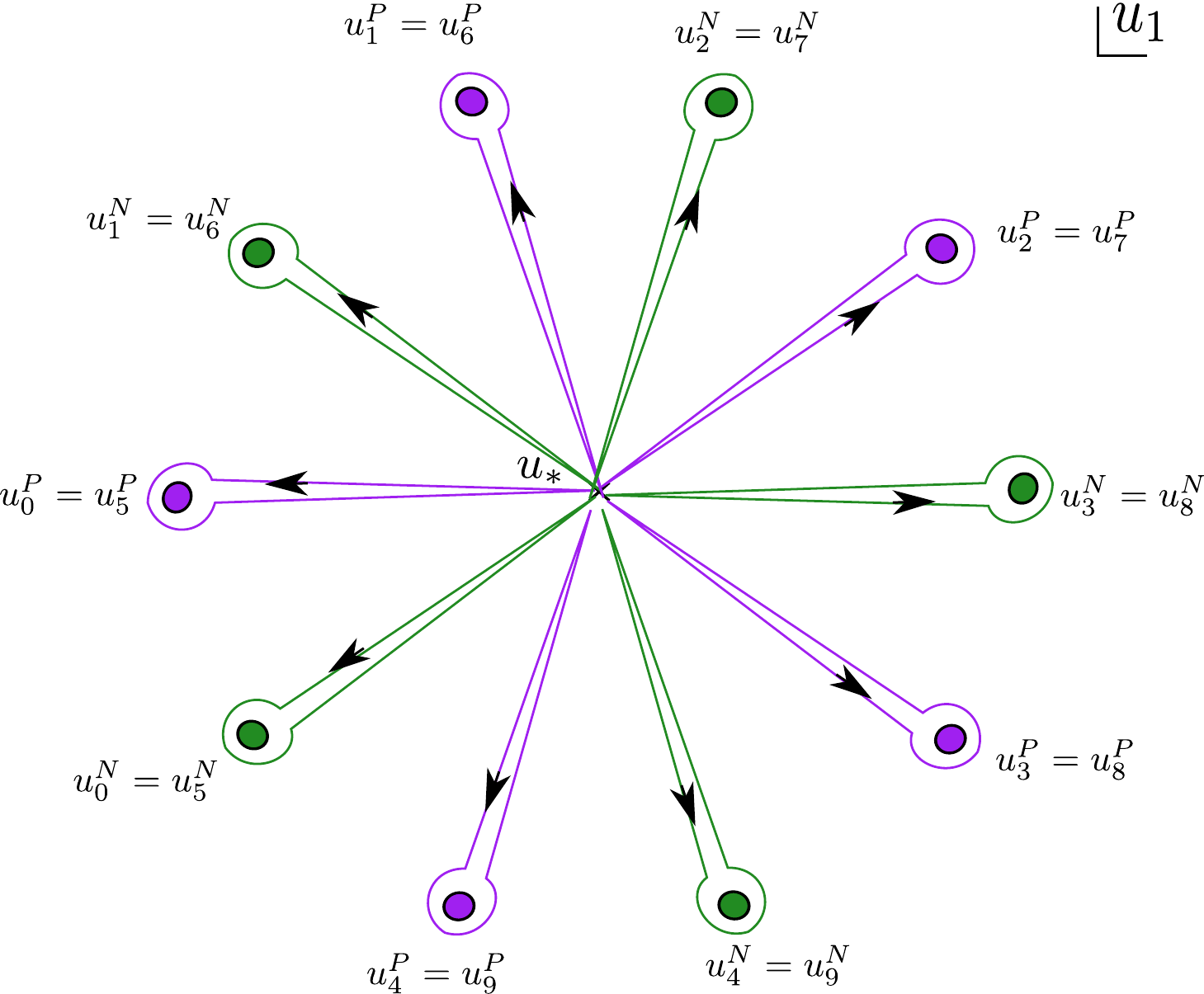} 
\caption{Vanishing cycles of $SU(10)$ curve are related to massless BPS dyons. On a moduli slice of the $u_1$-plane given by $u_2= \cdots = u_{r-1}=0, u_r = const$, we see $2(r+1)$ singular points all collide pairwise to form $r+1$ double-points, for odd $r$. Here $r=9$ case is drawn. This happens for any value of $u_r$, and it is to be contrasted with {\bf figure \ref{rank6SU7u1Iur}}. It is a little similar to  {\bf figure \ref{rank6SU7u1zerour}}, but note the difference in coloring.  
 As we go around each singular double-point on the $u_1$-plane, we have two vanishing cycle on the $x$-plane as shown in {\bf figure \ref{rank9SU10xplane}}. As denoted by subscripts here, $\nu^{P,N}_i $ and $\nu^{P,N}_{i+5}$ vanish at the same time on the $x$-plane. }
 \label{rank9SU10u1}
\end{center}  
\end{figure}

The question now is how does the above trick apply to odd rank case?
For $
u_2= \cdots = u_{r-1}=0$ and odd rank,  \eqref{fpm} reduces to
\begin{equation}  
  f_{\pm} \equiv  x^{r+1} +  u_1 x^{r-i} + u_r  \pm \Lambda
^{2r+2}.
\end{equation}   
All the terms have even powers in $x$. All the roots of $f_+$ will appear in pairs which add up to zero. When $f_+$ takes a double root, it will necessarily take another double root at a value opposite in sign on the $x$-plane. On the $u_1$-plane, we will observe 
{\bf figure \ref{rank9SU10u1}}, very similar to  {\bf figure \ref{rank6SU7u1zerour}}. Note the difference that for odd rank case, this happens at a generic value of $u_r$, while for even rank case it is only for $u_r=0$.

\begin{figure}[htb]
\begin{center}
\includegraphics[
width=4.5in
]{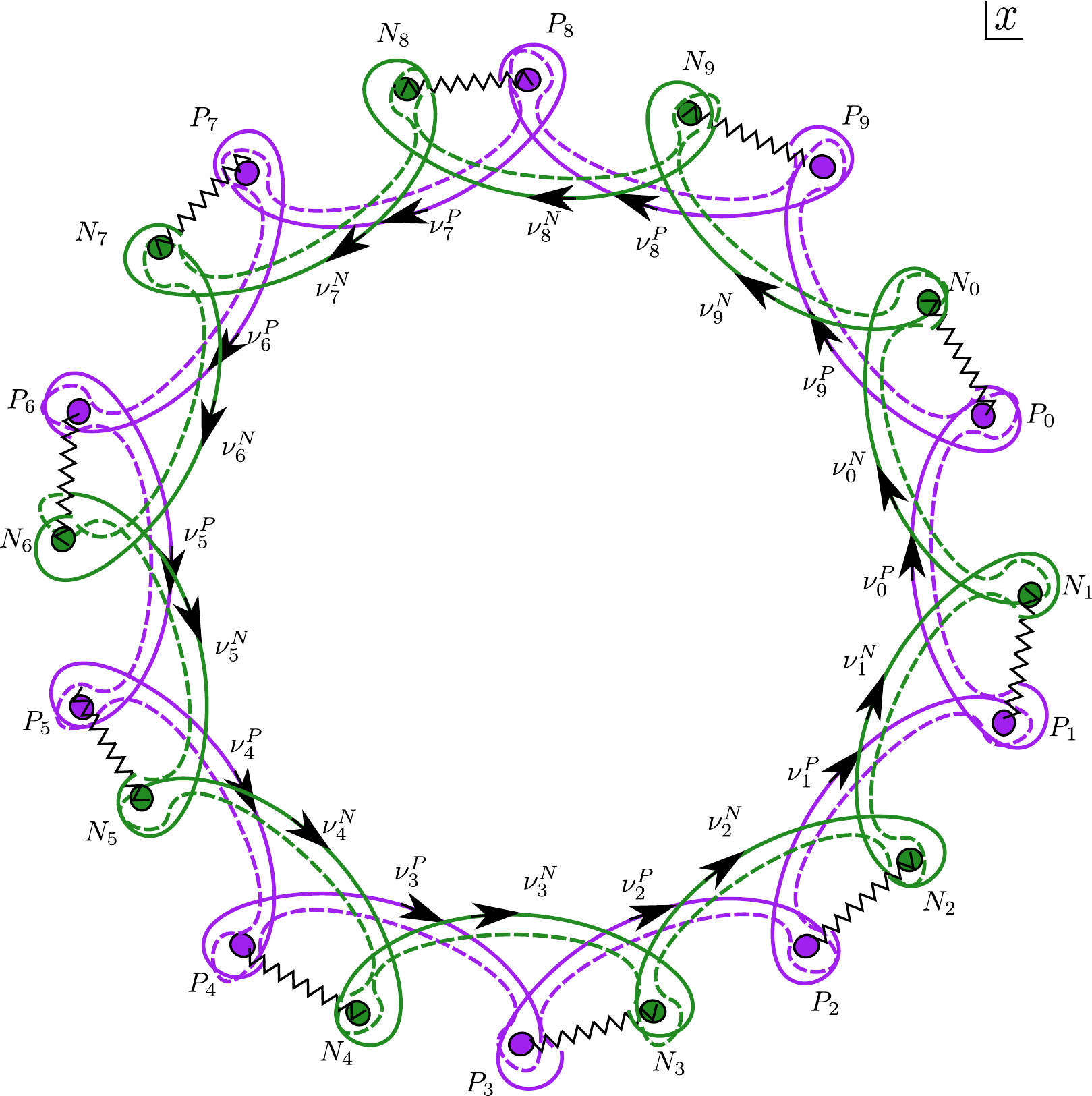} 
\caption{Drawn here are vanishing cycles for $SU(10)$ at a moduli slice, a $u_{8}$-plane given by $ u_1=u_2= \cdots = u_{7}=0$ and $\frac{ u_9}{\Lambda^{10} }\in {\mathbb{I}} \cup \{ 0\}$, or a $u_1$-plane given by $ u_1=u_2= \cdots = u_{8}=0$ and $\frac{ u_9}{\Lambda^{10} }\in {\mathbb{I}} \cup \{ 0\}$. Note similarity with {\bf figure \ref{su7flower}}.}
\label{rank9SU10xplane}
\end{center}  
\end{figure}

In {\bf figure \ref{rank9SU10u1}}, when we go around a singular point of $u^P_0 = u^P_{(r+1)/2}$, 1-cycles $\nu^P_0 $ and $\nu^P_{(r+1)/2}$ will vanish simultaneously on the $x$-plane as in {\bf figure \ref{rank9SU10xplane}}. Again it is no accident that each pair of simultaneously vanishing cycles are 180 degrees phase rotation of each other on the $x$-plane.
 \begin{figure}[htb]
\begin{center}
\includegraphics[
width=3.5in
]{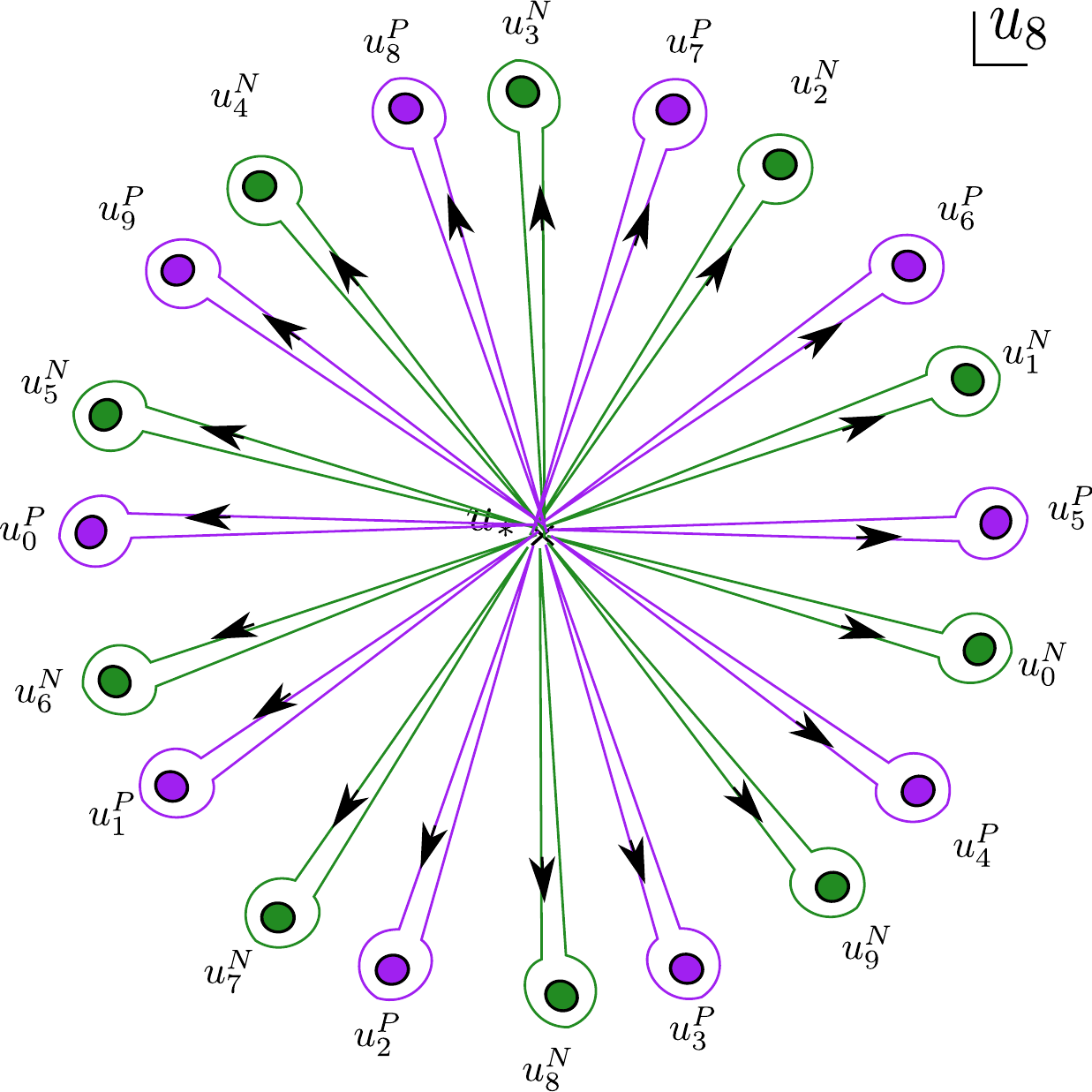}
\caption{ Singularity structure of $SU(10)$ curve on a moduli slice of the $u_8$-plane given by $u_1= u_2=\cdots = u_{r-2}=0, u_r = 0$. 
 Here $r=9$ case is drawn. 
 As we go around each singular point on the $u_8$-plane, we have a vanishing cycle on the $x$-plane as shown in {\bf figure \ref{rank9SU10xplane}}.}
\label{rank9SU10u8theta0}
\end{center}  
\end{figure} 

Thus we see that this is an alternative way to see the singularity structure.
  
 For odd $r$ on a moduli slice in \eqref{1plane}, all the roots of $f_\pm$ have degeneracy (order of vanishing) 2 as seen below:
  \begin{equation} \Delta_x f_{\pm} = (-4)^{\left[ \frac{r+1}{2} \right]}   (u_r  \pm \Lambda^{r+1})^{r-2} \left(   \left( \frac{1-r}{2}\right)^{\frac{r-1}{2}}  u_1^\frac{r+1}{2} + \left(\frac{r+1}{2}\right)^{\frac{r+1}{2}} (u_r  \pm \Lambda^{r+1}) \right)^2. \label{oddr} \end{equation} 
 For each singular point on the $u_1$ plane, we will have two cycles vanishing at the same time, which is non-generic higher singularity, which will be discussed later near {\bf figures \ref{sp4mlC}} and {\bf \ref{sp4mlQ}}.

Now what do we do for odd rank cases? The idea is to try the other $u_i$. We know that $u_r$ isn't a good coordinate on the moduli space, so let us try the $u_{r-1}$-plane.

For odd $r$, \eqref{1plane} gives a higher singularity, where all the vanishing cycles are paired up such that at any singular locus, we will see two mutually local cycles vanish simultaneously. Therefore we choose another slice of moduli space of \eqref{r1plane} which works for even and odd ranks, to give the same configuration of vanishing cycles as \eqref{SUinter}.
 On a hypersurface in a moduli space given as 
 \begin{equation}
 u_1=u_2= \cdots = u_{r-2}=0, \qquad u_r = {\rm const}, \label{r1plane}
 \end{equation}
 for any rank $r$, the following factors 
 \begin{equation} \Delta_x f_{\pm} = (-1)^{\left[ \frac{r}{2} \right]}   r^{r}  (u_{r-1})^{r+1} +  (-1)^{\left[ \frac{r+1}{2} \right]} (r+1)^{r+1} (u_r  \pm \Lambda^{r+1})^r  \label{anyRsu} \end{equation} of discriminant gives us $2(r+1)$ solutions on the $u_{r-1}$-plane as in {\bf figure \ref{rank9SU10u8theta0}}, and they correspond to singularities related to vanishing 1-cycles.  
 
   On the $u_{r-1}$-plane, choose $u_{r-1}=0$ as a reference point: Starting from here, we make a non-contractible loop surrounding each singular point on $u_{r-1}$-plane. At the reference point, $P_i$ (and $N_i$) points are ${\mathbb{Z}}_{r+1}$ symmetric among themselves on the $x$-plane. 
Demand $\frac{ u_r}{\Lambda^{r+1} }\in {\mathbb{I}} \cup \{ 0\}$ so that all $P_i$'s are ${\mathbb{Z}}_2$ symmetric to $N_i$'s on the $x$-plane\footnote{Note that if we allowed for real part of $u_r$, it will break $Z_2$ symmetry of \eqref{symPN}.}. Therefore all the branch points are on a circle at the reference point, as seen in {\bf figure \ref{rank9SU10xplane}}. As we vary $u_{r-1}$, we observe the vanishing cycles on the 
$x$-plane as in {\bf figure \ref{rank9SU10xplane}}.  Their non-vanishing intersection numbers are only
\begin{equation}
\nu_{i}^P  \cap  \nu_{i+1}^P=\nu_{i}^N \cap \nu_{i+1}^N= 1, \quad  \nu_{i}^N  \cap  \nu_{i}^P=  \nu_{i}^P  \cap  \nu_{i+1}^N= 2.
\end{equation}

\begin{figure}[htb]
\begin{center}
\includegraphics[
width=4in
]{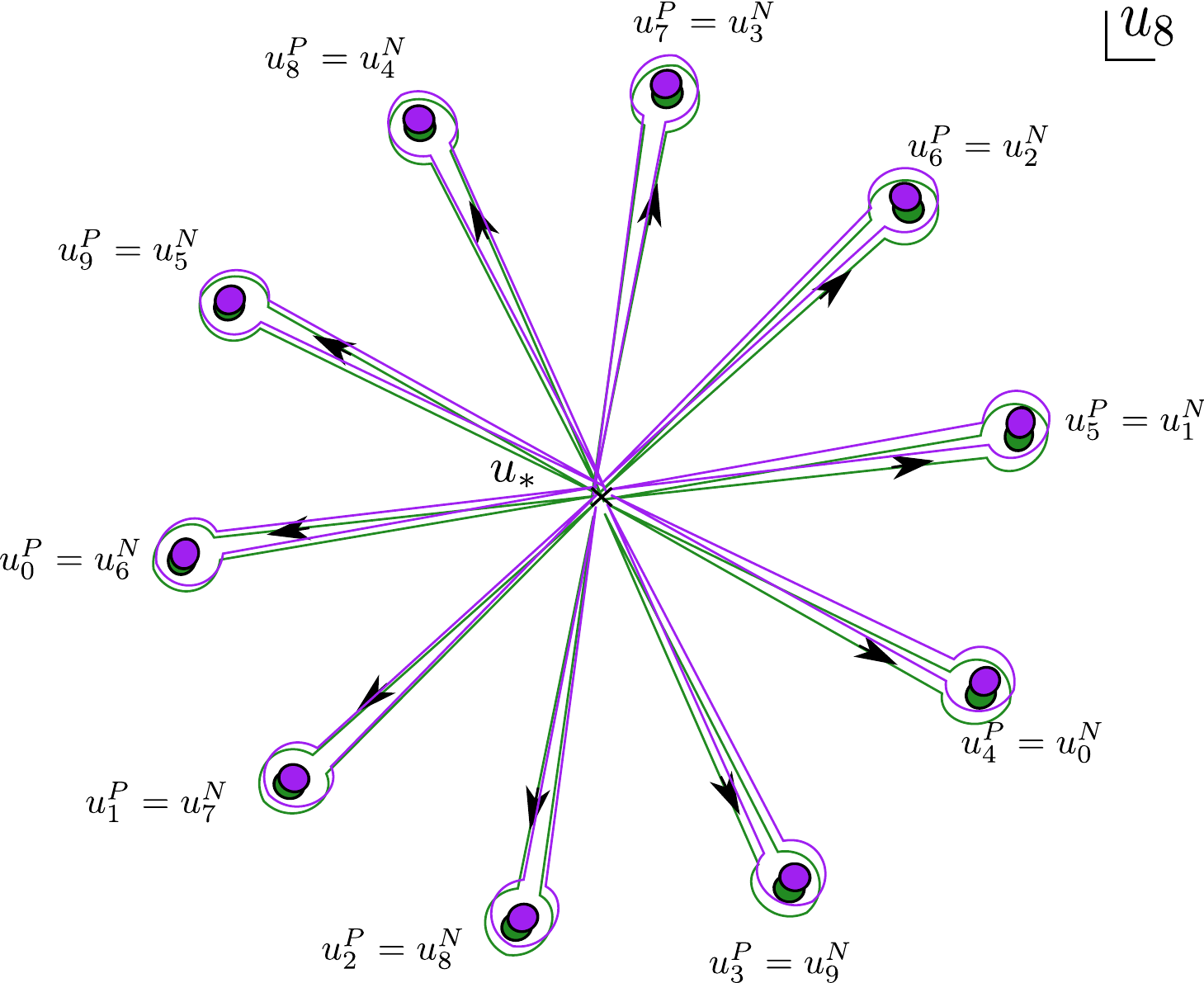}
\caption{Singularity structure of the $SU(10)$ curve on a moduli slice of the $u_8$-plane given by $u_1= u_2=\cdots = u_{7}=0$ and $u_9 =const $ with $\frac{ u_9}{\Lambda^{10} }\in {\mathbb{I}} \cup \{ 0\}$. From {\bf figure \ref{rank9SU10u8theta0}}, we vary $u_9$ such that $\theta \equiv \tan^{-1} \mathrm{Im} \left[ \frac{ u_r}{\Lambda^{r+1} }\right] =\frac{\pi}{18}$. Since this satisfies \eqref{oddrdoublepoint}, original $2(r+1)$ singular points all collide pairwise to form $r+1$ double-points, for odd $r$. 
As we go around each singular double-point on the $u_8$-plane, we have two vanishing cycle on the $x$-plane as shown in {\bf figure \ref{rank9SU10xplane}}. As denoted by subscripts here, $\nu^{P}_i $ and $\nu^{N}_{i+6}$ vanish at the same time on the $x$-plane. }
\label{rank9SU10u8thetaPi18}
\end{center}  
\end{figure}

\begin{figure}[htb]
\begin{center}
\includegraphics[
width=4in
]{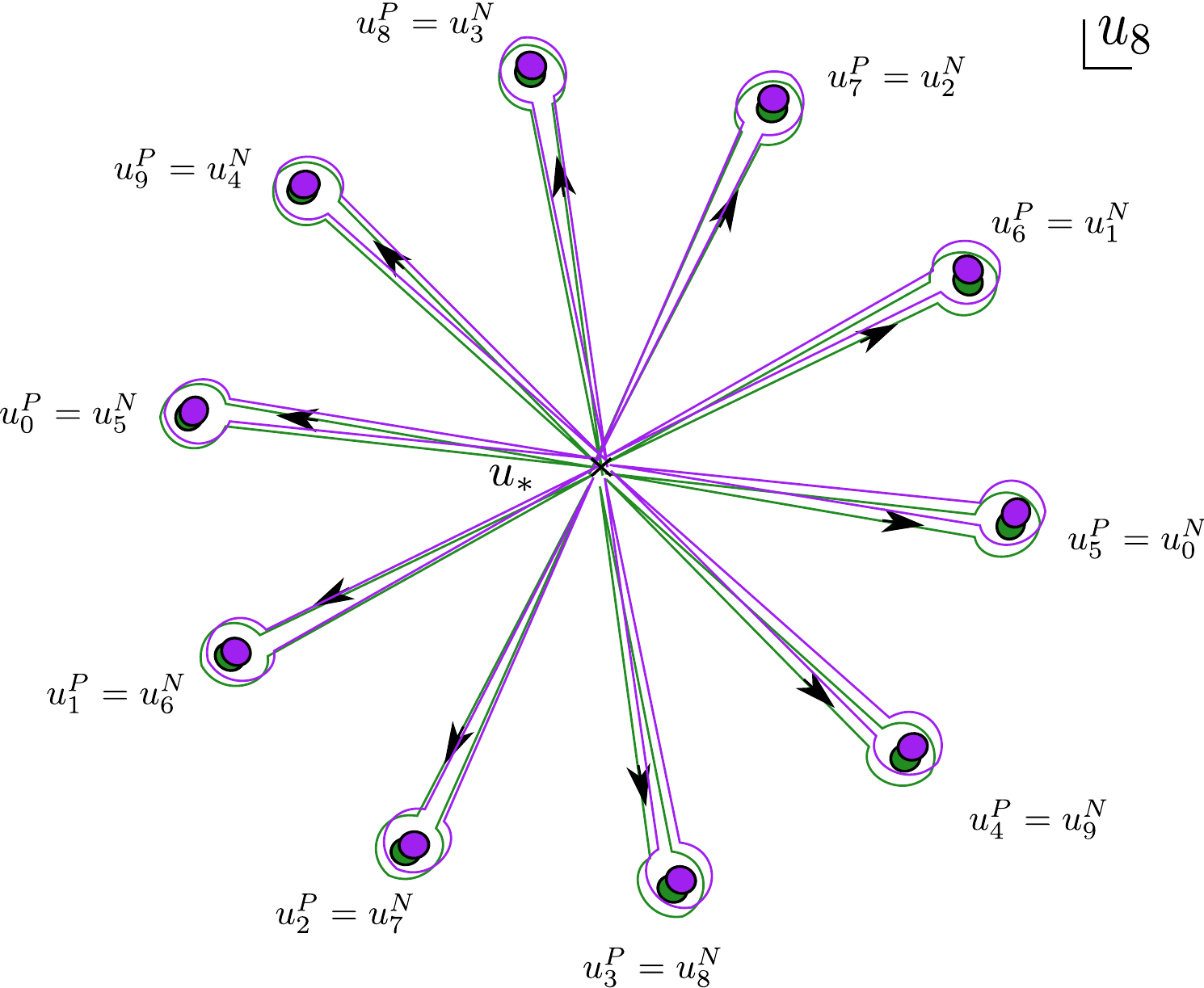}
\caption{Singularity structure of $SU(10)$ curve on a moduli slice of the $u_8$-plane given by $u_1= u_2=\cdots = u_{7}=0$ and $u_9 =const $ with $\frac{ u_9}{\Lambda^{10} }\in {\mathbb{I}} \cup \{ 0\}$. From {\bf figure \ref{rank9SU10u8theta0}}, we vary $u_9$ such that $\theta \equiv \tan^{-1} \mathrm{Im} \left[ \frac{ u_r}{\Lambda^{r+1} }\right] =-\frac{\pi}{18}$. Since this satisfies \eqref{oddrdoublepoint}, original $2(r+1)$ singular points all collide pairwise to form $r+1$ double-points, for odd $r$. 
As we go around each singular double-point on the $u_8$-plane, we have two vanishing cycle on the $x$-plane as shown in {\bf figure \ref{rank9SU10xplane}}. As denoted by subscripts here, $\nu^{P}_i $ and $\nu^{N}_{i+5}$ vanish at the same time on the $x$-plane. The $u_8$-plane behaviour is similar to {\bf figure \ref{rank9SU10u8thetaPi18}}, but on the $x$-plane, we have different choice of pairing of cycles which vanish simultaneously.}
\label{rank9SU10u8thetaminusPi18}
\end{center}  
\end{figure}

Again note the $\mathbb{Z}_2 \times \mathbb{Z}_{r+1}$ symmetry, where $\mathbb{Z}_2$ is for flipping $+ \leftrightarrow -$ (or $P$ $\leftrightarrow$ $N$) 
and $\mathbb{Z}_{r+1}$ is for changing index of $P_i$'s and $N_i$'s. Note that all roots are same distance from the origin which, in turn, means tht all roots are of equal 
lengths, i.e 
\begin{equation}
|P_i| = |N_j|. \label{symPN}
\end{equation}
 As we vary $u_r$, we will see that $u^P$ and $u^N$ points rotate on the $u_{r-1}$-plane, but in opposite directions.
Therefore, the singular points on the $u_{r-1}$-plane of {\bf figure \ref{rank9SU10u8theta0}} can collapse into $r+1$ points as in {\bf figures \ref{rank9SU10u8thetaPi18}} and {\bf \ref{rank9SU10u8thetaminusPi18}} for $\theta \equiv \tan^{-1} \mathrm{Im} \left[ \frac{ u_r}{\Lambda^{r+1} }\right]= \pm \frac{\pi}{18}$ respectively. More details will be given in the following subsection \ref{doublepointsing}.
 
\begin{figure}[htb]
\begin{center}
\includegraphics[
width=3.5in
]{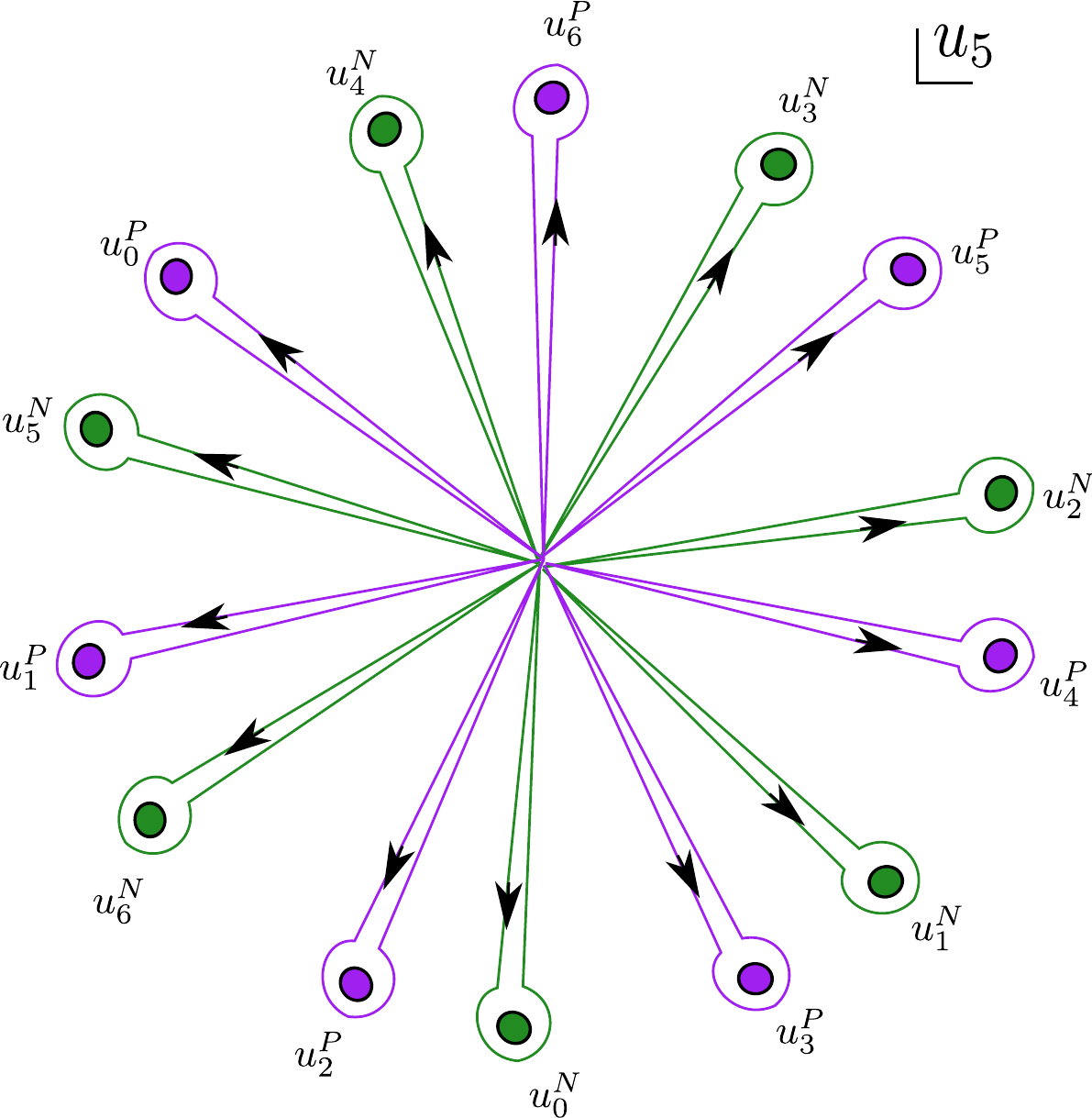}
\caption{Singularity structure of $SU(7)$ curve on a moduli slice, a $u_{5}$-plane given by $ u_1=u_2= \cdots = u_{4}=0$ and $\frac{ u_6}{\Lambda^{7} }\in {\mathbb{I}} \cup \{ 0\}$. We choose $ \theta \equiv \tan^{-1} \mathrm{Im} \left[ \frac{ u_r}{\Lambda^{r+1} }\right]$ to be in this range $0<\theta<\frac{\pi}{6}$, in order to dissatisfy \eqref{evenrdoublepoint}. As we go around each singular point on the $u_5$-plane, we have a vanishing cycle on the the $x$-plane as shown in {\bf figure \ref{su7flower}}. }
\label{rank6SU7u5smalltheta}
\end{center}  
\end{figure}

Less singular cases were given for rank 9 with $\theta=0$ in {\bf figure \ref{rank9SU10u8theta0}} and for rank 6 case with $0<\theta<\frac{\pi}{6}$ in {\bf figure \ref{rank6SU7u5smalltheta}}, where all singular points are separated on the $u_{r-1}$-plane.

\begin{figure}[htb]
\begin{center}
\includegraphics[
width=3.5in
]{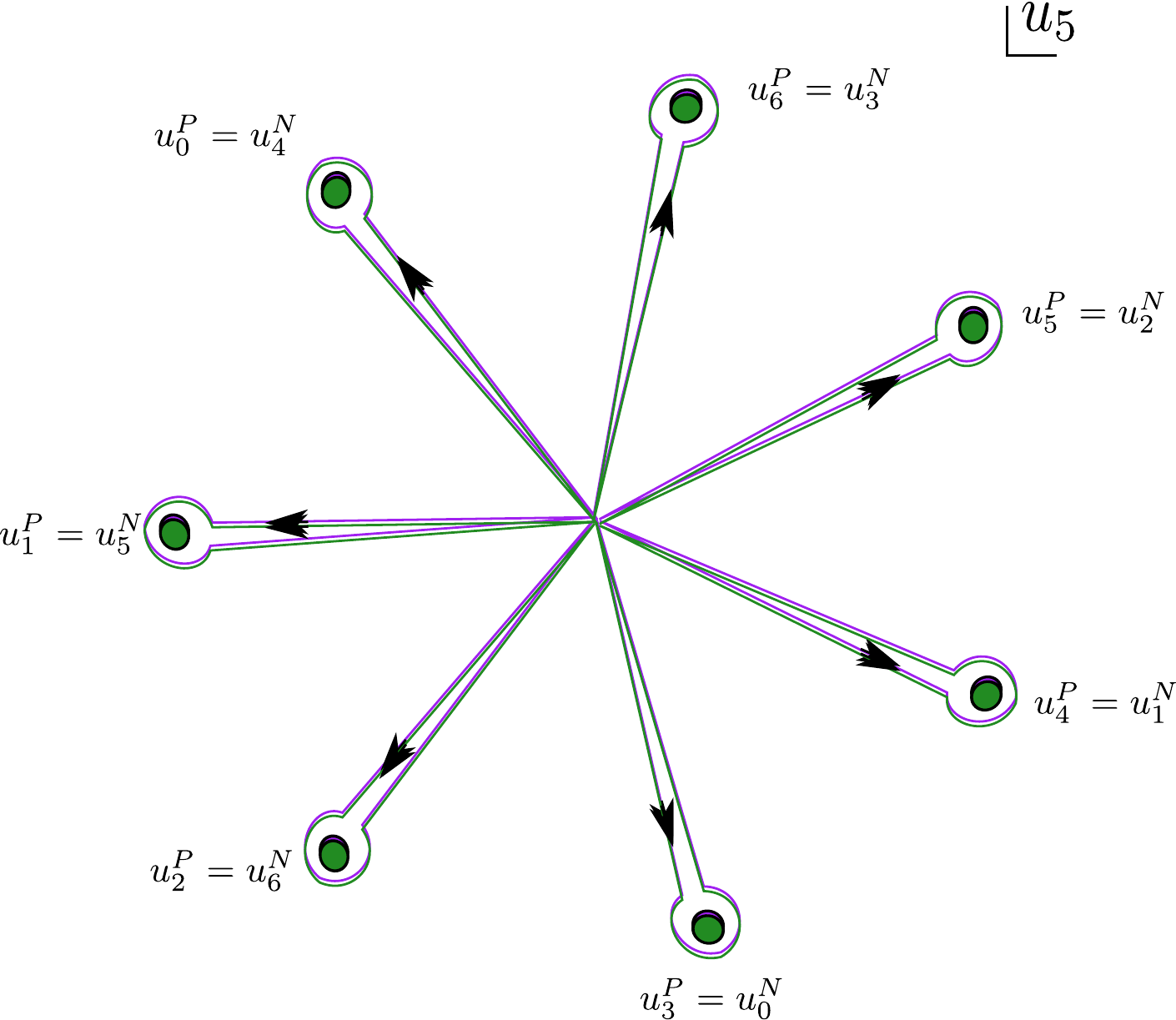}
\caption{Singularity structure of $SU(7)$ curve on a moduli slice, a $u_{5}$-plane given by $ u_1=u_2= \cdots = u_{4}=0$ and $ { u_6}=0$. Since this satisfies \eqref{evenrdoublepoint}, original $2(r+1)$ singular points all collide pairwise to form $r+1$ double-points, for even $r$. As we go around each singular double-point on the $u_5$-plane, we have two vanishing cycle on the $x$-plane as shown in {\bf figure \ref{su7flower}}. As denoted by subscripts here, $\nu^{P}_i $ and $\nu^{N}_{i+4}$ vanish at the same time on the $x$-plane.}
\label{rank6SU7u5zerour}
\end{center}  
\end{figure}

Before ending this subsection let us ask what does the $u_{r-1}$ plane give for even rank.
We will see that the underlying phenomena is similar.
Indeed the $u_{r-1}$ plane is like {\bf figure \ref{rank6SU7u5smalltheta}}, but at some discrete values of $\theta$, for example for $\theta=0$, we have 
{\bf figure \ref{rank6SU7u5zerour}} instead with pairwise collision of singular points.

\subsubsection{Shared roots between $\Delta_x f_\pm$ \label{doublepointsing}}

As we saw before, $f_\pm$ do not share roots for the $SU$ curves. However this does not prevent us from having repeated roots for $\Delta_x f_\pm$. When it happens, we will have two cycles each with $P$ and $N$ type 
that become massless at the same time (see discussions near {\bf figure \ref{sp4mlQC}} for more details). For example, we can ask whether \eqref{evenRsu}, \eqref{oddr}, and \eqref{anyRsu} have those singularities.

The answer to these questions, as we now know, lies in the discriminant. 
By observing the vanishing discriminant \begin{equation} \Delta_u \left[ (u^n+c_+) (u^n+c_-) \right] = n^{2n} c_+^{n-1} c_-^{n-1} (c_+-c_-)^{2n} =0,
\end{equation}
we know that 
$u^n+c_+$ and $u^n+c_-$ share roots only when $c_+-c_-=0$.
All the three relations, namely \eqref{evenRsu}, \eqref{oddr}, and \eqref{anyRsu} have the above forms with 
\begin{eqnarray}
{\bf I.} &&  u=u_1, \quad n=r+1, \\ 
 && c_\pm =\left((r+1)^{r+1} (u_r  \pm \Lambda^{r+1})^2  \right)/ \left(  2^2 (r-1)^{r-1}   \right) , \nonumber\\
{\bf II.} && u=u_1, \quad n=\frac{r+1}{2}, \\ 
  && c_\pm =      \left(\frac{r+1}{2}\right)^{\frac{r+1}{2}} (u_r  \pm \Lambda^{r+1})/\left( \frac{1-r}{2}\right)^{\frac{r-1}{2}}   , \nonumber\\ 
  {\bf III.} && u=u_{r-1}, \quad n=r+1, \\
 && c_\pm =\left( (-1)^{\left[ \frac{r+1}{2} \right]} (r+1)^{r+1} (u_r  \pm \Lambda^{r+1})^r  \right) / \left(  (-1)^{\left[ \frac{r}{2} \right]}   r^{r}    \right) \nonumber
\end{eqnarray}
 respectively.  (We restrict ourselves to $u_r / \Lambda^{r+1} \in \mathbb{I} \cup \{0 \} $
  and $\Lambda \ne 0$, therefore $u_r  \pm \Lambda^{r+1} \ne 0$ and $c_\pm \ne 0$.) 
 We now make the following observations.
\vskip.1in 

\noindent {\bf I.} ~ For \eqref{evenRsu}, $c_+=c_-$ can happen when $u_r=0$. In other words, for rank even case, on the $u_1$-plane (with $u_2=u_3= \cdots =0$), $u^P$ and $u^N$ points collide only at $u_r=0$ and we saw it already earlier near {\bf figure \ref{rank6SU7u1zerour}}.

\vskip.1in

\noindent {\bf II.} ~ For \eqref{oddr}, $c_+=c_-$ never happens.  In other words, for rank odd case, on the $u_1$-plane (with $u_2=u_3= \cdots =u_{r-1}=0$), $u^P$ and $u^N$ points never collide with each other on the $u_1$-plane.  
 
\vskip.1in

\noindent {\bf III.}~ For \eqref{anyRsu}, $c_+=c_-$ can happen when $(u_r  + \Lambda^{r+1})^r= (u_r  - \Lambda^{r+1})^r$. 

\vskip.1in

With
$ \theta \equiv \tan^{-1} \mathrm{Im} \left[ \frac{ u_r}{\Lambda^{r+1} }\right]$, it is equivalent to having 
 \begin{equation} \theta=\frac{ k \pi}{r }  \label{evenrdoublepoint} \end{equation} for even $r$, or
\begin{equation} \theta=\frac{ \pi }{ 2r } (2k+1) \label{oddrdoublepoint}\end{equation} for odd $r$, with $k \in \mathbb{Z}$.
In other words, with $u_1=u_2=u_3= \cdots =u_{r-1}=0$ and $\frac{ u_r}{\Lambda^{r+1} } \in \mathbb{Z}$, we will see $u^P$ and $u^N$ colliding on the $u_{r-1}$-plane, when above relations hold. For example, for rank 6 case, $\theta=0$ was given in {\bf figure \ref{rank6SU7u5zerour}}, and rank 10 case, $\theta = \pm \frac{pi}{18}$ cases were given in 
{\bf figures \ref{rank9SU10u8thetaPi18} and \ref{rank9SU10u8thetaminusPi18}} respectively. All these corresponded to having pairwise collision of $u^P$ and $u^N$ points on the $u_{r-1}$-plane.
 
Unlike the case of $u_r=0$ in {\bf figure \ref{rank6SU7u1zerour}}, each pair of simultaneously vanishing 1-cycle pairs are no longer simple 180 degrees rotation of each other in general.  
For $SU(10)$, at 16 discrete values of $\theta$ we will have 1-cycles vanishing simultaneously. As given in the formula below
\begin{eqnarray}
\theta = -\frac{\pi}{18}+\frac{k}{9}+n \pi \leftrightarrow u^P_{i+5} = u^N_{i+k}, \quad -3 \le k \le 4, \quad i, n \in \mathbb{Z} \label{su10doublesing}
\end{eqnarray}
at each choice of $\theta= -\frac{\pi}{18}+\frac{k}{9}+n \pi$, $\nu^P_{i+5}$ and $\nu^N_{i+k}$ will vanish at the same time on a point on the $u_8$-plane given by $u_8 = u^P_{i+5} = u^N_{i+k}$. Two examples are those given already in {\bf figures \ref{rank9SU10u8thetaPi18}} and {\bf \ref{rank9SU10u8thetaminusPi18}}. Other examples can be given simply by re-labelling of singular double-points on the $u_8$-plane.

Similar to the $SU(10)$ case in \eqref{su10doublesing}, now for $SU(7)$, 1-cycles vanish simultaneously at 10 discrete values of $\theta$ given below
\begin{eqnarray}
\theta =  \frac{k}{6}+n \pi \leftrightarrow u^P_{i+3} = u^N_{i+k}, \quad -2 \le k \le 2, \quad i, n \in \mathbb{Z}. \label{su7doublesing}
\end{eqnarray}
At each choice of $\theta =  \frac{k}{6}+n \pi$, $\nu^P_{i+3}$ and $\nu^N_{i+k}$ vanish at the same time on a point on the $u_5$-plane given by $u_5=u^P_{i+3} = u^N_{i+k}$. One example is given already in {\bf figures \ref{rank6SU7u5zerour}}. Other examples will be just simply given by re-labelling of singular double-points on the $u_5$-plane.

As we vary $\theta$, they get close to each other. One might worry that we will end up having mutually non-local pair 
to vanish simultaneously, however this does not happen. It happens only in $u_r >> \Lambda $ limit. As long as $\Lambda \ne 0$, we don't have to worry about that limit. More explicitly, note that in \eqref{su10doublesing} and \eqref{su7doublesing}, $\theta = \pm \frac{\pi}{2}$ is not included, and for all choices of $i, k$ the simultaneously vanishing cycles are mutually local.

\subsubsection{Dyon charges of massless BPS states}

We have now shown that we can always choose a moduli slice where all the vanishing cycles are arranged such that the intersection numbers are like \eqref{SUinter}, as the {\bf figures \ref{su7flower} and \ref{rank9SU10xplane}}. To read the dyon charges, all we now need are the vanishing cycles. These can be easily read off from the symplectic basis. 

\begin{figure}[htb]
\begin{center}
\includegraphics[
width=0.8\textwidth]{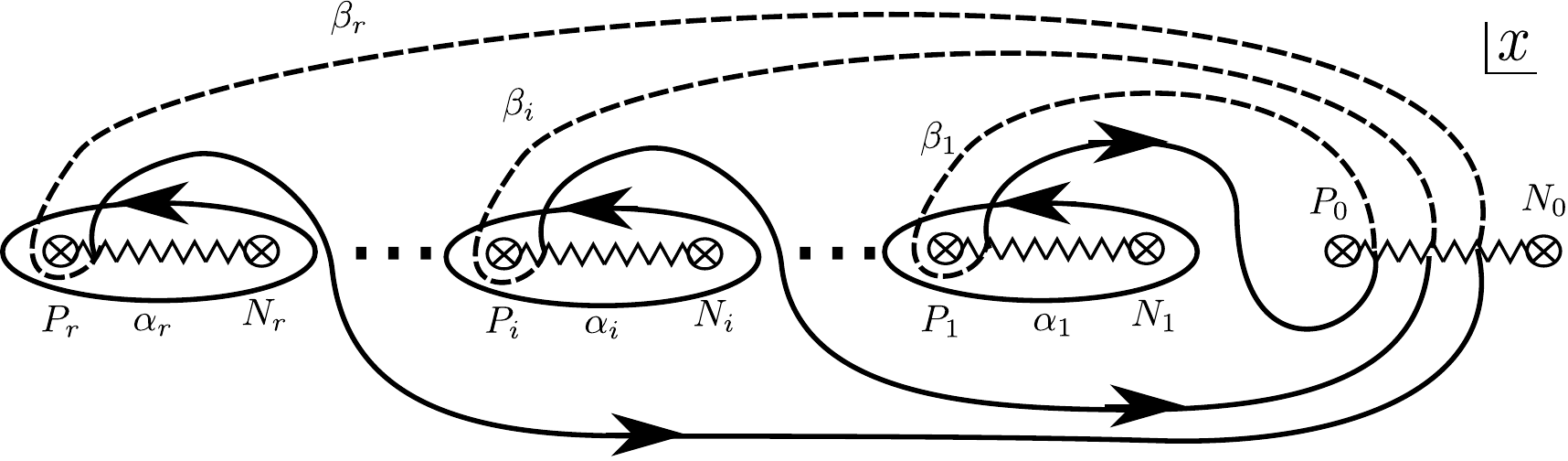}
\caption{A particular choice of symplectic basis cycles for $SU(r+1)$ curve}
\label{arcyclesbasis}
\end{center}  
\end{figure}

For example a symplectic basis given in {\bf figure \ref{arcyclesbasis}},
we can write down the cycles as:
\begin{eqnarray}    
\nu_{0}^{P}&=&\beta_{1}, \qquad   \nu_{0}^{N}=\beta_{1}+\sum_{i=1}^{r}\alpha_{i}+\alpha_{1}, \qquad  \nu_{r}^{P}=-\beta_{r}+\sum_{i=1}^{r-1}\alpha_{i}, \qquad \nu_{r}^{N}=-\beta_{r}-2\alpha_{r},  \nonumber   \\
\nu_{i}^{P} &=&\beta_{i+1}-\beta_{i}-\alpha_{i}, \qquad  
\nu_{i}^{N}=\beta_{i+1}-\beta_{i}+\alpha_{i+1}-2\alpha_{i}, \qquad i=1,\cdots,r-1. \label{sucharges}  
\end{eqnarray} 
In the {\bf figure \ref{arcyclesbasis}},
to make it convenient to draw and also to generalize to arbitrary ranks, we rearranged the branch cuts on the $x$-plane. However only the topological information is to be read off from here. This particular choice of symplectic basis is for convenience, and just matter of convention. We are choosing $\alpha_i$ cycles to go around each branch cut connecting $P_i$ and $N_i$ branch points. We are choosing $\beta_i$ cycles to connect between $P_0$ and $P_r$ branch points. However, as we learned in subsection \ref{whytrajectory} earlier, the trajectory is also important, not just the information of which two branch points are connected.

\subsection{How to reduce rank for $SU(r+1)$ \label{reduceSUrank}}

The motivation of this subsection is the following: imagine we make a brane construction in terms of D3-branes to realize this system. Then the number of branes in a stack will correspond to the rank. As we move a brane far away to infinity, we will see that the rank of the gauge group going down by one, without necessarily changing the type of gauge group. 

When the rank goes down what happens to the vanishing cycles and branch cuts? For sure, a branch cut has to be removed, because number of branch cut is $r+1$. One could imagine moving it far to infinity, or closing it down. There may not be a unique answer to this question, but here we will propose one method using 
the procedure of closing a branch cut by sewing it back. The procedure works in the following way.  
Looking at {\bf figure \ref{rank9SU10xplane}}, one could pick a branch cut connecting branch points $P_i$ and $N_i$, and for the vanishing cycles involved with this cut, we can merge them appropriately (in neighboring pairs), so that cycle starts or ends from those branch points. We can use 
\begin{eqnarray}
\nu^N_{i-1} + \nu^N_i&=&(\beta_{i }-\beta_{i-1}+\alpha_{i }-2\alpha_{i-1})+(\beta_{i+1}-\beta_{i}+\alpha_{i+1}-2\alpha_{i})\nonumber \\
&=& \beta_{i+1}-\beta_{i-1} +\alpha_{i+1} -2\alpha_{i-1}- \alpha_{i},\label{sumN} \\
\nu^P_{i-1} + \nu^P_i&=& (\beta_{i }-\beta_{i-1}-\alpha_{i-1})+(\beta_{i+1}-\beta_{i}-\alpha_{i})\nonumber \\
&=& \beta_{i+1} -\beta_{i-1}-\alpha_{i-1} -\alpha_{i},\label{sumP}
\end{eqnarray}
 as new vanishing cycles, instead of each one of the four cycles: $\nu^N_{i-1},\nu^N_i,\nu^P_{i-1} , \nu^P_i$. Note that now $\beta_i$ disappears, and once we set $\alpha_i=0$ by closing that branch cut down, \eqref{sumN} and \eqref{sumP} reduces back to the form of \eqref{sucharges}\footnote{If we are to use this result for brane construction of the 
$SU(r+1)$ SW theory, we will have to figure out what it means physically to set  $\alpha_i=0$.}. 
  
This method works smoothly to bring the rank from arbitrary high down to rank 2. From rank 2 to rank 1, there's a small subtlety that can be easily explained from {\bf figure \ref{su3cycles}}. Imagine here again we close down the branch cut connecting $P_1$ and $N_1$, and then make new vanishing cycles by merging $\nu_0 + \nu_1$. However it turns out that 
$\nu_0 + \nu_1=-\nu_2$ for both $P$ and $N$ types. 
In other words,
$SU(r+1)$ has $2(r+1)$ vanishing cycles. Each time as we go down in the rank, we lose 2 cycles. Then it appears that $SU(2)$ should have 4 vanishing cycles, 
but as it turns out, each of them are equal to each other
(up to sign) in the following way: 
\begin{equation}
\nu_{0}^{P}=\beta_{1},\qquad\nu_{0}^{N}=\beta_{1}+2\alpha_{1}, \qquad \nu_{1}^{P,N}=-\nu_{0}^{P,N}.
\end{equation}
At the end of the day $SU(2)$ only has 2 vanishing cycles. This difference for the rank 1 case was explained in \eqref{urbad}.
We will discuss similarly story for the $Sp$ case in subsection \ref{reduceSp}.

\newpage 

\section{Monodromies of pure ${\mathcal N}=2$ $Sp(2r)$ theories \label{spSec}}

In the last section we presented a detailed study of pure $SU(r+1)$ SW theories and computed their monodromies, singularity structures and the Argyres-Douglas loci. Here we
will extend the analysis to incorporate $Sp(2r)$ SW theories. The physics of $Sp(2r)$ theories is sufficiently different from their $SU$ counterparts rendering a detailed 
independent study possible. In fact for the $SU(r+1)$ theories we did not make much connections to the wall-crossing phenomena. It is now time therefore to make this 
connection more precise. Furthermore $Sp(2r)$ theories have an additional advantages of being represented in terms of branes in F-theory, a story that was elaborated 
in some details in \cite{DSW}. Here we will not venture towards that direction and restrict ourselves mostly to analysing the corresponding curves in these theories. 

 \subsection{Seiberg Witten curve for pure $Sp(2r)$ theories and root structure \label{sprootstructure}}
   
By taking no-flavor limit of \cite{ArgyresShapere}, as explained in detail in {\bf appendix \ref{PUREfromAS}}, we obtain the SW curve and SW 1-form for pure $Sp(2r)$ theory
in the following way:
\begin{equation}    
 y^{2}=f_{Sp(2r)}= f_C f_Q , \quad  \lambda   =  a \frac{dx}{2\sqrt{x}}  \log \left( \frac{   x f_C + f_Q+ 2\sqrt{x} y   }{  x f_C + f_Q -2\sqrt{x} y  } \right), \label{sp1curve1form}
\end{equation}    
with $f_C$ and $f_Q$ defined as:
\begin{equation}   
f_C  \equiv \prod_{a=1}^{r}\left(  x-\phi_{a}^{2}\right), \qquad f_Q \equiv  x f_C  +  16 \Lambda^{2r+2}.  \label{fCfQdef}
\end{equation}  
Let us make a few comments. As for the $SU$ case, one would expect an underlying $\mathbb{Z}_h$ symmetry here in the curve equation for the $f_Q$ part. However 
in contrast with the $SU$ case where both factors $f_\pm$ of hyperelliptic equation depend on $\Lambda$ as seen in \eqref{fpm}, 
note that for the $Sp$ case only $f_Q$ has $\Lambda$ dependence, while $f_C$ has no $\Lambda$ dependence. 
Additionally, 
since $f_{Sp(2r)}(x)$ is a $2r+1$-degree polynomial in $x$, $ y^{2}=f_{Sp(2r)}$ is a genus-$r$ Riemann surface. It has total $2r+2$ branch points on the $x$-plane: $2r+1$ branch points are finite roots to $f_{Sp(2r)}(x)=0$, and one extra branch point is at infinity on the $x$-plane.\footnote{Later in subsection \ref{inverseX}, we will discuss the same curve in an alternative equivalent form $ y^{2}=f_{Sp(2r)}^\prime$, with $x \rightarrow 1/x$ transformation. The polynomial $f_{Sp(2r)}^\prime$ will have a degree $2r+2$ in $x$, and the branch point at infinity of $ y^{2}=f_{Sp(2r)}$ will be brought back to origin of the $x$-plane.} Since $f_{Sp(2r)}$ factorizes, among $2r+1$ finite branch points, we see that $r$ branch points are $r$ solutions to $f_C(x) =0$, and $r+1$ branch points are $r+1$ roots to $f_Q (x)=0$, which we denote by $C_i$'s and $Q_i$'s
as implicitly given in \footnote{The origin of this notation is following: C and Q stand for classical and quantum. As clear from \eqref{fCfQdef}, $C_i$'s do not depend on $\Lambda$, that is only $Q_i$'s which depend on $\Lambda$.} 
\begin{equation} 
f_C =  \prod_{i=1}^{r} (x-C_i), \qquad f_Q=  \prod_{i=0}^{r} (x-Q_i).
\end{equation}
We also want to introduce $r$ gauge invariant moduli $u_i \in \mathbb{C}$ ($i=1,\cdots, r$) for these curves as implicitly given in
\begin{equation}   
\prod_{a=1}^{r}\left(  x-\phi_{a}^{2}\right) =  x^r + u_1 x^{r-1} + u_2 x^{r-2} +\cdots + u_r  .
\label{prodsumu}
\end{equation}
Or more explicitly, 
\begin{equation}
u_1=\left(  \sum_{a=1}^{r}-\phi_{a}^{2}\right), \qquad
u_2= \left(  \sum_{a=1}^{r}\phi_{a}^{2}\phi_{b}^{2}\right), \qquad   \cdots, \qquad
u_r =\prod_{a=1}^{r}\left(  -\phi_{a}^{2}\right)   .
\end{equation}
 When it is not ambiguous, we may use notation $u=u_1, v=u_2, w=u_3, \cdots$ for simplicity.

Observe in \eqref{fCfQdef} that $f_C=f_Q=0$ is possible only if $\Lambda=0$.  
In a quantum theory we demand $\Lambda\neq0$, therefore $f_C$ and $f_Q$ can never share a root: For any choices of moduli, $f_C$ and
$f_Q$ can never vanish at the same time. This allows the discriminant of the curve to factorize, 
somewhat similar to what we saw in the $SU(r+1)$ case, as: 
 \begin{eqnarray}   
  \Delta_x f_{Sp(2r)}& = &  (16 \Lambda^{2r+2})^{2r}  \left( \Delta_x f_{C}  \right) \left( \Delta_x f_{Q} \right). 
\end{eqnarray}  
We can study multiplicity of zero for $f_C$ and $f_Q$ separately without worrying about their roots getting mixed.\footnote{The branch points coming from $f_C$ and $f_Q$ never collide with the branch point at infinity before they collide with each other first, as will be discussed in subsection \ref{infAlone}. We want to consider only stable singularity - of two branch points colliding with each other - as discussed in \cite{SeibergWittenWithMatter}.} Again, just as in the $SU$ case, when we draw vanishing cycles and collision of branch points, we can use binary coloring. The branch points and vanishing cycles are all grouped into two mutually exclusive groups (for $Q$ and $C$ respectively.). 
 
The discriminants are polynomials in the moduli $u_i$'s, from the degree in each moduli, we can read off the number of solutions to the vanishing discriminant. Following arguments near \eqref{DiscDeg}, basically it is given by the power of the polynomial, except for the coefficient which is a constant term (just as $u_r$ for $f_C$, but not for $f_Q$ since $u_r$ appears as $x$-term in $f_Q$). 
\begin{eqnarray}   
\Delta_x f_C & =& \#  \left( u_{i}^{r} + \cdots \right) , \quad i\ne r \nonumber \\
   &=&  \# \left( u_r^{r-1} + \cdots \right)   , \nonumber \\
     \Delta_x f_Q  &=& \#  \left( u_{i }^{r+1} + \cdots \right)     .
    \label{urbadQ} 
\end{eqnarray}  
As similar to \eqref{urbad}, we again borrowed the notation from \eqref{DiscDeg} to denote the top degree in terms of each modulus. 

For each of $\Delta_x f_{C}$, there are $r$ solutions for any $u=u_i$ ($i\ne r$) which is not a constant piece.
However, for $u=u_{r}$ one degree lower as in first line of \eqref{urbadQ}. This is similar to the $SU$ case in \eqref{urbad}
See \eqref{degree2} for the explanation. However for the second line of \eqref{urbadQ}, the number of the solutions is always $r+1$ for any modulus. The reason is because $u_r$ is not a constant term of $f_Q$, while it is a constant term for $f_C, f_{\pm}$. Inside $f_Q$, $u_r$ is no longer blind to phase rotation on the $x$-plane.

Therefore for $r>1$, we expect $  {\color{forestgreen}r}+( {\color{purple}r+1})$ vanishing cycles for $Sp(2r)$, for bringing two {\color{forestgreen}C} (or {\color{purple}Q}) points together on $x$-plane. They are $2r+1$ solutions to the vanishing discriminant from the first and third equalities of \eqref{urbadQ}, corresponding to $2r+1$ massless dyons. For $r=1$, we cannot have the first equality of \eqref{urbadQ} any more, because there is only one modulus, and there is no solution to $\Delta_x f_C$. We can only satisfy the third equalities, with two possible roots. Therefore for $r=1$, we have only $(1-1)+(1+1)=2$ solutions. Since rank 1 case is studied much in past, we 
will therefore focus only on $r>1$ cases here.
 
At a generic point in $r_{\mathbb{C}}$-dimensional moduli space, $\Delta_x f \ne 0$ and all the branch points are separated on the $x$-plane. As we bring branch points together, we will go to a subspace with lower dimension. For example, it will eat up $1_{\mathbb{C}}$ degree of freedom to bring two branch points together on the $x$-plane. Each time we demand a branch point to collide with another, we lose $1_{\mathbb{C}}$ degree of freedom.  

As derived in {\bf appendix \ref{PUREfromAS}} using \cite{ArgyresShapere}, SW 1-form for the $Sp(2r)$ theory without matter is given by:
\begin{eqnarray}
\lambda  & =& a \frac{dx}{2\sqrt{x}}  \log \left( \frac{   x\prod \left(  x-\phi_{a}^{2}\right) + 8\Lambda^{2r+2} +\sqrt{x} y   }{  x\prod \left(  x-\phi_{a}^{2}\right) + 8\Lambda^{2r+2} -\sqrt{x} y  } \right), \label{OneForm}
\end{eqnarray}
for further convenience, this can be written in terms of $f_C$ and $f_Q$ as,
\begin{eqnarray}
\lambda  & =& a \frac{dx}{2\sqrt{x}}  \log \left( \frac{   x f_C + 8\Lambda^{2r+2} +\sqrt{x} y   }{  x f_C + 8\Lambda^{2r+2} -\sqrt{x} y  } \right)  \label{OneForm1} \\
& =& a \frac{dx}{2\sqrt{x}}  \log \left( \frac{  f_Q - 8\Lambda^{2r+2} +\sqrt{x} y   }{ f_Q -8\Lambda^{2r+2} -\sqrt{x} y  } \right) \nonumber.
\end{eqnarray}
Let us now observe how the 1-form behaves near a vanishing cycle. First we will restrict in moduli space of the curve, so that we are near a vanishing discriminant locus of the curve. Secondly, we restrict further along the Riemann surface, meaning that along the Riemann surface, we go to a neighborhood near a vanishing cycle. More explicitly we take $\Delta=0$ and $y=0$. For a given hyperelliptic curve, we will require the RHS to have a double root of $x$, and by going near that region, we require $y=0$, which means $f_C=0$ or $f_Q=0$. Plugging in $y=0,\ f_C=0$ and $y=0, \ f_Q=0$ in the first and the second line of \eqref{OneForm1} respectively, we confirm that the 1-form vanishes near a vanishing 1-cycle. Therefore, a singularity of the Seiberg-Witten curve will survive as a singularity of the Seiberg-Witten theory, with a possibility still remaining that Seiberg-Witten 1-form might add, but not subtract singularity. Singularity of the SW curve is a subset of the SW theory, and in this paper, we study the former.

  Branch points of the SW curve are the $r$ roots of $f_C$, $r+1$ roots of $f_Q$, and a point at infinity. Transforming $x\rightarrow
1/x$, the branch point at infinity comes to the origin, and subsection \ref{infAlone} explains that  stable singularity does not involve the branch point at infinity. Therefore, the branch point at infinity marked with $O_\infty$ does not participate in any of the vanishing cycles.

At discriminant loci $ \Delta_x f_{Sp(2r)}=0$, near the corresponding vanishing 1-cycle, the 1-form of \eqref{sp1curve1form} becomes infinitesimally small \cite{DSW}, far from becoming a delta function. This confirms that a singularity of the SW curve is indeed a singularity of the SW theory.  
At a generic point in $r_{\mathbb{C}}$-dimensional moduli space, $\Delta_x f \ne 0$ and all the branch points are separated. As we bring branch points together, we will go to a subspace with lower dimension. Maximal Argyres-Douglas theories occur at $r+1$ points computed in \cite{DSW}, where all the roots of $f_Q$ collide together\footnote{Scaling behaviour at maximal Argyres-Douglas points for pure ABCDE SW theory were studied in \cite{EHIY} and \cite{EH}, and there are two such points in moduli space for ADE groups. For B and C, number of maximal Argyres-Douglas points are $2r-1$ and $r+1$, and scaling behaviour is being studied \cite{Seo}.}.

\subsubsection{$x\rightarrow
1/x$ transform to bring a branch point at infinity to origin \label{inverseX}}
 
When we deal with hyperelliptic curve with odd degree, we necessarily have a branch point at infinity. It can be a hassle to keep track of the motion of the branch points with relation to the point at infinity. A simpler and more transparent way is that, we perform $x\rightarrow
1/x$ transform to bring a branch point at infinity to the origin.

Perform following change of variables, which corresponds to spherical projection on $x$ coordinate and rescaling of $y$ coordinate: 
\begin{eqnarray}
&& x \rightarrow -\frac{1}{\widetilde{x}}, \qquad
\widetilde{x} =\frac{ax+b}{cx+d}=-\frac{1}{x}, \nonumber \\
&& y \rightarrow \frac{\widetilde{y}}{(-\widetilde{x})^{r+1}}, \qquad
\widetilde{y} =\frac{y}{x^{r+1}}=(-\widetilde{x})^{r+1}y,
\end{eqnarray}
and applying on the odd-degree hyper-elliptic curve for pure $Sp(2r)$, we get the following series of algebraic manipulations: 
\begin{eqnarray}
 && y^{2} =\left( \ \prod_{a=1}^{r}\left( x-\phi _{a}^{2}\right) \right)
\left( x\prod_{a=1}^{r}\left( x-\phi _{a}^{2}\right) +16\Lambda
^{2r+2}\right)  \\
&& \frac{\widetilde{y}^{2}}{(\widetilde{x})^{2r+2}} =\left( \
\prod_{a=1}^{r}\left( -\frac{1}{\widetilde{x}}-\phi _{a}^{2}\right) \right)
\left( -\frac{1}{\widetilde{x}}\prod_{a=1}^{r}\left( -\frac{1}{\widetilde{x}}
-\phi _{a}^{2}\right) +16\Lambda ^{2r+2}\right)  \\
&& \widetilde{y}^{2} =\left( \widetilde{x}\ \prod_{a=1}^{r}\left( -1-\phi
_{a}^{2}\widetilde{x}\right) \right) \left( -\prod_{a=1}^{r}\left( -1-\phi
_{a}^{2}\widetilde{x}\right) +16\Lambda ^{2r+2}\widetilde{x}^{r+1}\right)  \nonumber \\
&& ~~~~~~~ = - \left( 1 + \sum_{i=1}^{r} \widetilde{u_i} \widetilde{x}^i  \right)  \widetilde{x}  \left( 1 + \sum_{i=1}^{r} \widetilde{u_i} \widetilde{x}^i  +(-1)^{r+1} 16\Lambda ^{2r+2}\widetilde{x}^{r+1}\right), \label{inverseXu}
\end{eqnarray}
which converts ($y, x$) to ($\widetilde{y}, \widetilde{x}$). 
In the last line of \eqref{inverseXu}, 
we rewrote in terms of Weyl-invariant moduli $\widetilde{u_i} = u_i (-1)^i $, the product becomes summation just as in \eqref{prodsumu}. 

Now absorbing phases into $y, \Lambda, u$'s, and dropping tildes for simplicity, 
 we obtain an even-degree hyper-elliptic curve expressed as
\begin{eqnarray}
y^{2}  &=&\left( 1 + \sum_{i=1}^{r}   u_i  {x}^i  \right)   {x}  \left( 1 +\sum_{i=1}^{r} u_i  {x}^i  +16\Lambda ^{2r+2} {x}^{r+1}\right) .  \label{evencurve}
\end{eqnarray} 
Let us now try another set of ($x, y$) transformations.  
Start from \eqref{sp1curve1form},
\begin{eqnarray}  
 y^{2} & = & \left(   \prod_{a=1}^{r}\left(  x-C_a\right)  \right)
\left(  x\prod_{a=1}^{r}\left(  x-C_a\right)  + 16 \Lambda
^{2r+2}\right)  ,
\end{eqnarray} 
 and demand $x\rightarrow 1/x, ~~ y \rightarrow y/x^{r+1}  {\prod_{i=1}^r C_i}$ with $C_a = 1/c_a, ~~ Q_a = 1/q_a$ to obtain
\begin{eqnarray}  
\prod_{i=1}^r C_i^2 y^{2} & = & x  \left(   \prod_{a=1}^{r}\left(  1 - x /c_a\right)  \right)
\left(   \prod_{a=1}^{r}\left(  1-x/c_a\right)  + 16 \Lambda
^{2r+2} x^{r+1} \right)   \nonumber \\
  y^{2} & = & x  \left(   \prod_{a=1}^{r}\left(  c_a - x  \right)  \right)
\left(   \prod_{a=1}^{r}\left(  c_a-x \right)  + 16 \Lambda
^{2r+2} \left( \prod_{i=1}^r c_i \right) x^{r+1} \right)  \nonumber   \\
  & = & x  \prod_{a=1}^{r}\left(  x -c_a  \right)   
   \prod_{a=0}^{r}\left( x- q_a  \right)   \nonumber \\
  & = & \#  x f_c f_q .    \label{inversecq}
\end{eqnarray} 
The results in \eqref{inversecq} and \eqref{evencurve} are equivalent, just like the two expressions given before $x$ inversion in \eqref{sp1curve1form}, \eqref{fCfQdef}, and \eqref{prodsumu}. (The only difference is writing down the polynomial in terms of roots or in terms of Weyl invariant moduli.)

\subsubsection{Stable singularity does not involve the branch point at infinity \label{infAlone}} 

In principle, multiple branch points can come arbitrarily close to each other on the $x$-plane. However, the correct question to ask is, which two come closer first. That is because 
we only want to consider stable singularity $-$ of two branch points colliding with each other. Rank 1 case is discussed in \cite{SeibergWittenWithMatter}: on the $x$-plane, two branch points approach each other faster than they approach the singularity (another branch point) at infinity.

Here we want to argue for general ranks, that the branch points coming from $f_C$ and $f_Q$ never approach with the branch point at infinity before they approach with each other first. In a language of the subsection \ref{inverseX}, it is same as showing that the branch points collide with each other before hitting the branch point at the origin. Starting from the \eqref{inversecq}, among whose roots $c_i$ and $q_i$, let us assume
\begin{equation}
  |q_r| \ll  |c_i|, |q_i|,   \qquad i=1,\cdots, r-1, \label{approx}
\end{equation}
and we will show that 
\begin{equation}
|q_r -c_r| \ll |c_r|, |q_r|. \label{proverel}
\end{equation} 
From \eqref{inversecq}, we demand $f_q=0$ for $x=q_r$, namely
\begin{equation}
   \prod_{a=1}^{r}\left(  c_a-q_r \right)  + 16 \Lambda
^{2r+2} \left( \prod_{i=1}^r c_i \right) q_r^{r+1} =0 ,
\end{equation}
which can be approximated using \eqref{approx} as
\begin{eqnarray}
-16 \Lambda
^{2r+2} \left( \prod_{i=1}^r c_i \right) q_r^{r+1} & = &  \prod_{a=1}^{r}\left(  c_a-q_r \right) \nonumber \\
& = &  \left(  c_r-q_r \right) \prod_{a=1}^{r-1}\left(  c_a-q_r \right) \nonumber \\
& \sim &  \left(  c_r-q_r \right) \prod_{a=1}^{r-1}   c_a    ,
\end{eqnarray}
so that finally we obtain
\begin{eqnarray}
 \left(  c_r-q_r \right)   \sim -16 \Lambda
^{2r+2}  c_r q_r^{r+1}  ,
\end{eqnarray} which proves \eqref{proverel}. While two branch points approach infinity, they approach each other faster. Therefore, we can ignore the branch point at infinity (or at the origin, if $x\rightarrow 1/x$ was performed) from the degeneration of the branch points.

\subsection{Monodromies for $Sp(2r)$ \label{monodromySp}}
Here we study monodromies and singularity structures of Seiberg
Witten curves for pure $Sp(2r)$ theory given by 
\begin{equation}
y^2 =  x \left(1+\sum_{i=1}^r u_i x^i \right)\left(1+\sum_{i=1}^r u_i x^i +x^{r+1} \right) 
\end{equation}
after appropriate coordinate transformations given in subsection \ref{inverseX} and setting $\Lambda=1$ without loss of generality.
In the light of the wall-crossing
phenomena, we do not expect the massless BPS dyon spectrum to be invariant.
Rather depending on where we are in the moduli space, we will have different
spectra.

\subsubsection{Vanishing cycles of $Sp(2r)$ curve in a moduli slice for $r \le 6$ and conjecture for higher ranks \label{conjectSp}}

\begin{figure}[htb]
   \begin{center}
        \includegraphics[width=\textwidth]{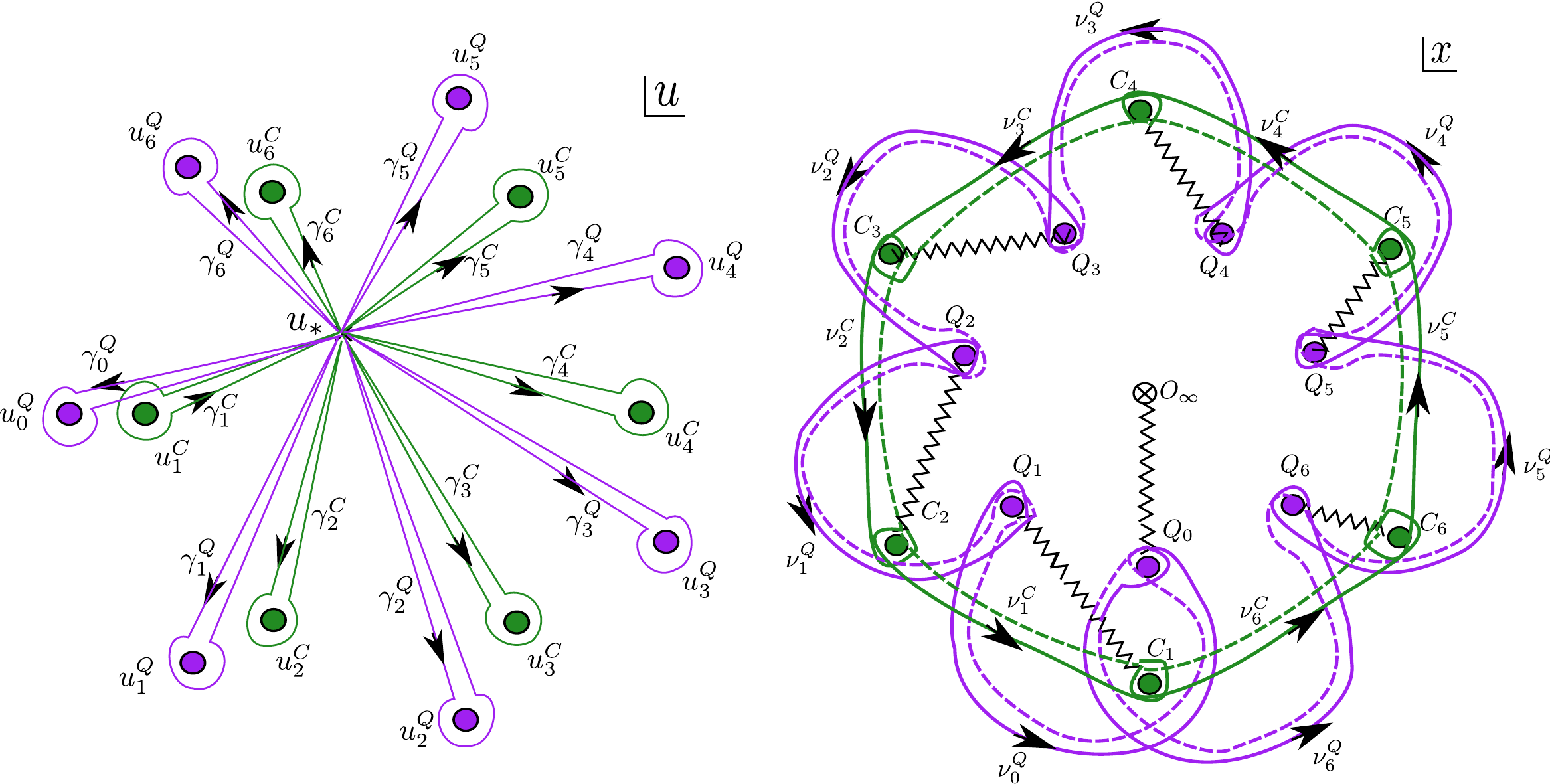}
    \end{center}
    \caption{Vanishing cycles of $Sp(12)$ curve at a moduli slice, which is given by a $u_1$-plane of $u_2=\cdots=u_{r-1}=0, u_r=1/9$}
    \label{sp6flower}
   \end{figure}

Restrict to a $u_1$-plane of the moduli space by fixing $u_2=\cdots=u_{r-1}=0$ and setting $u_r$ to be a fixed small number. Choose $u_1=0$ as a reference point. Up to rank 6, if we choose $u_r$ to be small enough\footnote{For example, choosing $u_r$ to be $-1/2$ (for rank $r=2$), $1/5$ ($r= 3$), $1/7$ ($r=4,5$), and $1/9$ ($r=6$), we obtain the similar configuration on the $x$-plane. All with units of $16 \Lambda^{2r+2}=1$.}, then branch points on the $x$-plane are arranged such that all the $Q_i$'s are surrounding origin $O_\infty$, and all the $C_i$'s are surrounding all the $Q$ points. Vanishing cycles have the following non-zero intersection numbers 
\begin{eqnarray}
\nu_{i}^Q\cap \nu_{i+1}^Q &=& \nu_{i}^C \cap \nu_{i+1}^C=-1,\quad \nu_{i}^Q\cap \nu_{i}^C= \nu_{i}^C \cap \nu_{i-1}^Q= 2  , \quad i = 1, \cdots, r \nonumber \\
\nu_{r}^Q\cap \nu_{0}^Q&=&-3, \quad \nu_{0}^Q\cap \nu_{r}^C=\nu_{1}^C \cap \nu_{r}^Q=2.  \label{spIntNum}
  \end{eqnarray}
{\bf Figure \ref{sp6flower}} shows such a configuration for the rank 6 case. 
We conjecture that for any rank $r$, it is always possible to choose $u_r$ to be small enough such that all the $Q_i$ points are inside the $\nu^C$ cycles, such that \eqref{spIntNum} hold.

Unlike the $SU(r+1)$ case, it is very difficult (if not impossible) to find an exact method to obtain the vanishing cycles for arbitrary high ranks of $Sp(2r)$. Instead, we compute the vanishing cycles in some patches of the moduli space for low ranks, and read off a pattern to conjecture for general ranks.

\begin{figure}[htb]
\begin{center}
\includegraphics[
width=.9\textwidth
]{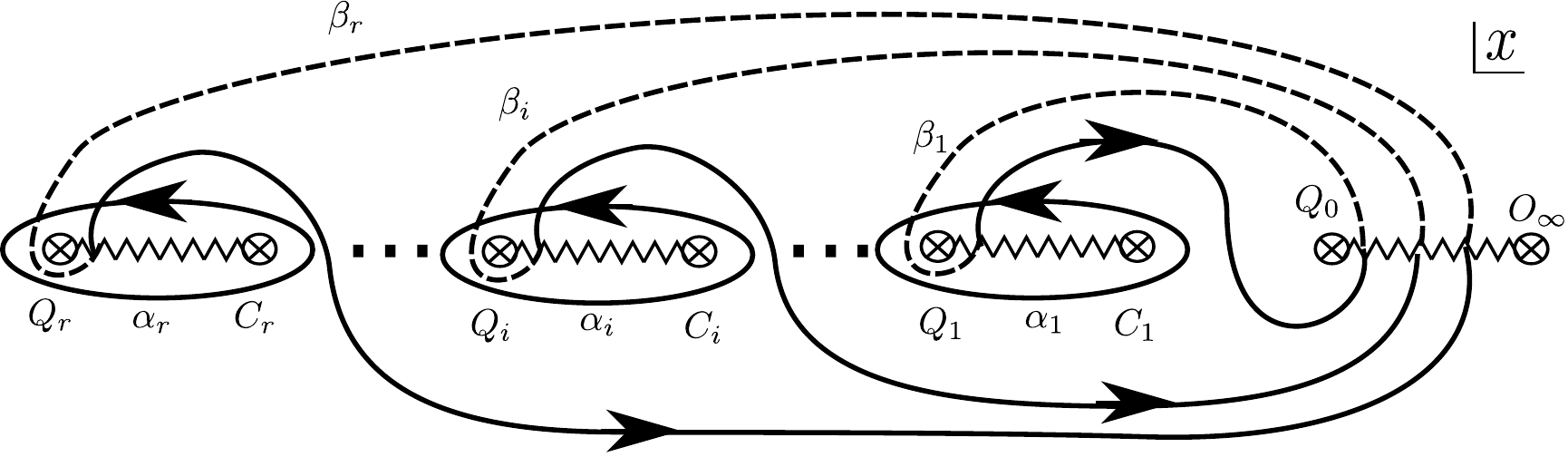}
\end{center}
\caption{A particular choice of symplectic basis cycles for the $Sp(2r)$ curve}
\label{sprcyclesbasis}
\end{figure}

   We write down these $2r+1$ vanishing cycles in terms of symplectic basis given in {\bf figure \ref{sprcyclesbasis}}. To make it convenient to draw and also to generalize for arbitrary ranks, we rearranged the branch cuts on the $x$-plane. However only the topological information is to be read off from here. This particular choice of symplectic basis is for convenience, and is just a matter of convention. We are choosing $\alpha_i$ cycles to go around each branch cut connecting $Q_i$ and $C_i$ branch points. We are choosing $\beta_i$ cycles to connect between $Q_0$ and $Q_r$ branch points. However, as we learned in subsection \ref{whytrajectory} earlier, the trajectory is also important, not just the information of which two branch points are connected. The various cycles now are:
    \begin{eqnarray}  
   \nu_{i}^{C}=-\beta_{i}+\beta_{i+1}+\alpha_{i+1} , \quad
 \nu_{i}^{Q}=   \nu_{i}^{C}  -\alpha_{i}+ \alpha_{i+1},     \quad
   i=1,\cdots,r-1  , \nonumber \\
 \nu_{r}^{C}=\beta_{1} -\beta_{r} -\sum_{i=2}^{r}\alpha_{i}, \quad
 \nu_{0}^{Q}=\beta_{1} + 2\alpha_1, \quad  
 \nu_{r}^{Q}= \nu_{r}^{C}+ \beta_1  + \alpha_1 -\alpha_{r}. \label{spcharges}
  \end{eqnarray}
It is interesting to note that all the C cycles add up to zero, i.e 
\begin{equation}
\sum_i^r \nu_{i}^{C} =0.
\end{equation}
This is no accident, and the result is 
independent of the choice of the symplectic basis. Having a linear relation among $2r+1$ cycles is not strange. We have $2r+1$ vectors on $2r$ dimensional vector space, and we expect to have at least one relation among them. However, for rank 2, case this translates into $\nu_{1}^{C}=-\nu_{2}^{C}$, and we will have $2r$ vanishing cycles (instead of $2r+1$) for rank 2, in this region of the moduli space.

\subsubsection{$M_\infty$ for rank 1 does not exist for higher ranks $r \ge 2$. \label{MinftyHighrank}}

We discussed earlier that a vanishing cycle on the SW curve is coming from a non-contractible loop around singularity in the moduli space. In order to have a non-contractible loop, the singular region needs to have a complex codimension 1. 
In this subsection, we will see that the singularity associated with monodromy at infinity is complex dimension 0 locus of the moduli space, or a complex codimension $r$ loci. Therefore this can give monodromy only for $r=1$ case. 
Alternatively, we could think of it as an intersection loci of various other complex codimension 1 singular loci.

 In the limit of small $\Lambda$, or more precisely 
 \begin{equation}  
\prod_{a=1}^{r} |C_a|  \gg \left|\Lambda ^{2r}\right|, \qquad \mathrm{or} \qquad  |u_i|  \gg \left|\Lambda ^{2r}\right| , \label{largeu}
\end{equation} 
the roots
can be expanded as below: 
\begin{eqnarray}
Q_{a} &\sim &C_{a}  ,\qquad a=1,\cdots,r ,\nonumber  \\
Q_0 &\sim&\frac{16\Lambda ^{2r+2}}{\prod_{a=1}^{r}\left( -C_a\right) }=\frac{16\Lambda ^{2r+2}}{u_r}.
\end{eqnarray}
On the $x$-plane, the branch cuts connecting $Q_a$ and $C_a$ become arbitrarily small, and a cut connecting $Q_0 \sim 0$ and the singularity at infinity become a semi-infinite line. 

Does this regime of \eqref{largeu} gives us a monodromy? The answer turns out to be affirmative for rank 1, but negative for rank $r>1$ cases. In order to count the dimension of the region given by \eqref{largeu}, we perform a $u \rightarrow 1/u$ transformation, the large $u_i$ region of \eqref{largeu} becomes a neighborhood near origin $u_i^\prime =1/u_i \ll 1$, a complex codimension $r$ locus. For rank 1 case, a singular locus near the origin of the moduli space does provide a non-contractible loop around the locus, therefore we obtain $M_\infty$. However, for higher ranks, we have more directions in the moduli, so any closed loop can be contracted without having to wrap around the neighborhood near the origin. 
A generic $1_\mathbb{C}$ dimensional surface in the moduli space will not intersect with this region \eqref{largeu}.

Since the analysis is purely based on rank and dimension, we expect similar behaviour for other gauge groups as well. We will therefore not elucidate this case-by-case 
basis but instead go to an explicit example where we can argue for the wall-crossing phenomena.

\subsection{Singularity structure of $Sp(4)=C_2$: Jumping BPS spectra and wall-crossing \label{Sp4}}

\begin{figure}[htb]
\begin{center}
\includegraphics[width=0.8\textwidth]{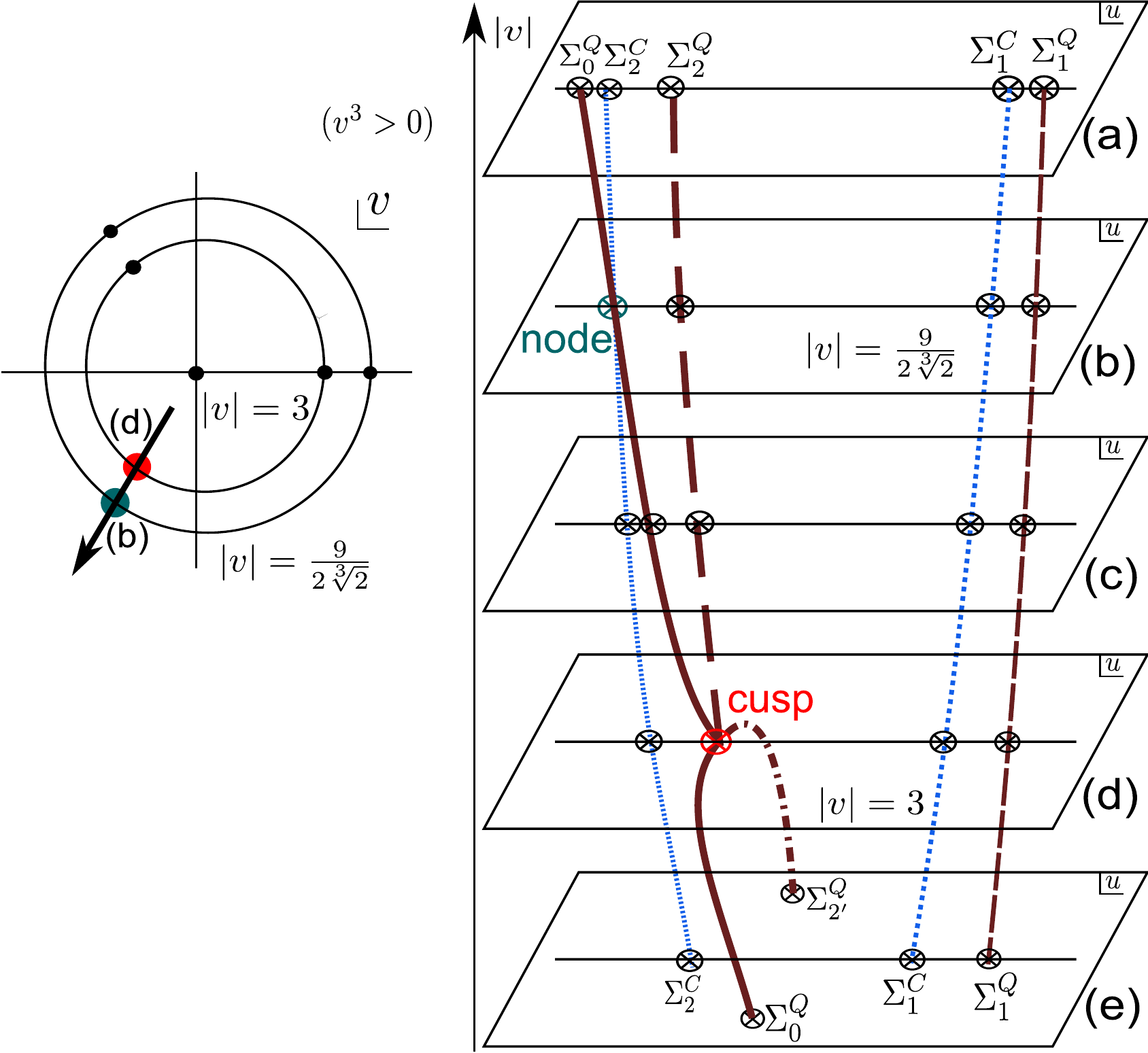}
\end{center}
\caption{A slice of moduli space for pure $Sp(4)$ Seiberg-Witten theory.
At $v=0$, $\Sigma^C_1$ and $\Sigma^C_2$ intersect. $\Sigma_2^Q$ change into $\Sigma_{2^\prime}^Q$ at a cusp point (note different types of dashing)}
\label{sp4jihye1}
\end{figure}

So far we studied vanishing cycles at the intersection of vanishing discriminant and a dim-$1_{\mathbb{C}}$ hypersurface of the moduli space (at singular points on the various $u$-planes). Here we will see how vanishing cycles change as we vary the choice of hypersurface. 
We present here a detailed example of jumping BPS spectra for the $Sp(4)$ SW curve. Its 2 complex dimensional moduli space is spanned by $\{u\equiv u_1, v\equiv u_2 \} \subset \mathbb{C}$. Let us take a $3$ real dimensional slice of the
moduli space for viewing convenience\footnote{A similar study was done in \cite{KLYTmndrmSU} for $SU(3)$ and in \cite{LPGg2} for $G_2$ gauge group, where they also take a slice in moduli space to show the discriminant loci and their intersection.}. For example, to reduce one real degree of freedom, we will fix the
phase of $v$ such that $v^3 \in \mathbb{R}$, as in the left side of
{\bf figure \ref{sp4jihye1}}.
One might ask why we are choosing this particular moduli section. We will soon see that this is where we observe various exotic phenomena of having multiple mutually local and non-local BPS states becoming massless. 

Each of the five $u$-planes, marked by (a) through (e) are slices at
different magnitude of $v$. The blue and brown curves denoted by $\Sigma$'s are where we have at
least one massless dyons. These ${\rm codim}_{\mathbb{C}}$-$1$ singular loci are captured by the vanishing
discriminant of the curve. When $\Sigma$'s intersect, we have a worse singularity:
massless dyons coexist at these ${\rm codim}_{\mathbb{C}}$-$2$ loci. The SW curve degenerates into either a cusp or a node form. The shape of intersection loci of $\Sigma$'s also take cusp or node form respectively, each leading to different
kind of singularity (mutually non-local and local). Similar phenomena occur for pure $SU(3)$ theory \cite{KLYTsimpleADE}.

A careful reader might have noticed that when $\Sigma$'s intersect each other, it is seen as colliding of singular points on the corresponding $u$-plane. For example, on the $u$-planes marked by (a,c,e), there are five singular points where $\Sigma$'s pierce through. On the $u$-planes marked by (b) and (d), two of the singular points are on top of each other. 

From the earlier part of the paper (see subsection \ref{EXandDisc}), one might recall that collision of branch points on the $x$-plane was captured by vanishing of discriminant operator with respect to $x$, $\Delta_x$. For example, $\Delta_x f=0$ is where we have branch points colliding on the $x$-plane. Similarly, when the singular points collide on the $u$-plane, it is captured by another discriminant operators with respect to now a new variable $u$, $\Delta_u$. For example, see the right side of
{\bf figure \ref{sp4jihye1}} for an example for  $Sp(4)=C_2$. Various $\Sigma$'s meet in the moduli space, and they are captured by making the $u$-plane slices. We can count number of intersection points between $\Sigma$'s and each $u$-planes. When this number reduces, this gives a candidate for singularity. We will explain more in this direction near {\bf figure \ref{figure8}} later. 

 In the case of $Sp(4)=C_2$, these are captured by vanishing double discriminant as below:
\begin{eqnarray}   
 \Delta_u \Delta_x f_{Sp(4)}&=& 2^8 v (v^3-3^3)^3 (2^4 v^3- 3^6)^2 , 
   \end{eqnarray}    
   whose roots 
   \begin{equation}
v=\Bigg\{ 0, 3 \alpha_3^i, \frac{9}{2 \sqrt[3]{2}} \alpha_3^j \Bigg\}
   \end{equation}
   are marked by seven dots in the left of the {\bf figure \ref{sp4jihye1}}.  Some of these (all six except for $v=0$) correspond to having two massless BPS dyons. In subsection \ref{exteriordVSdd} we will discuss how $d\Delta_x f=\Delta_x f=0$ is equivalent to having two massless BPS dyons. $v=0$ does not satisfy that relation, but the rest 6 does (with the proper choice of $u$ value). 
 When $v=0$, on the $u$-plane two $\Sigma_C$'s collide, but it does not translate into having two massless dyons, instead just one ($\nu_C$). 
Note the degeneracy of roots to the double discriminant. The roots $3 \alpha_3^i$ and $ \frac{9}{2 \sqrt[3]{2}} \alpha_3^j $ each have degeneracy 3 and 2. And in next section we will learn that this is a universal criteria for having mutually non-local and local massless dyons. This double discriminant method was used in \cite{APSW} and \cite{ACSW,AW} etc, respectively for rank 2 and rank 1 with flavors. Actually vanishing double discriminant is a necessary condition for having multiple massless dyons, but in subsection \ref{exteriordVSdd} we will see that it is not a sufficient condition. Asking for vanishing double discriminant will give all the candidates of singularity related to having multiple BPS states. However, not all candidates will survive, as we will give more details near {\bf figure \ref{figure8}} later.

\begin{figure}[htb]
        \begin{center}
\includegraphics[height=19.3cm]{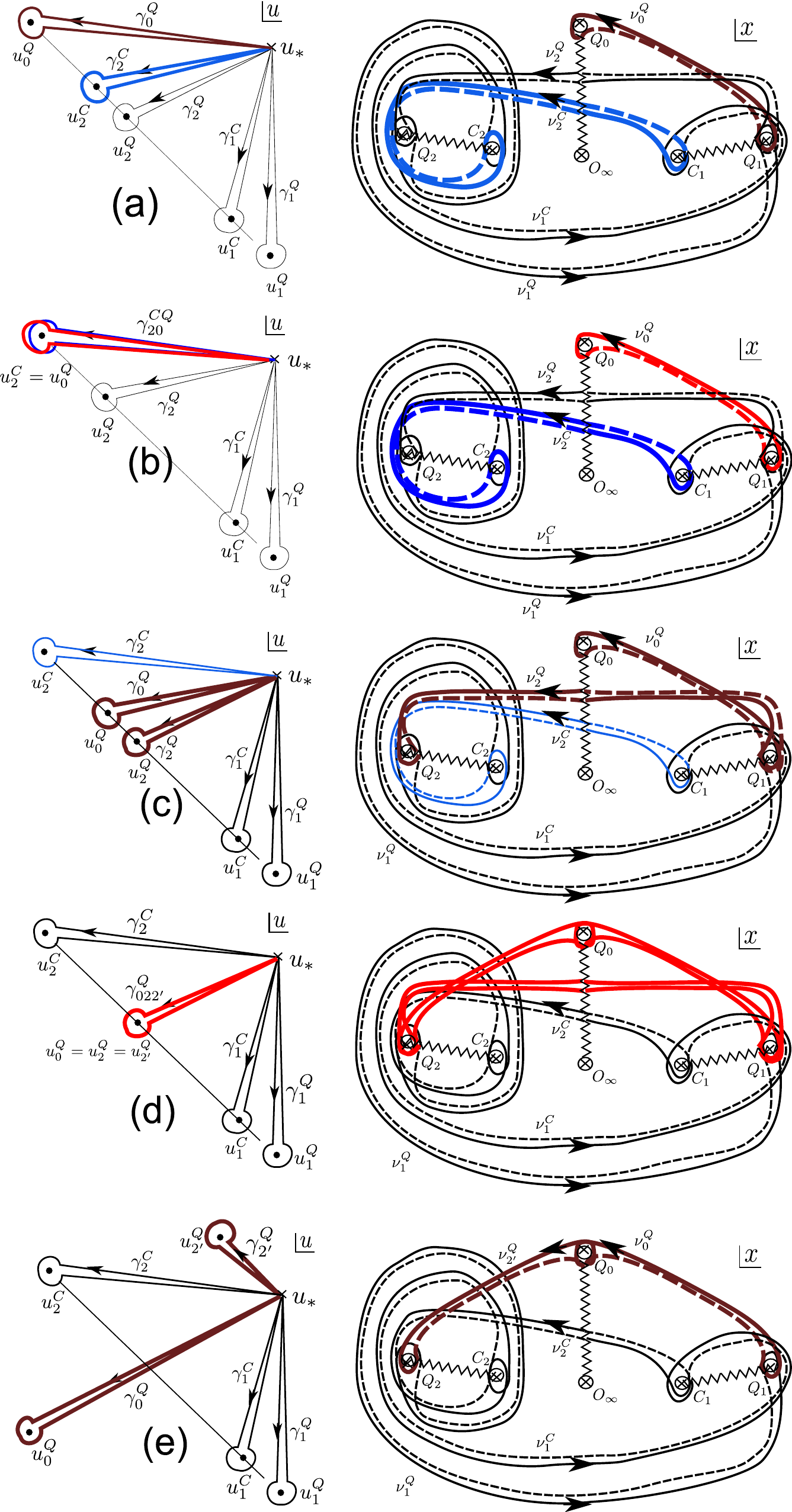}
        \end{center}
        \caption{Singularity structures at slices of the moduli space for pure $Sp(4)$ Seiberg-Witten theory.}\label{sp4jihye2}
        \end{figure}

On the right of {\bf figure \ref{sp4jihye2}}, each of the five $u$-planes, marked with (a) to (e), are slices of the moduli space at
different magnitude of $v$. ${\rm Codim}_{\mathbb{C}}$-1 discriminant loci of the SW curve, $\Delta_x f=0$, are marked by $\Sigma$'s, where at
least one dyon becomes massless. They intersect at more singular ${\rm codim}_{\mathbb{C}}$-$2$ loci specified by $\Delta_x f=d(\Delta_x f)=0$: two dyons become massless simultaneously. Here {\emph {double discriminant}} also vanishes $\Delta_u \Delta_x f=0$, as seen by colliding points on the $u$-planes marked (b) and (d) in the right of the {\bf figure \ref{sp4jihye2}}, which is as expected from \cite{APSW}. 

When $\Sigma$'s intersect, they can intersect like node and cusp, as in (b) and (d) of {\bf figure \ref{sp4jihye2}} respectively. When it happens, the SW curve itself degenerates into either node or cusp, each leading to a different
kind of singularity - mutually local and and mutually non-local (Argyres-Douglas) respectively. And this can be captured by order of vanishing (2 and 3 respectively) of the double discriminant\footnote{We obtained this result rather empirically while studying vanishing cycles, but later learned that it was used implicitly in \cite{ACSW,AW} to classify rank 2 curves.}.

For each $u$-planes marked by (a) to (e) of {\bf figure \ref{sp4jihye1}}, we
have drawn the corresponding $x$-planes in {\bf figure \ref{sp4jihye2}} displaying the vanishing cycles
on the $x$-plane, for each slice.  
For lack of space, we won't give the full details of how we read off vanishing cycles on the $x$-plane, but partial explanation and an example for slice (e) is given in subsection \ref{RelativeChoreo} near {\bf figure \ref{concise}}.
Let us have a closer look staring from the
top slice marked as (a).

\begin{description}
\item[(a)] At first, the given $u$-plane and $\Delta_x f_{Sp(4)} =0$ loci (marked as $\Sigma^{C,Q}_{i}$'s in {\bf figure \ref{sp4jihye1}}) intersect at five points $u^{C,Q}_{i}$'s. We call them singularity points because they are associated with vanishing 1-cycles on the $x$-plane. Each of the 5 singular points on the $u$-plane ($u^{C,Q}_{i}$'s) is responsible for one vanishing 1-cycle ($\nu^{C,Q}_{i}$'s with corresponding choice of sub-/super- scripts). By observing the relative trajectory of branch points on the 
$x$-plane, we can read
off the 5 vanishing cycles as below:
\begin{eqnarray} 
\nu_{0}^{Q}=\beta_{1} , \quad
  \nu_{1}^{Q}=-\beta_1 + \beta_{2} +\alpha_{1} +4 \alpha_2    , \quad  
  \nu_{2}^{Q}=-\beta_1-\beta_{2}-\alpha_{1}-2\alpha_{2}, \nonumber \\
  \nu_{1}^{C}=-\beta_1 + \beta_{2} +2\alpha_{1} +3 \alpha_2  , \quad 
    \nu_{2}^{C}=-\beta_1 - \beta_{2} - \alpha_2 .  \label{cycle}
  \end{eqnarray} 
  
All the discriminant loci $u^{C,Q}_{i}$'s are separated on the $u$ plane here.  
  Especially, two vanishing cycles $\nu_{0}^{Q}$ (brown) and $\nu_{2}^{C}$ (blue) vanish at two different moduli loci 
 $u_{0}^{Q}$ and $u_{2}^{C}$ respectively. 

\item[(b)] As we change the moduli $|v|$ to go from (a) to (b), singular points on the $u$-plane, $u^Q_0$ and $u^C_2$, now collide, and corresponding 1-cycles $\nu^Q_0$ and $\nu^C_2$ vanish simultaneously at the same place in the moduli space.   
However these two cycles are mutually local, and responsible branch points on the $x$-plane collide pairwise ($C_1 \leftrightarrow C_2$, $Q_0 \leftrightarrow Q_1$). The SW curve degenerates into a node form, $y^2 \sim (x-C)^2 (x-Q)^2 \times \cdots$. 
Singularity loci (of vanishing $\Delta_x f $) $\Sigma_{0}^{Q}$ and $\Sigma_{2}^{C}$ intersect at $|v|=\frac{9}{2\sqrt[3]{3}}$, with node-like
crossing.  

\item[(c)] We can change the moduli $|v|$ further to reach the configuration depicted in (c) i.e $u^Q_0$ and $u^C_2$ are separated again as in (a). Dyon charges of the vanishing cycles  for (a), (b), and (c) remain unchanged as \eqref{cycle}, when we go through node-like singularity of (b).  
Note that two reds points $u_{0}^{Q}$ and $u_{2^{\prime }}^{Q}$
are separated but they are running toward each other, in preparation for the next part of the scenario.

\item[(d)] As we change the magnitude of the moduli $v$, discriminant loci $\Sigma _{0}^{Q}$ and $\Sigma _{2}^{Q}$ intersect on the $u$-plane of  
 $|v|=3$, forming a cusp-like singularity. Two singular points $u_{0}^{Q}$ and $u_{2}^{Q}$ collide on the $u$-plane, and the vanishing cycles $\nu
_{0}^{Q}$ and $\nu _{2}^{Q}$ {\it merge}. In other
words, three branch points $Q_{0,1,2}$ collide at the same time on the $x$-plane, unlike the case (b) of {\bf figure \ref{sp4jihye2}} where four branch points collide pairwise. Two cycles $\nu_{0}^{Q}$ and $\nu_{2}^{Q}$ become massless at the same time, but they are mutually {\emph non-local}. The SW curve
degenerates into a cusp form $y^{2}\sim (x-a)^{3}\times \cdots $, giving
Argyres-Douglas theory with $SU(2)$ singularity, with two mutually non-local massless BPS dyons. 
\item[(e)] After passing Argyres-Douglas loci, in (e), we again have 5 separated singular points on the $u$-plane. Especially $u_{0}^{Q}$ and $u_{2}^{C}$ are separated. However note that the BPS dyon charges of vanishing cycles changed  from the cases of (a), (b), and (c) given in  \eqref{cycle} as we go through cusp-like (or Argyres-Douglas) singularity of (d). Instead of $\nu^Q_2$, we have a new vanishing cycle
\begin{equation}
 \nu_{2^\prime}^{Q}=-\beta_{2}-\alpha_{1}-2\alpha_{2} = \nu_{0}^{Q}+\nu_{2}^{Q}. \label{chargeJump}
 \end{equation}
Considering that Argyres-Douglas loci live inside the wall of marginal stability \cite{ShapereVafa}, this jump of BPS charge is refreshing to observe. 
\end{description}

One might ask why such exotic phenomena occur at different values of $v$. Is there a systematic method to locate the loci where multiple dyons become massless? This will be answered in the following section \ref{doublediscsection} where we will discuss the method for both $SU(r+1)$ and $Sp(2r)$ groups.  
 
Lastly, we can comment on the degeneracy and charge conservation. Our method a priori does not say anything about degeneracy, however, considering charge conservation, we could see some constraint near an Argyres-Douglas point.
Degeneracies in each segment $d^Q_{0a} = d^Q_{0c}$ and $d^C_{2a} = d^C_{2c}$ may take any values, because at (b), dyon changes do not change. However, from \eqref{chargeJump}  and charge conservation
\begin{equation}d^Q_{0c} \nu_{0}^{Q}+d^Q_{2c} \nu_{2}^{Q} = d^Q_{0e} \nu_{0}^{Q}+d^Q_{2^\prime e}  \nu_{2^\prime}^{Q},  \end{equation}
 one has
\begin{equation}d^Q_{0c} \nu_{0}^{Q}+d^Q_{2c} \nu_{2}^{Q} = d^Q_{0e} \nu_{0}^{Q}+d^Q_{2^\prime e}   \nu_{0}^{Q}+d^Q_{2^\prime e}  \nu_{2}^{Q}  . \end{equation}
Coefficients of  $\nu_{0}^{Q}$ and $\nu_{2}^{Q}$ must match from the above two equations, giving us the following two relations 
\begin{equation}
d^Q_{0c}   = d^Q_{0e}  +d^Q_{2^\prime e}, ~~~~~~~ d^Q_{2c} =  +d^Q_{2^\prime e} . \end{equation}
Therefore combining everything together, we have  
\begin{equation}
d^Q_{0c} = d^Q_{0e}+d^Q_{2c}, ~~~~~~~
d^Q_{2c} = d^Q_{2^\prime e} . \end{equation}

\subsection{How to reduce rank for $Sp(2r)$ \label{reduceSp}}

As motivated in subsection \ref{reduceSUrank} for the $SU$ case, we again consider the scenario of reducing the rank, with some possible brane pictures in mind.
We will see what happens to the vanishing cycles when the rank of the gauge group goes down by one, without changing the type of gauge group.  

 Again, there may not be a unique answer, but here we propose one method.
Looking back at {\bf figure \ref{sp6flower}}, one could pick a branch cut connecting branch points $Q_i$ and $C_i$, and for the vanishing cycles involved with this cut, we can merge them appropriately (in neighboring pairs), so that cycle starts or ends from those branch points. We can use 
\begin{eqnarray}
\nu^Q_{i-1} + \nu^Q_i&=& (\nu_{i-1}^{C}  -\alpha_{i-1}+ \alpha_{i})+(\nu_{i}^{C}  -\alpha_{i}+ \alpha_{i+1})\nonumber \\
&=&   \nu_{i-1}^{C}+\nu_{i}^{C}   -\alpha_{i-1} + \alpha_{i+1}, \label{sumQ} \\
\nu^C_{i-1} + \nu^C_i&=& (-\beta_{i-1}+\beta_{i}+\alpha_{i})+(-\beta_{i}+\beta_{i+1}+\alpha_{i+1}) \nonumber \\
&=&-\beta_{i-1}+\beta_{i+1}+\alpha_{i}  +\alpha_{i+1}, \label{sumC}
\end{eqnarray}
 as new vanishing cycles, instead of each one of the 4 cycles: $\nu^Q_{i-1},\nu^Q_i,\nu^C_{i-1} , \nu^C_i$. Note that now $\beta_i$ disappears, and once we set $\alpha_i=0$ by closing that branch cut down, \eqref{sumQ} and \eqref{sumC} reduces back to the form of \eqref{spcharges}\footnote{We can again express similar concern that we had for the 
$SU$ case, namely: If we are to use this result for brane construction of the $Sp(2r)$ SW theory, we will have to figure out what it means physically to set  $\alpha_i=0$. This will be discussed elsewhere.}.
 
This method works smoothly to bring the rank down to rank 3. When we reduce from rank 3 to 2 with the above method we obtain 
    \begin{eqnarray}  
   \nu_{1}^{C}&=&- \nu_{2}^{C}=-\beta_{1}+\beta_{2}+\alpha_{2} ,     \nonumber \\
 \nu_{0}^{Q}&=&\beta_{1} + 2\alpha_1, \quad   \nu_{1}^{Q}=   -\beta_{1}+\beta_{2}   -\alpha_{1}+ 2\alpha_{2}, \quad
 \nu_{2}^{Q}= 2\beta_{1} -\beta_{2}   + \alpha_1 -2\alpha_{2}. \label{sp2charges}
  \end{eqnarray}
From the $\nu^Q$'s, we can make the combination $  \nu_{1}^{Q}+\nu_{2}^{Q} =   \beta_{1}   $ and $\nu_0^Q$ which are same as the two vanishing cycles for rank 1 case: the famous monopole and dyon as given by
   $\beta $, $\beta +2\alpha $.    
   
\newpage
\section{Novel tools to capture higher singularities of the SW curves \label{doublediscsection}}  
  
In the previous sections we have been toying with three different techniques to locate singular loci in the moduli space of SW theories, namely the double discriminant, 
discriminant loci and the exterior derivative. It is time now to clarify the roles of each and point out what combinations of the three techniques that would serve as the 
most useful guide in locating singular loci in the moduli space. 

One immediate distinction between the exterior derivative and the double discriminant is a bit more obvious:
the exterior derivative can pinpoint us to the moduli loci of having multiple massless BPS dyons, while double-discriminant 
shows how to determine whether they are mutually local or not. We will elaborate this more as we go along. 
  
In the previous sections we saw for the 
 $A_2=SU(3)$ and $C_2=Sp(4)$ examples that some special singularities occur at special points (i.e codimension 2 loci) in the moduli space.  
We observed that $\Delta_x f=0$ gives one complex relation. Its vanishing loci is codimension 1, and we have a vanishing cycle there. 
Similarly, we observed that at some codimension 2 loci, $\Delta_x f=0$ loci intersect each other, giving two vanishing cycles there. Now if we want to find a moduli region 
where $\Delta_x f=0$ itself becomes singular then this locus can be captured by taking an exterior derivative.  
In other words, by demanding $ \Delta_x f=d \Delta_x f=0$ we have more singular theories. 

 As explained in subsection \ref{EXandDisc}, the exterior derivative $d$ can be written in terms of the partial derivatives with respect to all the coordinates. Therefore the exterior derivative inside the moduli space is given as $d= \sum_{i=1}^{r} du_i \frac{\partial  }{\partial u_i} $.
 One might think that demanding the $d =0$ actually reduces $r$ degrees of freedom, since we demand all the $r$ partial derivatives to vanish. However, $ \Delta_x f=d \Delta_x f=0$ indeed contains codimension 2 solutions (instead of codimension $r+1$).
 
When the $ \Delta_x f=0$ loci becomes singular (where  $ \Delta_x f=d \Delta_x f=0$ holds), double discriminant also vanishes  $\Delta_u \Delta_x=0$. In other words, singularity is seen from the moduli slices as well. However, just because it looks singular in some moduli slices (of lower dimension), it does not mean that it is singular in the full moduli space. The former is captured by vanishing double discriminant, and the latter is captured by vanishing exterior derivative. Therefore the latter implies the former, but not the other way around. In other words, $ \Delta_x f= \Delta_u \Delta_x f=0$ is a necessary but not sufficient condition for having $ \Delta_x f=d \Delta_x f=0$. We will ask how fast double discriminant vanishes, i.e., its order of vanishing $-$ this will help us distinguish Argyres-Douglas points from mutually local points. 
     
\subsection{Exterior derivative can pinpoint to the moduli loci of having multiple massless BPS dyons \label{exteriordVSdd}}

Let us now make the distinction between exterior derivative and double discriminant clearer. The starting point is the simple observation that the
vanishing discriminant condition of the SW curve $\Delta_x f=0$ defines an algebraic variety $\Sigma$. Since $\Delta_x f$ is written only in terms of moduli $u_i$'s and without $x$ and $y$, $\Sigma$ is an algebraic variety embedded inside the moduli space, denoting moduli loci of massless BPS states. When this algebraic variety $\Sigma$ self-intersects, 2 or more BPS states become massless. How is this captured? Our study of pure $Sp(4)$ SW curve and the {\bf figure \ref{sp4jihye2}} might suggest that we can take the $u$-plane slices of $\Sigma$ loci and see where singular points collide on the 
$u$-plane. This is captured by vanishing double discriminant, namely $\Delta_x f=\Delta_u \Delta_x f =0$, leading one to conclude that this relation may capture the true singular loci of the theory. This conclusion would be too premature as we will argue below that, although the 
double discriminant technique captures all the singularities, it also comes with extra solutions that aren't true singularities. The correct technique then would be 
to use the vanishing
exterior derivative $d$, which is a standard tool to look for singularities of algebraic varieties. Therefore we demand $\Delta_x f=d \Delta_x f =0$ as our tool for searching the singular loci in the theory. To illustrate this, let us 
first present a quick heuristic example, where we show the difference between these two methods. We will go to a more detailed elaboration later.

\begin{figure}[htb]
\begin{center}
\includegraphics[
width=2.5in]{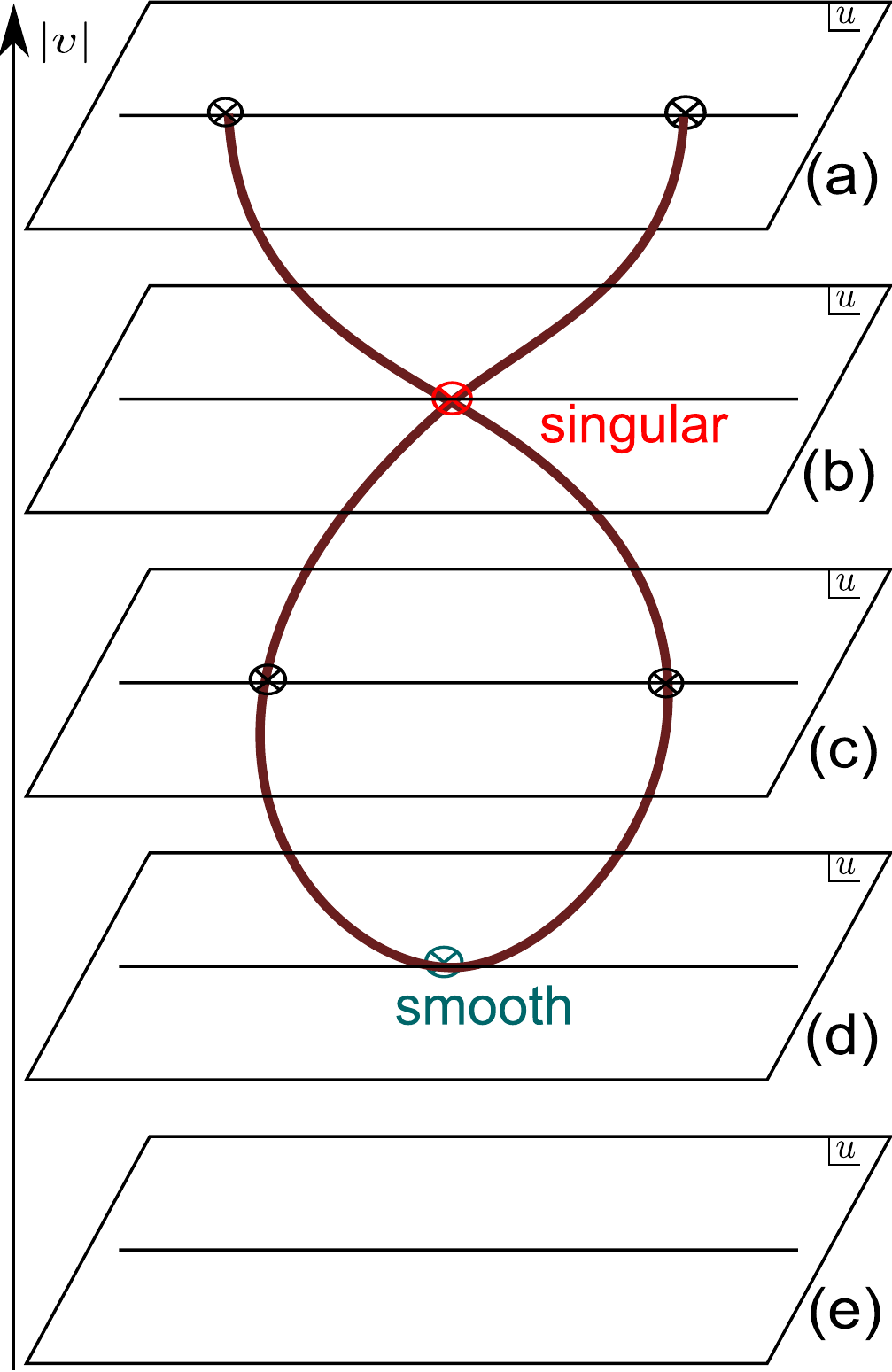}
\end{center}
\caption{A heuristic example showing difference between vanishing double discriminant and a vanishing exterior derivative. Vanishing double discriminant will single out slices (b) and (d), where as the exterior derivative will only take the point in (b) as a singularity. Only (b) is the true singularity. Vanishing double discriminant is a necessary but not a sufficient condition for a singularity. Compare this picture with {\bf figure \ref{sp4jihye1}}.}
\label{figure8}
\end{figure}
 
The example that we have in mind is depicted 
in {\bf figure \ref{figure8}}. Inside a moduli subspace, we draw a figure-eight-like object, which is analogous to $\Delta_x f=0$ loci as in {\bf figure \ref{sp4jihye1}}. We mark various $u$-planes with (a) to (e). The number of singular points changed on each $u$-planes. When the singular points collide on the $u$-plane we have $\Delta_u\Delta_x f=0$, for example at slices (b) and (d). Slice (b) is true singularity, however (d) is smooth\footnote{In slice (e), it appears that there are no solutions. With complex numbers this won't happen, namely the number of solutions to vanishing discriminant does not change by moving in the moduli space, that is always $2r+2$ and $2r+1$ for $SU(r+1)$ and $Sp(2r)$.}. Thus we intuitively see 
that $\Delta_x f=\Delta_u \Delta_x f =0$ and $\Delta_x f=d \Delta_x f =0$ have different sets of solutions. The latter is a subset of the former. The result can be 
succinctly expressed in the following way: 
\begin{eqnarray}
S^d &\equiv& \{ \vec{u} | \Delta_x f=d \Delta_x f =0 \} , \nonumber \\
S^i & \equiv & \{ \vec{u} | \Delta_x f=\Delta_{u_i} \Delta_x f =0 \} , \nonumber \\ 
S^\ast & \equiv  & \{ \vec{u} | \Delta_x f=\Delta_{u_1} \Delta_x f =\Delta_{u_2} \Delta_x f =\cdots =\Delta_{u_r} \Delta_x f =0 \} = S^1 \cap S^2 \cap \cdots \cap S^r , \nonumber \\
S^d  &\subset& S^\ast \subset S^i .\label{sets}
\end{eqnarray}
In the next couple of subsections we 
will provide examples for rank 2 and 3 for the 
$SU$ and $Sp$ groups. It turns out that $S^d = S^\ast \not\subset S^u, S^v$ for rank 2 and $S^d \not\subset S^\ast $ for rank 3 cases.
We have denoted $u = u_1, v = u_2 , w=u_3$ as before.

\subsection{$SU(r+1)$ examples \label{suexample}}

Here we will see explicitly how the exterior derivative and double discriminant distinguishes certain singularities of the pure $SU$ SW theory. We will see that as we go to 
higher ranks, exterior derivative is important to distinguish the loci with multiple massless BPS states.

For clarity and to fix the notations, let us reproduce the equation of the curve again here: 
 \begin{eqnarray}  
 y^{2}=f_{SU(r+1)}&\equiv& \left( x^{r+1} + u_1 x^{r-1} + u_2 x^{r-2} +\cdots + u_r \right)^2 - \Lambda
^{4r+4} \nonumber \\   &=& f_{+} f_{ -} ,  \label{SUcurveAgain} \end{eqnarray}
with $f_\pm$ defined in the usual way as:
\begin{equation}
   f_{\pm} \equiv  x^{r+1} + u_1 x^{r-1} + u_2 x^{r-2} +\cdots + u_r  \pm \Lambda
^{2r+2}, \end{equation} whose roots can be denoted by 
 \begin{equation}f_{+} \equiv \prod_{i=0}^{r} (x-P_i) \qquad f_{-} \equiv \prod_{i=0}^{r} (x-N_i).\end{equation}

\subsubsection{The rank 2 case: $SU(3)$}

Using the curve \eqref{SUcurveAgain} the $SU(3)$ curve is given by
 \begin{eqnarray}  
 y^{2}=f_{SU(3)}&\equiv& \left(  x^{3} + u x + v \right)^2 - \Lambda
^{12}\nonumber \\   &=& f_{+} f_{ -}  \label{SU3curve}  , \end{eqnarray}
with $f_\pm$ now defined in the following way
\begin{equation}
   f_{\pm} \equiv  x^{3} + u x + v     \pm \Lambda
^{6}. \end{equation}   
The vanishing of the discriminant and the exterior derivatives imply the following relation that we discussed earlier:
\begin{equation} 
\Delta_x f_{SU(3)}=d \Delta_x f_{SU(3)} =0, \label{SU3sing}
\end{equation} whose solutions now make a set 
 \begin{eqnarray}  
S^d_{SU(3)}&=&\Bigg\{ \left( \frac{u}{\Lambda^4},  \frac{v}{\Lambda^6} \right)\Big|   (0,1) , (0,-1), \nonumber \\
& &  \left( \frac{-3}{2^{2/3}},0 \right), \left(\frac{-3}{2^{2/3}}\omega_3,0\right), \left( \frac{-3}{2^{2/3}}\omega_3^2,0\right) \Bigg\}, \label{SU3cusp}
\end{eqnarray}
with $\omega_m$ being $m$'th root of unity.
  In \cite{KLYTsimpleADE, KLYTmndrmSU} the monodromy is studied, and the first line of \eqref{SU3cusp} are the cusp locations, and the second line of \eqref{SU3cusp} are the node locations (where both the discriminant loci and the SW curve develop cusp and node like singularities respectively, as seen in \cite{KLYTsimpleADE, KLYTmndrmSU}). 
More explicitly, \eqref{SU3sing} can also be written as 
  \begin{equation} 
\Delta_x f_{SU(3)}=\frac{\partial}{\partial u} \Delta_x f_{SU(3)} =\frac{\partial}{\partial v} \Delta_x f_{SU(3)} =0 . \label{SU3singExp}
\end{equation} 
It turns out that even if we demand a partial relation of the form 
  $\Delta_x f_{SU(3)}=\frac{\partial}{\partial u} \Delta_x f_{SU(3)} =0$ or 
  $ \Delta_x f_{SU(3)} =\frac{\partial}{\partial v} \Delta_x f_{SU(3)} =0 $, the solution set is the same as in \eqref{SU3cusp}. In other words, while $\Delta_x f_{SU(3)}=0$ holds,
$\frac{\partial}{\partial u} \Delta_x f_{SU(3)} =0$ is equivalent to $ \frac{\partial}{\partial v} \Delta_x f_{SU(3)} =0$. Therefore vanishing exterior derivative reduces only 1 degree of freedom, instead of $r$.

How are these reflected in the behaviour of double discriminant? We see that $S^d_{SU(3)} =S^\ast_{SU(3)} \not\subset S^i_{SU(3)} $. 
A set of solutions to $\Delta_x f_{SU(3)}=\Delta_u \Delta_x f_{SU(3)} =0$, $S^u_{SU(3)}$ has extra elements such as \begin{equation}
   \left(-3 \omega_3^i,1 \omega_2^j \right) \in S^u_{SU(3)} - S^d_{SU(3)} \ne \varnothing , \quad i, j \in \mathbb{Z} . 
   \end{equation}
Similarly, a set of solutions to $\Delta_x f_{SU(3)}=\Delta_v \Delta_x f_{SU(3)} =0$, $S^v_{SU(3)}$  has extra elements such as \begin{equation}
  \left(\frac{3}{2^{2/3}}\omega_3^i,2 \omega_2^j \right) \in S^v_{SU(3)} - S^d_{SU(3)} \ne \varnothing, \quad i, j \in \mathbb{Z} . 
   \end{equation}
However demanding all together  $\Delta_x f_{SU(3)}=\Delta_u \Delta_x f_{SU(3)} =\Delta_v \Delta_x f_{SU(3)} =0$ gives the same solutions as \eqref{SU3cusp}, therefore 
\begin{equation} 
S^d_{SU(3)} =S^\ast_{SU(3)}. \label{SU3same}
\end{equation}
  
\subsubsection{The rank 3 case: $SU(4)$}

Again using the curve \eqref{SUcurveAgain}, the $SU(4)$ curve becomes
 \begin{eqnarray}  
 y^{2}=f_{SU(4)}&\equiv& \left( x^{4} + u x^{2} + v x +   w \right)^2 - \Lambda
^{16} \label{SU4curve} = f_{+} f_{ -} , \end{eqnarray}
with 
\begin{equation}
   f_{\pm} \equiv  x^{4} + u x^{2} + v x +   w  \pm \Lambda
^{8}. \end{equation}  Setting $\Lambda =1$ with 
$\left( \frac{v}{\Lambda^6},  \frac{w}{\Lambda^8} \right) = (0,1)$ satisfies almost all the relations
\begin{eqnarray}
&&\Delta_x f_{SU(4)} = \Delta_u \Delta_x f_{SU(4)}  =\Delta_v \Delta_x f_{SU(4)} = 0, \nonumber\\
&& \Delta_w \Delta_x f_{SU(4)} = \frac{\partial}{\partial u} \Delta_x f_{SU(4)} =\frac{\partial}{\partial v} \Delta_x f_{SU(4)} =0,
\end{eqnarray} 
except one relation of the form
\begin{equation}
\frac{\partial}{\partial w} \Delta_x f_{SU(4)} =2^{11}   u^4 (64 - 16 u^2 +   u^4) \ne 0
\end{equation}
for generic values of $u$. In other words, even if we demand all the double discriminant to vanish, the exterior derivative may not vanish. This means 
   \begin{equation}
  (u,0,1) \in S^\ast_{SU(4)} - S^d_{SU(4)} \ne \varnothing,  \quad u (64 - 16 u^2 +   u^4)\ne 0.  \label{SU4diff}
   \end{equation}
   Therefore double discriminant themselves cannot give necessary and sufficient condition for having 2 vanishing 1-cycles. (They are necessary but not sufficient.)

\subsection{$Sp(2r)$ examples \label{spexample}}

The story by now should be clear from the $SU(r+1)$ examples. Double discriminant technique is not the most efficient way of capturing the singularity loci in the 
moduli space of SW theories. For completeness, let us see how the picture appears for the $Sp(4)$ and $Sp(6)$ examples.

\subsubsection{The rank 2 case: $Sp(4)$}

 From the $Sp(2r)$ curve the $Sp(4)$ curve can be easily extracted as
\begin{equation}
y^2 =  x \left(1+ u x + v x^2 \right)\left(1+u x + v x^2 +x^{3} \right) .
\end{equation}
The solution set for the vanishing of the discriminant and the exterior derivative: $\Delta_x f_{Sp(4)}=d \Delta_x f_{Sp(4)} =0$
are 
\begin{eqnarray}  
S^d_{Sp(4)}&=&\left\{ (u,v)  |   (3 \omega_3^i ,3\omega_3^{2i}), \left(3 \sqrt[3]{2} \omega_3^j, \frac{9}{2 \sqrt[3]{2}} \omega_3^j \right) \right\} , \quad i, j \in \mathbb{Z}.
 \end{eqnarray}
 If we demand only some of the partial derivatives to vanish, we may/will get extra solutions.
For example,
$\Delta_x f_{Sp(4)}=\frac{\partial}{\partial u} \Delta_x f_{Sp(4)} =0 $ has $(u,v) =(0,0)$ as an extra solution, and $\Delta_x f_{Sp(4)}=\frac{\partial}{\partial v} \Delta_x f_{Sp(4)} =0 $ has no extra solution.
 Demanding all the double discriminant, we get the same solution as exterior derivative, for rank 2 case here.
$ S^d_{Sp(4)} =S^\ast_{Sp(4)}$

How are these reflected in the behaviour of double discriminant? We see that $S^d_{Sp(4)} =S^\ast_{Sp(4)} \not\subset S^i_{Sp(4)} $. 
A set of solutions to $\Delta_x f_{Sp(4)}=\Delta_u \Delta_x f_{Sp(4)} =0$, $S^u_{Sp(4)}$ has extra elements such as \begin{equation}
   (0,0) \in S^u_{Sp(4)} - S^d_{Sp(4)} \ne \varnothing . 
   \end{equation}
Similarly, a set of solutions to $\Delta_x f_{Sp(4)}=\Delta_v \Delta_x f_{Sp(4)} =0$, $S^v_{Sp(4)}$  has extra elements such as \begin{equation}
 \left(3,-\frac{15}{4}\right), \left(3,\frac{9}{4} \right) \in S^v_{Sp(4)} - S^d_{Sp(4)} \ne \varnothing . 
   \end{equation}
However demanding all together  $\Delta_x f_{Sp(4)}=\Delta_u \Delta_x f_{Sp(4)} =\Delta_v \Delta_x f_{Sp(4)} =0$ gives a set of solutions $S^\ast_{Sp(4)}$, which turns out to be 
\begin{equation}
S^d_{Sp(4)} =S^\ast_{Sp(4)}.
\end{equation}
Note the similarity with the $SU(3)$ case of \eqref{SU3same}.  

\subsubsection{The rank 3 case: $Sp (6)$}

The story for the $Sp(6)$ case is somewhat similar to the $SU(4)$ case discussed earlier. The $Sp(6)$ curve takes the following form
\begin{equation}
y^2 =  x \left(1+u x + v x^2 + w x^3 \right)\left(1+u x + v x^2 + w x^3  +x^{4} \right) .
\end{equation}
Note that now
$(u,v,w)= (8 \omega_4^i ,14 \omega_4^{2i}, 8 \omega_4^{3i})$ will make all the single and double discriminant vanish, but none of the partial derivative vanishes. Instead the partial derivative takes the following values
\begin{equation}
\left(\frac{\partial}{\partial u} \Delta_x f_{Sp(6)},\frac{\partial}{\partial v} \Delta_x f_{Sp(6)},\frac{\partial}{\partial w} \Delta_x f_{Sp(6)}\right) = 13312 (\omega_4^i, -1, \omega_4^{3i})
\end{equation}
where $\omega_4$ is the fourth root of unity. Thus the non-vanishing of the partial derivatives clearly means that the vanishing of the double discriminant is not enough to
capture the singular loci. This can be expressed more succinctly as
 \begin{equation}
 (8 \omega_4^i ,14 \omega_4^{2i},8 \omega_4^{3i}) \in S^\ast_{Sp(6)} - S^d_{Sp(6)} \ne \varnothing ,
   \end{equation}
    which is similar to the $SU(4)$ case in \eqref{SU4diff}.

\subsection{Order of vanishing of $\Delta_u \Delta_x f$ is high at the Argyres-Douglas loci. \label{oov3}}

Massless dyons coexist at a ${\rm codim}_{\mathbb{C}}$-$2$ loci $\{ u_i | \Delta_x f=d \Delta_x f=0 \}$, 
there the double discriminant also vanishes  $\Delta_u \Delta_x f =0$.  
 And the curve looks like either of following two: 
   \begin{center}
\begin{tabular}{|c|c|c|}\hline curve degen.n &$y^2 = (x-a)^{\color{red}3} \times \cdots $ &$y^2 = (x-a)^{\color{nodegreen}2} (x-b)^{\color{nodegreen}2}  \cdots $ \\ \hline
o.o.v of $\Delta_u \Delta_x$ & {\color{red}3} & {\color{nodegreen}2} \\ \hline
shape of curve & {\color{red}cusp} & {\color{nodegreen}node}\\ \hline
shape of $\Delta_x=0$ & {\color{red}cusp} & {\color{nodegreen}node} \\ \hline
intersection & mutually non local &    local  \\ \hline
name & Argyres-Douglas &    mutually local    \\ \hline
\end{tabular} 
\end{center}
 In the following subsections we will elaborate the content of the above table. Before moving ahead, however, two comments are in order. First,
the method that we use here may not look as sophisticated as the rank 2 classification of \cite{ACSW, AW}, where they demand various physical properties 
such as Z-consistency. However, our method can be easily generalised to higher ranks and easily passes more sophisticated tests coming from physics. One subtlety is the 
 scaling behaviour of the curve and the 1-form needs to be checked (see for example \cite{GST}) but we will leave this for future work \cite{Seo}.

Secondly, one might ask whether there is a physical meaning for any higher $\Delta^{n} f =0$ (such as $\Delta_{u_i}
\Delta_{u_j} \Delta_x f=0$). Since the order of vanishing for the double discriminant is $\ge 2$, 
taking any extra discriminant operator again will give zero automatically. So we don't have to go beyond $\Delta^2 f = 0$ (that is, $\Delta_u \Delta_x f=0$).

\subsubsection{Factorization of double discriminant, and order of vanishing: Argyres-Douglas loci \label{FactorDD}} 
 
As we saw in the above table,  
massless dyons coexist at a ${\rm codim}_{\mathbb{C}}$-$2$ loci of the vanishing discriminant and the exterior derivative, where the curve looks like either of following two:
\begin{itemize}
\item $y^2=(x-a)^3 \times \cdots  $
The curve has a cusp-like singularity ({\color{forestgreen}Argyres-Douglas}). Vanishing discriminant $\Delta_x f=0$ locus also intersects at $\Delta_u \Delta_x f=0$ ({\color{forestgreen}o.o.v$ = 3$}) with  {\color{forestgreen}cusp}-like singularity. 
Two massless dyons are mutually non-local (although the terminology becomes a little tricky because the spectra jumps across this point, for example as we saw for the 
$Sp(4)$ case earlier.) 
 \item $y^2=(x-a)^2 (x-b)^2 \times \cdots  $
The curve has a node-like singularity. Vanishing discriminant $\Delta_x f=0$ locus also intersects at $\Delta_u \Delta_x f=0$ ({\color{forestgreen}o.o.v $= 2$}) with \ {\color{forestgreen}node}-like singularity. Two massless dyons are mutually local.
\end{itemize}
  
%
Here we discuss roots to $\Delta_u \Delta_x f=0$. Under right circumstances, including $ \Delta_x f=d \Delta_x f=0$, each root to $\Delta_u \Delta_x f=0$ will correspond to two massless dyons with appropriate combinatoric meanings. Demanding only $\Delta_u \Delta_x f=  \Delta_x f=0$ does pick all the candidates, but it does not always correspond to having 2 massless BPS dyons. The former is a necessary but not the sufficient condition for the latter.
  
 Order of vanishing of each root of $\Delta_u \Delta_x f$ tells us the type of singularities. To see this note that the double discriminant also factorizes in the following 
way\footnote{Observe that the following two relations and the results in 
subsection \ref{curvedegen} work well for rank 4 and higher. In subsection \ref{blind} we present examples with smaller ranks.}: 
  \begin{eqnarray}   
 \Delta_u \Delta_x f_{Sp(2r)} =  \Delta_u \Delta_x(f_Q f_C)   &=&\# \left( v^{(2r+1)^2} + \cdots \right) \nonumber\\   
 &=& \#  \left( v^{r(r+1)} + \cdots \right)^2   \left( v^{r} + \cdots \right)^3 \left( v^{r+1} + \cdots \right)^3 \nonumber\\  & &\times  
 \left( v^{r(r-3)/2} + \cdots \right)^2 \left( v^{(r+1)(r-2)/2} + \cdots \right)^2 \nonumber\\
 & \equiv & \# ({QC_{II}})^2 ({C_{III}})^3 ({Q_{III}})^3 ({C_{II}})^2 ({Q_{II}})^2, \label{DDsp} \end{eqnarray}    
  \begin{eqnarray}   
  \Delta_u \Delta_x f_{SU(r+1)}= \Delta_u \Delta_x (f_+ f_-) &=&\# \left( v^{(2r+2)^2} + \cdots \right) \nonumber\\  
 &=& \#  \left( v^{(r+1)^2} + \cdots \right)^2  \left( v^{r+1} + \cdots \right)^3 \left( v^{r+1} + \cdots \right)^3 \nonumber\\  & &\times  
 \left( v^{(r+1)(r-2)/2} + \cdots \right)^2 \left( v^{(r+1)(r-2)/2} + \cdots \right)^2 \nonumber\\
 & \equiv & \# ({PN_{II}})^2 ({N_{III}})^3 ({P_{III}})^3 ({N_{II}})^2 ({P_{II}})^2,
 \label{DDsu}
  \end{eqnarray}   
for $u=u_1, v=u_2$\footnote{If we chose $u=u_i$, then this above formula seems to work for $v=u_{i\pm1}$. This may be related to the phase rotation on the $x$-plane and perhaps also that on the $u$-planes.}. We borrowed the notation from \eqref{DiscDeg} to denote the top degree in terms of each modulus.
 Here the notation for each factor is chosen in the following way. The subscripts with $II$ and $III$ correspond to the order of vanishing, note that they match the power of each factor. Factors $({Q_{II}}), ({Q_{III}})$ (factors $({C_{II}}), ({C_{III}})$ resp.) are present also in $\Delta_u \Delta_x(f_Q  )$ (in $\Delta_u \Delta_x(f_C)$ resp), while ${QC_{II}}$ does not appear in either  $\Delta_u \Delta_x(f_Q  )$ or $\Delta_u \Delta_x(f_C)$, but only when we consider $ \Delta_u \Delta_x f_{Sp(2r)} =  \Delta_u \Delta_x(f_Q f_C)$. Similarly, factors $({P_{II}}), ({P_{III}})$ (factors $({N_{II}}), ({N_{III}})$ resp.) are present also in $\Delta_u \Delta_x(f_+ )$ (in $\Delta_u \Delta_x(f_-)$ resp), while ${PN_{II}}$ does not appear in either  $\Delta_u \Delta_x(f_+)$ or $\Delta_u \Delta_x(f_-)$, but only when we consider $ \Delta_u \Delta_x f_{SU(r+1)}= \Delta_u \Delta_x (f_+ f_-)$. 

Note that each factor gives order of vanishing two or three - each corresponding to node and cusp like singularity. In this subsection, we will give each factor a combinatoric meaning.
We can consider various combinations: 

\begin{itemize}
\item[(a)] Two pairs of branch points colliding pairwise, and the number of choices for the branch points being reflected on the degree of the polynomial, i.e., number of roots. 
\item[(b)] Three branch points colliding all together, and the o.o.v. is higher i.e o.o.v. = 3. These are of course the Argyres-Douglas points. 
\end{itemize}

\subsubsection{Roots of $\Delta_u \Delta_x f_{Sp(2r), SU(r+1)}=0$ and curve degenerations \label{curvedegen}}

Now that we have set up the problem, let us analyse the singularity structures of \eqref{DDsp} and \eqref{DDsu}. 
Since the analysis of the singularity structures of \eqref{DDsu} and
 \eqref{DDsp} are similar, in terms of figures, we will provide examples for the $Sp(8)$ case only in this subsection.
However the pattern is clear for both the $SU$ and $Sp$ groups of arbitrary 
ranks\footnote{Observe that results here 
work well for rank 4 and higher. In subsection \ref{blind} we present examples with smaller ranks.}.

\vskip.1in

\begin{figure}[htb]
\begin{center}
\includegraphics[
width=2.3 in 
]{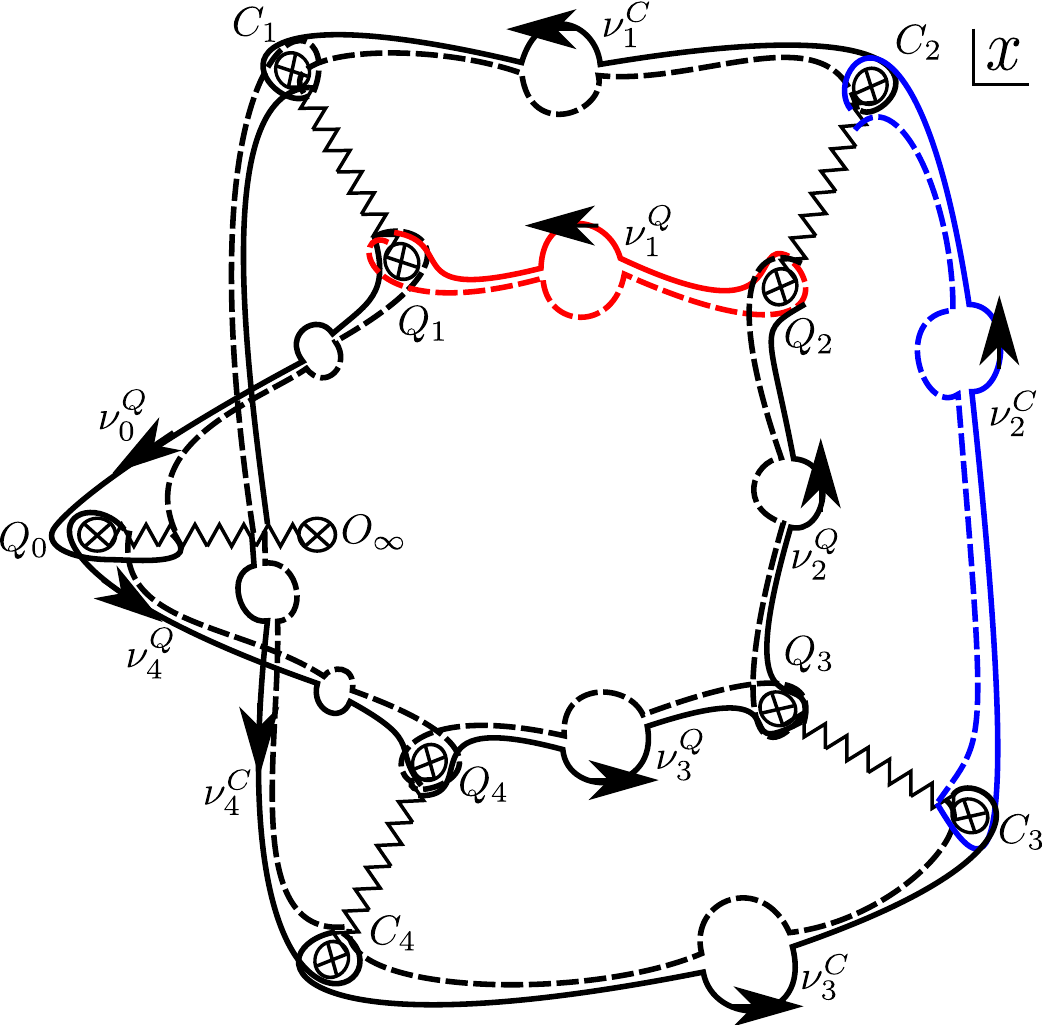}
\end{center}
\caption{An example with $Sp(8)=C_4$:   Two pairs of branch points on the $x$-plane collide each other pairwise. $ y^2 = (x-C_i)^2 (x-Q_j)^2  \times \cdots $}%
\label{sp4mlQC}%
\end{figure}

\noindent {\bf Case I:} ~ As an example, the first factor in second line of \eqref{DDsp} is 
     \begin{equation} ({QC_{II}}) \equiv \left( v^{r(r+1)} + \cdots \right). \label{qc2} \end{equation}
     This corresponds to two pairs of branch points on the $x$-plane collide each other pairwise, where each pair is $Q$ type and $C$ type respectively. The curve degenerates into a node-like singularity
     \begin{equation} y^2 = (x-C_i)^2 (x-Q_j)^2  \times \cdots .\end{equation}
Number of choices for choosing one pair of $C_i$'s and $Q_i$'s are given as: 
\begin{equation}
\left(
\begin{array}
[c]{c}%
r\\
1
\end{array}
\right)
\left(
\begin{array}
[c]{c}%
r+1\\
1
\end{array}
\right)= { r(r+1)},
 \end{equation} which is exactly the power of $v$ in \eqref{qc2}. Note that the number of ways to choose a pair of $C_i$'s to collide with each other is same as choice of a $\nu_C$ vanishing cycle which is $\left(
\begin{array}
[c]{c}%
r\\
1
\end{array}
\right) $, instead of all possible ways of choosing two $C_i$ points out of $r$ points which is $ \left(
\begin{array}
[c]{c}%
r\\
2
\end{array}
\right) $. {\bf Figure \ref{sp4mlQC}} shows an example for $C_4$ case. 

\vskip.1in

\noindent {\bf Case II:}~
Similarly, this can explain the first factor of second line in \eqref{DDsu}, which is
     \begin{equation} ({PN_{II}}) \equiv \left( v^{(r+1)^2} + \cdots \right). \label{pn2} \end{equation}
        This corresponds to two pairs of branch points on the $x$-plane collide each other pairwise, where each pair is $P$ type and $N$ type respectively. The curve degenerates into a node-like singularity
     \begin{equation} y^2 = (x-P_i)^2 (x-N_j)^2  \times \cdots .\end{equation}
Number of choices for choosing one pair of $P_i$'s and $N_i$'s are given as: 
\begin{equation}
\left(
\begin{array}
[c]{c}%
r+1\\
1
\end{array}
\right) 
\left(
\begin{array}
[c]{c}%
r+1\\
1
\end{array}
\right) 
 = { (r+1)^2},
 \end{equation} which is exactly the power of $v$ in \eqref{pn2}.

\vskip.1in

\begin{figure}[htb]
\begin{center}
\includegraphics[
width=2.3 in 
]{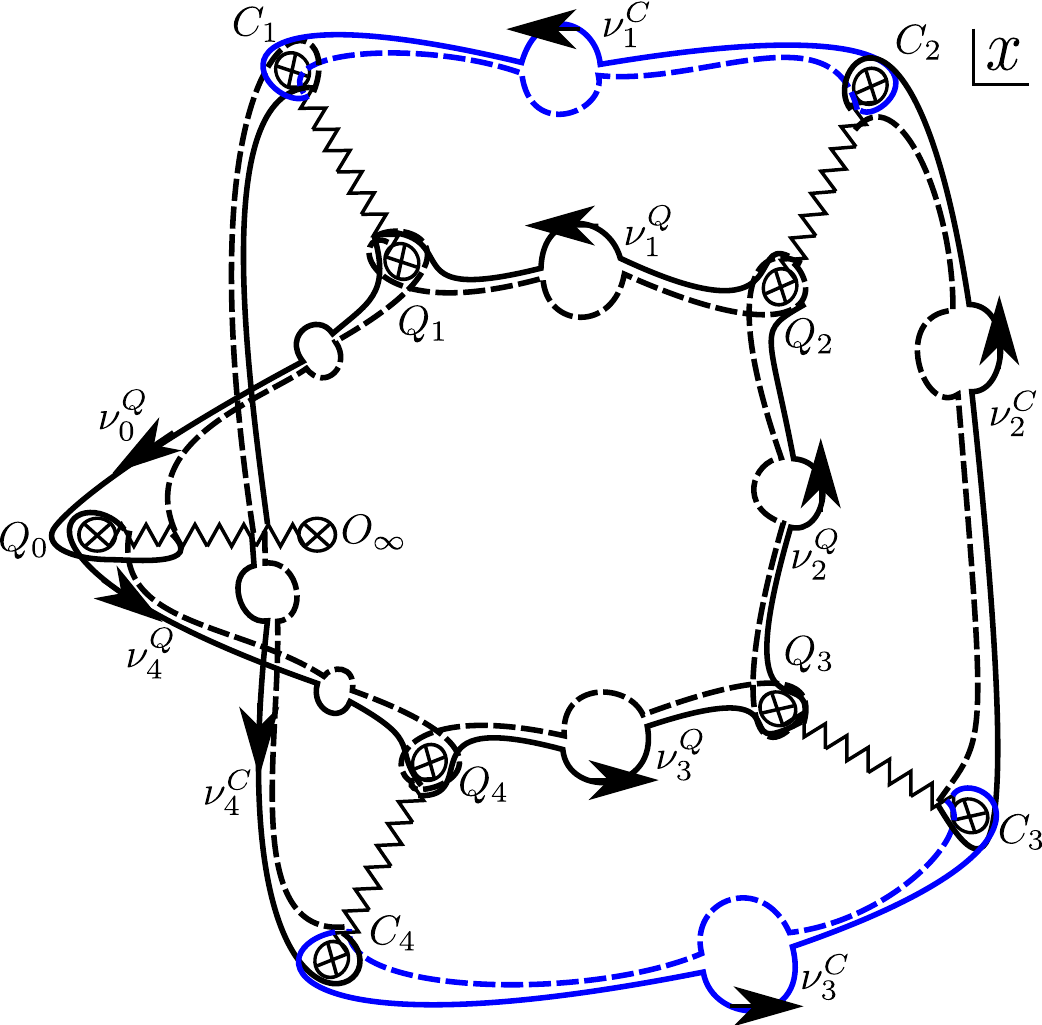}
\end{center}
\caption{An example with $Sp(8)=C_4$:  Two pairs of branch points on the $x$-plane collide each other pairwise.
  $  y^2 = (x-{C_i})^2 (x-{C_j})^2  \times \cdots $
}%
\label{sp4mlC}%
\end{figure}

\noindent {\bf Case III:}~
The first factor in the third line of \eqref{DDsp} is
  \begin{equation}  ({C_{II}}) \equiv \left( v^{{\color{forestgreen}r(r-3)/2}} + \cdots \right), \end{equation} 
and this corresponds to the scenario where two pairs of $C$-type branch points on the $x$-plane collide each other pairwise. The curve degenerates into a node-like singularity
    \begin{equation}  y^2 = (x-{C_i})^2 (x-{C_j})^2  \times \cdots  .\end{equation}
 Number of choices for $C_i$'s is same as number of choosing two $\nu_C$ cycles which do not share branch points.
   \begin{equation}  
  \frac{1}{2} \left(
\begin{array}
[c]{c}%
r \\
1
\end{array}
\right)  \left(
\begin{array}
[c]{c}%
r-3\\
1
\end{array}
\right)    = {\color{forestgreen}r(r-3)/2}. \label{c2comb}\end{equation}  
The first factor of \eqref{c2comb} corresponds to choosing one $\nu_C$ cycle out of $r$ choices. The second factor of \eqref{c2comb} corresponds to ways of choosing the second $\nu_C$ out of $r-3$ choices. $-3$ comes from removing the first cycle, and two other neighboring cycles which share branch points. The overall factor of half in \eqref{c2comb} comes because it does not matter the order of choosing two cycles. {\bf Figure \ref{sp4mlC}} shows an example for $C_4$ case. 

\vskip.1in

\begin{figure}[htb]
\begin{center}
\includegraphics[
width=2.3 in 
]{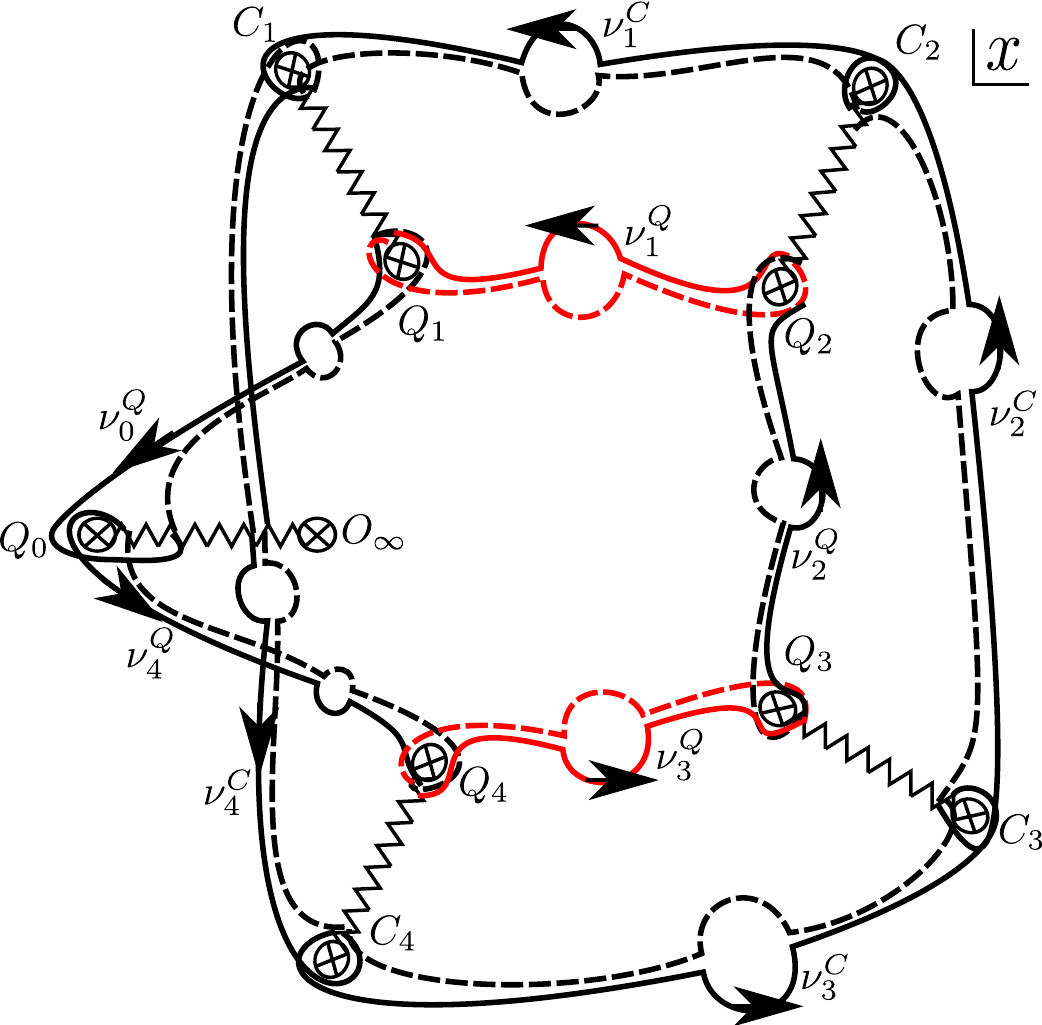}
\end{center}
\caption{An example with $Sp(8)=C_4$: Two pairs of branch points on the $x$-plane collide each other pairwise.
     $ y^2 = (x-{Q_i})^2 (x-{Q_j})^2  \times \cdots $ }%
\label{sp4mlQ}%
\end{figure}

\noindent {\bf Case IV:}~
The second factor in the third line of \eqref{DDsp}  
  \begin{equation}  ({Q_{II}}) \equiv \left( v^{{\color{forestgreen}(r+1)(r-2)/2}} + \cdots \right) , \end{equation} 
is about having two pairs of $Q$-type branch points on the $x$-plane collide each other pairwise. The curve degenerates into a node-like singularity
    \begin{equation}  y^2 = (x-{Q_i})^2 (x-{Q_j})^2  \times \cdots . \end{equation}
 Number of choices for $Q_i$'s is same as number of choosing two $\nu_Q$ cycles which do not share branch points, with similar combinatorics as in \eqref{c2comb}.
   \begin{equation} \frac{1}{2} \left( \begin{tabular}{ c }  $r+1$ \\ $ 1$ \end{tabular} \right)  \left( \begin{tabular}{ c }  $r -2$ \\ $ 1$ \end{tabular} \right)    = {\color{forestgreen}(r+1)(r-2)/2}. \label{q2comb}\end{equation}  
The first factor of \eqref{q2comb} corresponds to choosing one $\nu_Q$ cycle out of $r+1$ choices. The second factor of \eqref{q2comb} corresponds to ways of choosing the second $\nu_Q$ out of $(r+1)-3$ choices. {\bf Figure \ref{sp4mlQ}} shows an example for the $Sp(8)$ case. 
Similarly, we can understand $({P_{II}})$ and $({N_{II}})$ of \eqref{DDsu} and its power in $v$ in terms of combinatorics.

\vskip.1in

\begin{figure}[htb]
\begin{center}
\includegraphics[
width=2.3 in 
]{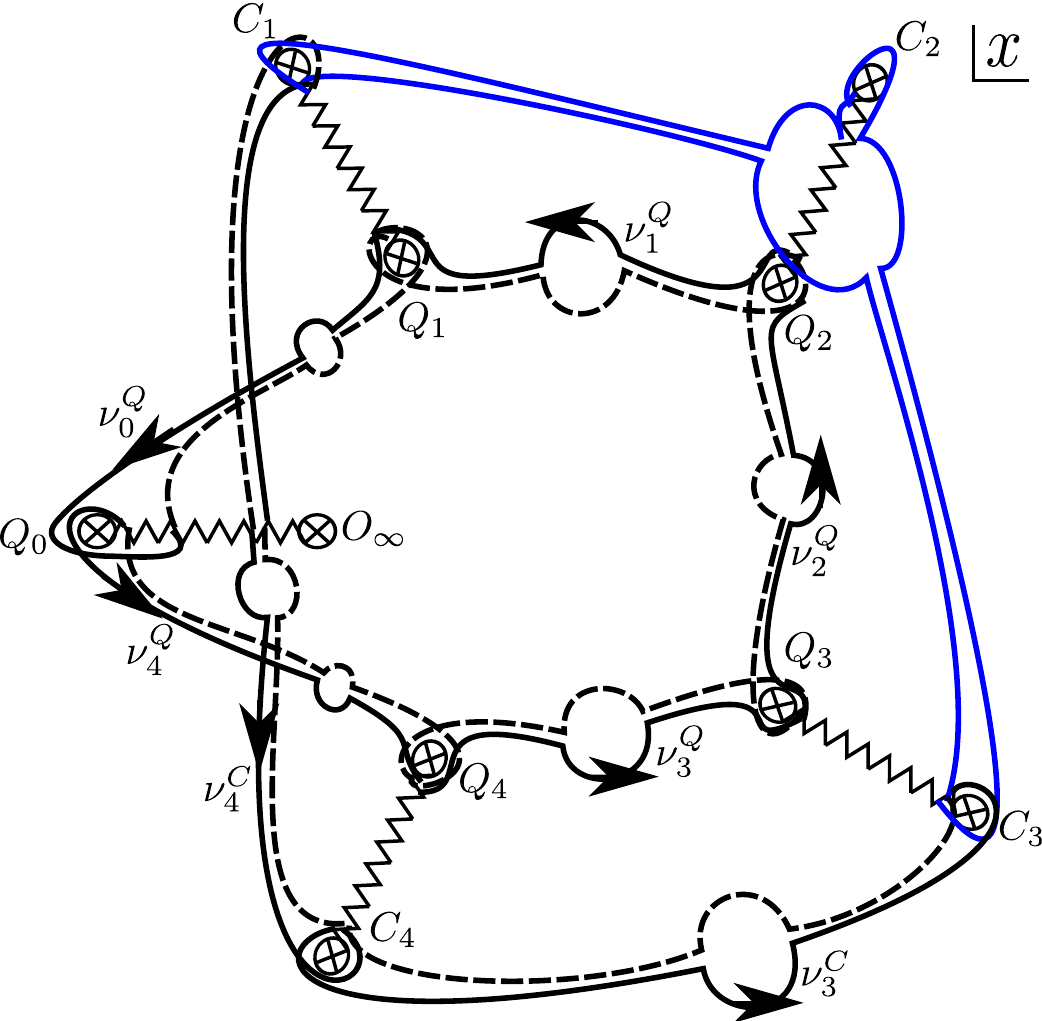}
\end{center}
\caption{An example with $Sp(8)=C_4$: Three branch points on the $x$-plane collide all together.
 $  y^2 = (x-{C}_i)^{3}  \times \cdots $
}%
\label{sp4adC}%
\end{figure}

\noindent {\bf Case V:}~
The second factor in the second line of \eqref{DDsp}  
  \begin{equation}  ({C_{III}}) \equiv \left( v^{r} + \cdots \right) \end{equation} 
is having three neighboring $C$-type branch points on the $x$-plane collide all together, as shown in {\bf figure \ref{sp4adC}}  for $C_4$ case. The curve degenerates into a cusp-like singularity
     \begin{equation}  y^2 = (x-{C_i})^{3}  \times \cdots  . \end{equation} 
 Number of choices for a chunk of $C_i$'s is same as number of choosing one $C_i$ point. 
   \begin{equation} \left(
\begin{array}
[c]{c}%
r \\
1
\end{array}
\right)    =r. \label{c3comb}\end{equation}   
   
\vskip.1in

\begin{figure}[htb]
\begin{center}
\includegraphics[
width=2.3 in 
]{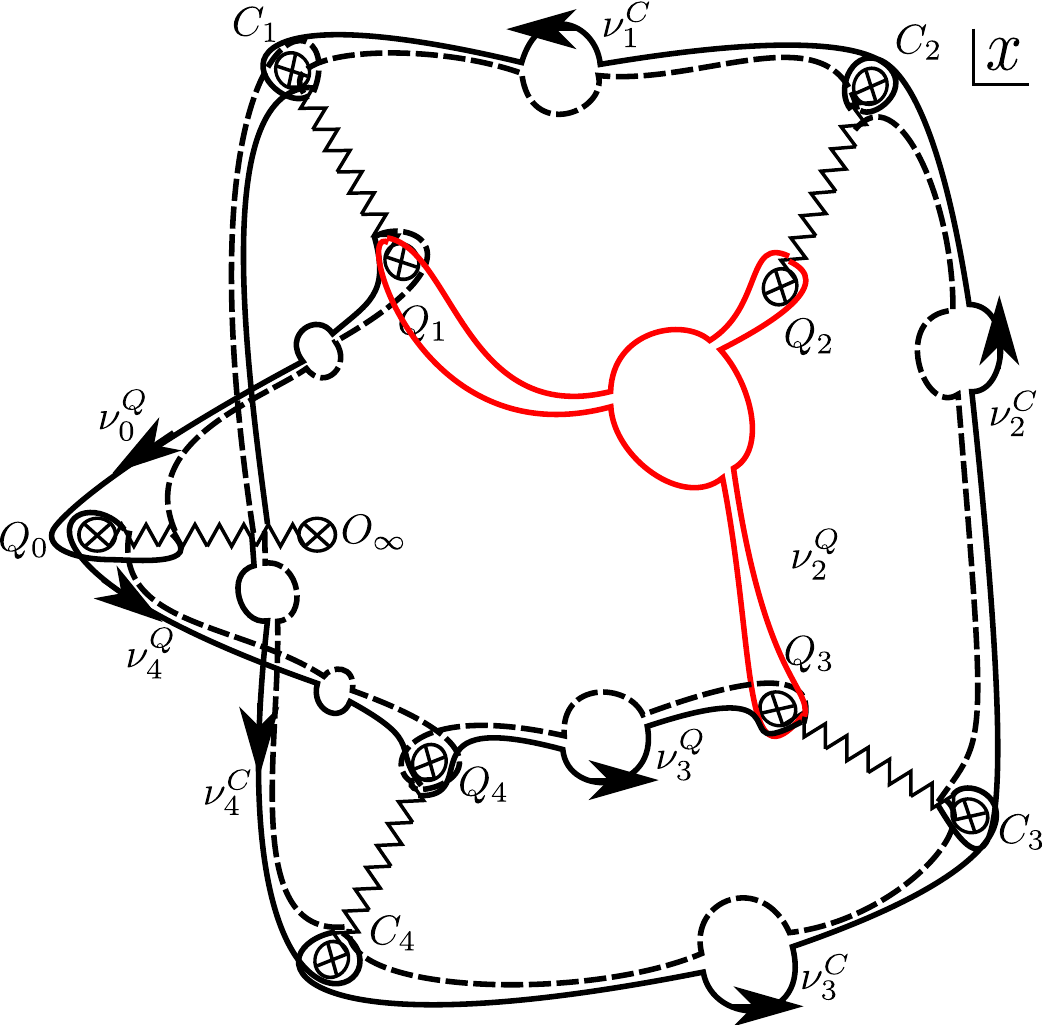}
\end{center}
\caption{An example with $Sp(8)=C_4$: Three branch points on the $x$-plane collide all together.
  $ y^2 = (x-{Q}_i)^{3}  \times \cdots $ }%
\label{sp4adQ}%
\end{figure}

\noindent {\bf Case VI:}~  
The third factor in the second line of \eqref{DDsp}  
  \begin{equation}  ({Q_{III}}) \equiv \left( v^{r+1} + \cdots \right) \end{equation} 
is about having three neighboring $Q$-type branch points on the $x$-plane collide all together. The curve degenerates into a cusp-like singularity
     \begin{equation}  y^2 = (x-{Q_i})^{3}  \times \cdots, \end{equation} 
 and again the number of choices for a chunk of $Q_i$'s is same as number of choosing one $Q_i$ point. 
   \begin{equation}\left(
\begin{array}
[c]{c}%
r+1\\
1
\end{array}
\right)   =r+1. \label{q3comb}\end{equation}   
   {\bf Figure \ref{sp4adQ}} shows an example for $C_4$ case. 
 Similarly, we can understand second and third factors of the second line of \eqref{DDsu}, $({N_{III}})$ and $({P_{III}})$.

\subsubsection{A blind modulus is handicapped: gives subtlety with low ranks \label{blind}} 
 
Our discussions in the previous sections 
may have hinted already one subtlety related to 
$u_r$, the moduli which appears in the constant term. When we rotate phase on the $x$-plane, all other moduli will change their phases, except for $u_r$. Therefore we call $u_r$ a blind modulus, i.e it is blind to the rotation of the phase on the $x$-plane. 
 
For the low ranks, the combinatorics of previous subsection doesn't work the same way, due to some accidental symmetry. 
For example, double discriminant of $Sp(4)=C_2$ factorizes as below:
\begin{eqnarray}   
 \Delta_u \Delta_x f_{Sp(4)}&=& 2^8 v (v^3-3^3)^3 (2^4 v^3- 3^6)^2 \nonumber \\
  & =& \# v ({Q_{III}})^3 ({QC_{II}})^2    .
  \end{eqnarray}    
Note that we do not have $({Q_{II}}), ({C_{II}}), ({C_{III}})$ because we don't have enough number of points of those kind. (For example, we cannot choose three $C$ points because there are only 2 of them in rank 2 case.) The solutions to $({QC_{II}})$ and $({Q_{III}})$ are the same values as in subsection \ref{spexample}.  
 The $v$-plane of {\bf figure \ref{sp4jihye1}} marks the roots to the double discriminant. As we approach $v=0$, two $\nu_C$ vanish at the same time. ($\Sigma_C$ intersect in {\bf figure \ref{sp4jihye1}}.) But here $d \Delta \ne 0$, and only 1 (instead of two) BPS dyon becomes massless.

As we go to the next rank $C_3=Sp(6)$ case, the subtlety associated with $v=0$ disappears. The double discriminant of $C_3$ factorizes in the following way:
\begin{eqnarray}   
 \Delta_u \Delta_x f_{Sp(6)}   &=&\#  \left( v^3 + \cdots \right)^3  
	 \left( v^4 + \cdots \right)^3  \left[ \left( v + \cdots \right)  \left( v + \cdots \right)\right]^2 \left( v^{12} + \cdots \right)^2 \nonumber \\
&=&\#    \left(w^2 + \cdots \right)^3 
	 \left( w^4 + \cdots \right)^3  \left[ \left( w^2 + \cdots \right)  \left( w^2 + \cdots \right) \right]^2  \left( w^{12} + \cdots \right)^2 \nonumber \\ 
 & =& \#  ({C_{III}})^3 ({Q_{III}})^3  ({Q_{II}})^2 ({QC_{II}})^2, \label{DDc3}
 \end{eqnarray}    
whose behaviour in the first line of \eqref{DDc3} is same as (first line of) \eqref{DDsp} except for missing factor of $({C_{II}})$ - which needs $r\ge 4$. When we wrote LHS down in terms of the blind modulus $w=u_3$ as in the second line of \eqref{DDc3} the combinatorics is less clear.
The higher $C_4$ cases work similarly.

For the $SU(r+1)$ case, we will discuss subtlety for $SU(4)$ and $SU(3)$. 
For $SU(4)$, the double discriminant factorizes as:
\begin{eqnarray}   
  \Delta_u \Delta_x f_{SU(4)}
 &=& \#   \left( v^{2} + \cdots \right)^2 \left( v^{2} + \cdots \right)^2   \left( v^{4} + \cdots \right)^3 \left( v^{4} + \cdots \right)^3    
 \left( v^{16} + \cdots \right)^2 \nonumber  \\
&=&  \# \left(v^4 \right)^2 \left( v^4 + \cdots \right)^3 \left( v^4 + \cdots \right)^3
\left( v^{16} + \cdots \right)^2 \nonumber \\
&=&\#  \left(v^4 \right)^2  \left( w^3 + \cdots \right)^3 \left( w^3 + \cdots \right)^3
\left( w^9 + \cdots \right)^2 \nonumber \\
 & = & \# ({N_{II}} \cdot {P_{II}})^2 ({N_{III}})^3 ({P_{III}})^3  ({PN_{II}})^2 , \end{eqnarray} 
 with a close agreement to \eqref{DDsu} except that here accidental symmetry gives $({N_{II}})=({P_{II}})$. This is related to the phenomena seen in subsection \ref{exactSUmonodromy} near {\bf figure \ref{rank9SU10u1}}, and probably it is because $v$ is blind to $Z_2$ part of the $x$-plane rotation.  All the roots to $({N_{II}})=0$ and $({P_{II}})=0$ happen to be equal to each other. The overall factor of $v^8$ signals node-like singularity however we will see that many singular points are colliding on the $u$-plane. In fact, 8 singular points on generic $u$-plane will collapse into 4 singular points\footnote{Roots to $\Delta_x f=0$ with $v=0$ is $u=\pm \sqrt{\pm 1 + w}$.} each with degeneracy 2 at $v=0$ $u$-plane. (Non-generic choice of $w$ can even worsen this.) The singularity is still node-like, but there will be four nodes forming simultaneously. For higher ranks, this does not happen generically, since generic values of other moduli will break this accidental symmetry.
 
For $SU(3)$, the double discriminant factorizes as:
\begin{eqnarray}   
\Delta_u \Delta_x f_{SU(3)} &=& (v-1)^4 (v+1)^4 (v^3)^2 \nonumber \\
 & = & \#  (N_{IV})^4 (P_{IV})^4 ({PN_{II}})^2 \end{eqnarray}   
The rank is too small to have any $({P_{II}})$ or $({N_{II}})$. All three roots to $({PN_{II}})$ happen to be equal to each other by accidental symmetry (there are not enough number of moduli left to break $Z_{r+1}$ symmetry). Similarly to the $SU(4)$ case, 6 singular points on generic $u$-plane will collapse into 3 singular points each with degeneracy 2 at $v=0$ $u$-plane. Three node-like singularity will happen simultaneously. The subtlety comes from the order of vanishing of maximal Argyres-Douglas points: o.o.v of $(N_{IV}), (P_{IV})$ is 4 instead of 3. This deviation is likely to be coming from the fact that $v$ is a constant piece in the moduli, blind to phase rotation on the $x$-plane, as explained in subsection \ref{blind}. As shown in \eqref{SU3cusp}, $v=1, u=0$ is where three branch points collide, forming the Argyres-Douglas points. We expect this type of subtlety to disappear in higher ranks.
 
 \newpage 
 
\section{Monodromies of pure ${\cal N} = 2$ $SO(2r+1)$ theories \label{demandCurveForm}}
Previously, we discussed locating singularity of having 1 or 2 massless dyons by solving various algebraic relations (such as vanishing discriminant and exterior derivative). It is nice and systematic, but it works only for up to having 2 massless dyons, and it does not single out the loci of 3 or more massless dyons. 

Here we will present another (perhaps simpler, if not always systematic) method, overcoming that disadvantage. Instead of solving algebraic equations, we parametrize the curve in terms of some complex parameters, such that the curve takes a certain form. The advantage here is that we can demand arbitrary number of massless dyons, and also we can control which of them are mutually local and non-local, by demanding the structure of branch point - by demanding which ones collide together etc.
This was implicitly used in \cite{DSW} for some singular loci of the $Sp(2r)$ case, and especially for maximal Argyres-Douglas for pure $Sp(2r)$ case. (Due to potential subtlety with scaling dimension etc, it may be safer to call them ``candidates'' for maximal Argyres-Douglas points) $r+1$ points in the moduli space were found there, where maximal Argyres-Douglas is expected and the curve takes the maximal cusp singularity. We will use this technique here to solve for the maximal Argyres-Douglas points for the $SO(2r+1)$ case. 

\subsection{$SO(2r+1)$, $2r-1$ maximal Argyres-Douglas points \label{SOoddADmax}}

To study the $SO(2r+1)$ examples, we will 
 start from the curves with the matter, and then take the pure limit. We will worry about the 1-form a little later in this subsection. 
For the $SO(2r)$ case, the SW curve with the flavors is given as (see for example \cite{ArgyresShapere}):
\begin{equation} y^2 = C_{SO(2r)}^2 - \Lambda^{2(2r-2-N_f)} x^4 (x^2 -m^2)^{N_f},\end{equation}
with 
\begin{equation}C_{SO(2r)}(x) \equiv   x^{2r} + s_2 x^{2r-2} +\cdots + s_{2r-2} x^2 +\tilde{s}_r^2.\end{equation}
In pure case, it simplifies further, such that  \cite{KLYTsimpleADE}):
\begin{equation} y^2 = C_{SO(2r)}^2 - \Lambda^{2(2r-2 )} x^4 =C_{SO(2r),+} C_{SO(2r),-}   \end{equation}
with\footnote{Note that both $C_{SO(2r),\pm}$ has $\Lambda$ dependence, just as in the $SU$ case where both factors $f_\pm$ of hyperelliptic equation depend on $\Lambda$ as seen in \eqref{fpm}. This is in contrast with the $Sp$ case where only one of two factors of hyperelliptic equation depend on $\Lambda$ as seen in \eqref{fCfQdef}.} \begin{equation} C_{SO(2r),\pm} = C_{SO(2r) }  \pm \Lambda^{ (2r-2 )} x^2 = x^{2r} + s_2 x^{2r-2} +\cdots + s_{2r-2} x^2 +\tilde{s}_r^2 \pm \Lambda^{ (2r-2 )} x^2.\end{equation} Some of the monodromy properties for pure $SO(2r)$ were studied in \cite{BLSo} with emphasis on the $SO(8)$ example. 

Maximal Argyres-Douglas points for the $SO(2r)$ will be two moduli points given by \cite{EHIY}
\begin{equation}
 s_{2r-2} =  \Lambda^{ \pm (2r-2 )}, \quad s_{2i}=0, ~~i\ne r-1 \label{SOevenMaxAD}
 \end{equation}
 which makes $C_{SO(2r),\mp} = x^{2r}$ into a maximal cusp form. Just as in the $SU$ case, it is straightforward to get 2 maximal Argyres-Douglas points, partially thanks to $\mathbb{Z}_2$ symmetric structure between $C_{SO(2r),\pm}$ and between $f_\pm$ in the SW curve, which is lacking in the $Sp(2r)$ case.

Although the curve for 
$SO(2r+1)$ is only slightly different from the $SO(2r)$ case, it is much harder to solve for the maximal Argyres-Douglas points. 
There are $2r-1$ solutions for the maximum AD points. In fact this number is equal to the dual coxeter number of $SO(2r-1)$, due to the 
underlying $\mathbb{Z}_{2r-1}$ symmetry present in the form of the pure curve. (See \cite{SO5DS} for more about this discrete symmetry and instanton corrections). 
 Curve with flavors is given in \cite{ArgyresShapere} 
 \begin{equation}
 y^2 = C_{SO(2r+1) }^2 - \Lambda^{2(2r-1-N_f)} x^2 (x^2 -m^2)^{N_f}
 \end{equation}
again with $C_{SO(2r+1) }(x) \equiv  x^{2r} + s_2 x^{2r-2} +\cdots + s_{2r-2} x^2 + {s}_{2r}$. 
In pure case, this reduces to \cite{KLYTsimpleADE,SO5DS}
 \begin{equation}
  y^2 = C_{SO(2r+1) }^2 - \Lambda^{2(2r-1 )} x^2  = C_{SO(2r+1) ,+}  C_{SO(2r+1), -}  \label{SOoddcurve} .
  \end{equation}
with \begin{equation} C_{SO(2r+1) ,\pm} = C_{SO(2r+1) }  \pm \Lambda^{ (2r-1 )} x = x^{2r} + s_2 x^{2r-2} +\cdots + s_{2r-2} x^2 + {s}_{2r}   \pm \Lambda^{ (2r-1 )} x . \end{equation} For the rest of the section, let us restrict ourselves to $SO(2r+1)$ only, and we will drop the subscript $SO(2r+1)$.
 
 To proceed now, first let us check some aspect of the behaviour of the SW 1-form 
\begin{equation}
\lambda =xd\log \left( \frac{C-y}{C+y}\right) .
\end{equation}
Using $d \log \left( C\pm y\right) =\frac{1}{C\pm y}d\left( C\pm y\right) $
it can be further simplified into 
\begin{equation}
d\log \left( \frac{C-y}{C+y}\right) =dC\left( \frac{1}{C-y}-\frac{1}{C+y}%
\right) -dy\left( \frac{1}{C-y}+\frac{1}{C+y}\right) =\frac{2ydC-2Cdy}{%
C^{2}-y^{2}}.
\end{equation}
Therefore, as $y\rightarrow 0$, we observe $\lambda \rightarrow \frac{-2xdy}{C}$, and the SW 1-form does not blow up near the vanishing cycles of the pure $SO(2r+1)$ SW curve. 

The question now is: under
 what condition does  $C_{\pm} $ take the maximal cusp form? Note that the number of degrees of freedom allows only the cusp of the following form: 
 \begin{equation}
 C_{+} = (x+b)^{r+1} (x^{r-1} + u_1 x^{r-2} + u_2 x^{r-3} + \cdots + u_{r-2} x + u_{r-1} ) \label{CmaxCusp}
 \end{equation}
for the proper choice of $s_{2i}$'s, which later turns out to be \eqref{svalueformAD}. The story is similarly for $C_{-}$, which is easily obtained by $x \rightarrow -x$. 
 
What is left to do now is to find the values of $s_{2k}$ which will allow \eqref{CmaxCusp}. For that purpose, we will rewrite $b$ in terms of $\Lambda$ and $r$, and then express the $s_{2k}$ in terms of $b$. We obtain this result by solving them for rank up to 12 from Mathematica, and then use them 
to guess the form of the ansatz for the higher orders. Plugging them in the above equations, we verify that \eqref{CmaxCusp} indeed hold for general rank $r$. 
An immediate benefit of this is that we can also check the total number of possible solutions, and show that it indeed equals $2r-1$, the number of ansatz we propose here.

The analysis will be a little detailed, therefore let us
first display the result. There are $2r-1$ possible values of $b$, which satisfy
\begin{equation} 
\Lambda^{2r-1} = (-1)^{r+1} \frac{(2r)!!b^{2r-1}}{(2r-3)!!}.
\end{equation}
In other words, \begin{equation}   b   = \omega_{2r-1}^k \left[ (-1)^{r+1} \frac{(2r-3)!!}{(2r)!!} \right]^{1/(2r-1)}\Lambda , \quad k\in \mathbb{Z}, \label{bvalueformAD}
 \end{equation} where $\omega_m$ is $m$'th root of unity.
Then it turns out that all the moduli of the curve $s_{2k}$'s are given by 
\begin{equation}
s_{2k} = (-b^2)^k \frac{(2r-1)}{(2r-2k-1)} \frac{r!}{k! (r-k)!}. \label{svalueformAD}
\end{equation} 
  
Our task is now to prove that these are indeed roots for arbitrary rank.  To start then we consider \eqref{CmaxCusp}, \begin{eqnarray}
g(x) & \equiv&   C+\Lambda ^{(2r-1)}x  \nonumber\\ 
&=&x^{2r}+s_{2}x^{2r-2}+\cdots +s_{2r-2}x^{2}+{s}_{2r}+\Lambda ^{(2r-1)}x \nonumber\\
&=&(x+b)^{r+1}(x^{r-1}+u_{1}x^{r-2}+u_{2}x^{r-3}+\cdots +u_{r-2}x+u_{r-1}),
\end{eqnarray}
and then shift $x \rightarrow (x-b)$ to get
\begin{eqnarray} g(x) &\equiv &(x-b)^{2r} + s_2 (x-b)^{2r-2} +\cdots + s_{2r-2} (x-b)^2 + {s}_{2r} + \Lambda^{ (2r-1 )} (x-b) \nonumber \\  &=& x^{r+1} (x^{r-1} + v_1 x^{r-2} + v_2 x^{r-3} + \cdots + v_{r-2} x + v_{r-1} ). \label{shiftb} \end{eqnarray}
We only need to show that low degree coefficients of LHS (or the first line of \eqref{shiftb}) vanishes. At this stage we do not care so much for the 
exact values of $u$ and $v$. They can be easily expressed in terms of each other.
 It is also clear that when we define $g_k$ in the following way:
\begin{equation}
g_k  \equiv \left. \frac{\partial^k}{\partial x^k} g(x) \right|_{x=0}, \qquad k=0,1,\cdots, r,
\end{equation}
they all {\it vanish} because of the exponent of $x$ in second line of \eqref{shiftb} is at most $r+1$ and so is always greater than $r$.
This would mean that the following $r+1$ relations 
\begin{eqnarray}
g_0 &\equiv& b^{2r} + s_2 b^{2r-2} + \cdots + s_2r - b \Lambda^{2r-1} =0 \nonumber  \\
g_1&\equiv &-(2r)b^{2r-1} - (2r-2)s_2 b^{2r-3} + \cdots - 2s_{2r-2} b + \Lambda^{2r-1} =0\nonumber \\
&\cdots& \nonumber \\
g_k &\equiv& \sum_i \frac{(2r-2i)!}{(2r-2i-k)!} (-b)^{2r-2i-k} s_{2i} 
\end{eqnarray}
will fix all $r$, $s_{2k}$ and $b$ in terms of $\Lambda$. It is then a simple exercise to show that 
\begin{equation} g(x) = \sum_{m=0}^{2r} \frac{g_m}{m!} x^m. \end{equation}
Furthermore it is straightforward to see that $g_0  + b g_1 , g_2 , g_3, \cdots, g_r$ only depend on $b$ and $s_{2k}$ with no dependence on $\Lambda$. Letting them all vanish gives $r$ relations, and in terms of $b$, they uniquely fix $s_{2k}$, since all the $s_{2k}$ appear linearly. (In other words, we can see the $r$ relations as linear equations for the $s_{2k}$ variables). 
The solution turns out to be the following 
\begin{equation}
s_{2k}=(-b^{2})^{k}\frac{(2r-1)}{(2r-2k-1)}\frac{r!}{k!(r-k)!} , \label{svalue}
\end{equation}
which will be justified soon. Since the number of solutions is fixed, justifying ansatz will be enough. 

Further requiring $g_1=0$, we can get a relation between $\Lambda$ and $b$, but since both appears with the power to the $2r-1$ in $g_1$, there is a phase ambiguity. The highest power of $b$ is $2r-1$, therefore we expect only $2r-1$ solutions consistent with our earlier ansatz. Thus they make a complete set of roots for a given value of $k$: 
 \begin{equation}   b   = \omega_{2r-1}^k \left[(-1)^{r+1} \frac{(2r-3)!!}{(2r)!!} \right]^{1/(2r-1)}\Lambda , \qquad  k\in \mathbb{Z} \label{bvalue} .
  \end{equation} 
It is now time to 
show that \eqref{svalue} and \eqref{bvalue} are indeed solutions. For this we will check the coefficients in powers of $x$ in \eqref{shiftb}. 
Introducing $s_0=1$ for shorthand, which satisfies \eqref{svalue} as well, and by plugging in \eqref{bvalue}, we can rewrite 
 \eqref{shiftb} in the following way: 
\begin{eqnarray}
g(x)&=&(x-b)^{2r}+s_{2}(x-b)^{2r-2}+\cdots +s_{2r-2}(x-b)^{2}+{s}_{2r}+\Lambda
^{(2r-1)}(x-b)\nonumber \\
&=&\sum_{k=0}^{r}s_{2k}(x-b)^{2r-2k}+\Lambda ^{(2r-1)}(x-b)\nonumber \\
&=&\sum_{k=0}^{r}s_{2k}\sum_{n=0}^{2r-2k}x^{n}(-b)^{2r-2k-n}\left( 
\begin{array}{c}
2r-2k \\ 
n%
\end{array}%
\right) +\Lambda ^{(2r-1)}(x-b)\nonumber \\
&=&\sum_{n=0}^{2r}\sum_{k=0}^{k\leq \left( r-\frac{n}{2}\right)
}s_{2k}x^{n}(-b)^{2r-2k-n}\left( 
\begin{array}{c}
2r-2k \\ 
n%
\end{array}%
\right) +\Lambda ^{(2r-1)}(x-b) \nonumber  \\
&=&\sum_{n=0}^{2r}x^{n}\sum_{k=0}^{k\leq \left( r-\frac{n}{2}\right) }\left[
(-b^{2})^{k}\frac{(2r-1)}{(2r-2k-1)}\frac{r!}{k!(r-k)!}\right]
(-b)^{2r-2k-n}\left( 
\begin{array}{c}
2r-2k \\ 
n%
\end{array}%
\right) +\Lambda ^{(2r-1)}(x-b)  \nonumber \\
&=&\sum_{n=0}^{2r}x^{n}\sum_{k=0}^{k\leq \left( r-\frac{n}{2}\right) }\left[
(-1)^{k}\frac{(2r-1)}{(2r-2k-1)}\frac{r!}{k!(r-k)!}\right] (-b)^{2r-n}\left( 
\begin{array}{c}
2r-2k \\ 
n%
\end{array}%
\right) +\Lambda ^{(2r-1)}(x-b) \nonumber \\ 
&=&\sum_{n=0}^{2r}(-b)^{2r-n}x^{n}\left[ \sum_{k=0}^{k\leq \left( r-\frac{n}{%
2}\right) }\frac{(2r-1)(-1)^{k}}{(2r-2k-1)}\frac{r!}{k!(r-k)!}\left( 
\begin{array}{c}
2r-2k \\ 
n%
\end{array}%
\right) \right] +\Lambda ^{(2r-1)}(x-b) \label{expandshiftb} .
\end{eqnarray}
We expect all the powers of $x$ to vanish up to exponent $r$ i.e upto $x^r$. We will use various binomial identities to evaluate the finite sums. These identities 
are proved in {\bf appendix \ref{binoid}}.


The rest of this subsection is straightforward albeit somewhat tedious. 
Starting again from the last line of \eqref{expandshiftb}, we can check the coefficients up to degree $r$ in $x$. 
First, the constant term of \eqref{expandshiftb}
 \begin{eqnarray}
{g_0}&=&  (-b)^{2r}\left[ \sum_{k=0}^{k\leq r}\frac{(2r-1)(-1)^{k}}{%
(2r-2k-1)}\frac{r!}{k!(r-k)!}\left( 
\begin{array}{c}
2r-2k \\ 
0%
\end{array}%
\right) \right] +\Lambda ^{(2r-1)}(-b) \nonumber   \\
&=&  (-b)^{2r}\left[ \sum_{k=0}^{k\leq r}\frac{(2r-1)(-1)^{k}}{%
(2r-2k-1)}\frac{r!}{k!(r-k)!}\right] +(-1)^{r+1}\frac{(2r)!!b^{2r-1}}{%
(2r-3)!!}(-b)\nonumber \\
&=&    (-1)^{2r}(b)^{2r}\left[ \sum_{k=0}^{k\leq r}\frac{%
(2r-1)(-1)^{k}}{(2r-2k-1)}\frac{r!}{k!(r-k)!}\right] +(-1)^{r+2}\frac{%
(2r)!!b^{2r-1}}{(2r-3)!!}(b) \nonumber \\
&=&(b)^{2r}    (-1)^{r\ }\left( \ \left[ \sum_{k=0}^{k\leq r}%
\frac{(2r-1)(-1)^{r-k}}{(2r-2k-1)}\frac{r!}{k!(r-k)!}\right] +\ \frac{%
(2r)!!\ }{(2r-3)!!}\ \right) \nonumber \\   &=&(b)^{2r}    (-1)^{r\ }\left( \ (2r-1)\left[
\sum_{k=0}^{k\leq r}\frac{(-1)^{k}}{(2k-1)}\frac{r!}{k!(r-k)!}\right] +\ 
\frac{(2r)!!\ }{(2r-3)!!}\ \right)\nonumber \\
&=&(b)^{2r}    (-1)^{r\ }\left( (2r-1)\left( -\frac{(2r)!!}{%
(2r-1)!!}\right) +\ \frac{(2r)!!\ }{(2r-3)!!}\right) =0  \label{constantterm}
\end{eqnarray}
vanish. 
In the fifth line of \eqref{constantterm}, a term 
$\sum_{k=0}^{k\leq r}\frac{(-1)^{k}}{(2k-1)}\frac{r!}{k!(r-k)!}$ simplifies into $\left( -\frac{(2r)!!}{%
(2r-1)!!}\right)$ thanks to \eqref{hyper0} and \eqref{hyper0eval}. For $x$ term's coefficient, it simplifies in the following way:
\begin{eqnarray}
 {g_1}&=&(-b)^{2r-1}x^{{}}\left[ \sum_{k=0}^{k\leq \left( r-\frac{1}{2}\right) }\frac{%
(2r-1)(-1)^{k}}{(2r-2k-1)}\frac{r!}{k!(r-k)!}\left( 
\begin{array}{c}
2r-2k \\ 
1%
\end{array}%
\right) \right] +\Lambda ^{(2r-1)}x \nonumber  \\
&=&(-b)^{2r-1}x^{{}}\left[ \sum_{k=0}^{k\leq \left( r-1\right) }\frac{%
(2r-1)(-1)^{k}}{(2r-2k-1)}\frac{r!}{k!(r-k)!}\left( 2r-2k\right) \right]
+(-1)^{r+1}\frac{(2r)!!b^{2r-1}}{(2r-3)!!}x \nonumber \\
&=&\left( \ 2r(2r-1)\left[ \sum_{k=0}^{k\leq \left( r-1\right) }\frac{%
(-1)^{r-k}}{(2r-2k-1)}\frac{\left( r-1\right) !}{k!(r-k-1)!}\ \right] +\frac{%
(2r)!!}{(2r-3)!!}\ \right) (-1)^{r+1}b^{2r-1}x \nonumber \\
&=&\left( \ 2r(2r-1)\left[ \sum_{k=1}^{k\leq r}\frac{(-1)^{k}}{(2k-1)}\frac{%
\left( r-1\right) !}{\left( r-k\right) !(k-1)!}\ \right] +\frac{(2r)!!}{%
(2r-3)!!}\ \right) (-1)^{r+1}b^{2r-1}x .  \label{xterm}
\end{eqnarray}
Note that in the last line of \eqref{xterm},  
$\sum_{k=1}^{k\leq r}\frac{(-1)^{k}}{(2k-1)}\frac{\left( r-1\right) !}{\left(
r-k\right) !(k-1)!}$ can be rewritten as 
\begin{equation}
\sum_{k=1}^{k\leq r}\frac{(-1)^{k}}{(2k-1)}\frac{\left( r-1\right) !}{\left(
r-k\right) !(k-1)!}=\sum_{k=0}^{k\leq \left( r-1\right) }\frac{(-1)^{k+1}}{%
(2k+1)}\frac{\left( r-1\right) !}{\left( r-k-1\right) !(k)!}  \label{xtermrewrite},
\end{equation}
and this further simplifies 
 thanks to \eqref{hyper1} and \eqref{hyper1eval}.   
Finally combining all these together we get
\begin{eqnarray}
 {g_1}&=&\ 2r(2r-1)\left[ \sum_{k=1}^{k\leq r}\frac{(-1)^{k}}{(2k-1)}\frac{\left(
r-1\right) !}{\left( r-k\right) !(k-1)!}\ \right] +\frac{(2r)!!}{(2r-3)!!} \nonumber \\
&=&2r(2r-1)\left[ -\frac{\ (2r-2)!!}{(2r-1)!!}\ \right] +\frac{(2r)!!}{%
(2r-3)!!}  =0 ,
\end{eqnarray} 
which means that the term linear in $x$ also vanishes. What about higher powers of $x$?  
It is straightforward to show that the $x^2$ term 
\begin{eqnarray}
\frac{g_2}{2!}&=&(-b)^{2r-2}x^{2}\left[ \sum_{k=0}^{k\leq \left( r-1\right) }\frac{%
(2r-1)(-1)^{k}}{(2r-2k-1)}\frac{r!}{k!(r-k)!}\left( 
\begin{array}{c}
2r-2k \\ 
2%
\end{array}%
\right) \right]\nonumber \\
&=&(-b)^{2r-2}x^{2}\left[ \sum_{k=0}^{k\leq \left( r-1\right) }\frac{%
(2r-1)(-1)^{k}}{(2r-2k-1)}\frac{r!}{k!(r-k)!}\frac{\left( 2r-2k\right)
\left( 2r-2k-1\right) }{2}\right]\nonumber \\
&=&(-b)^{2r-2}x^{2}\left[ \sum_{k=0}^{k\leq \left( r-1\right) }\frac{%
(2r-1)(-1)^{k}}{1 }\frac{r!}{k!(r-k-1)!}\ \right]\nonumber \\
&=&(-b)^{2r-2}x^{2}r(2r-1)\left[ \sum_{k=0}^{k\leq \left( r-1\right) }\frac{%
(-1)^{k}}{1 }\frac{\left( r-1\right) !}{k!(r-k-1)!}\ \right]\nonumber \\
&=&(-b)^{2r-2}x^{2}r(2r-1)\left[ \sum_{k=0}^{k\leq \left( r-1\right) }\frac{%
(-1)^{k}}{1 }\left( 
\begin{array}{c}
r-1 \\ 
k%
\end{array}%
\right) \ \right]\nonumber \\
&=&(-b)^{2r-2}x^{2}r(2r-1)\left[ \left( 1-1\right) ^{r-1}\ \right] =0 
\end{eqnarray}
also vanishes.  
 %
%
In fact all the $x^n$ terms with 
$ 2\leq n\leq r$ can be shown to vanish. To argue for this,
 we start by simplifying the coefficients in the following way:
\begin{eqnarray}
\frac{g_n}{n!}&=& (-b)^{2r-n}x^{n} \left[ \sum_{k=0}^{k\leq \left( r-\frac{n}{2}%
\right) }\frac{(2r-1)(-1)^{k}}{(2r-2k-1)}\frac{r!}{k!(r-k)!}\left( 
\begin{array}{c}
2r-2k \\ 
n%
\end{array}%
\right) \right] \nonumber \\
&=&(-b)^{2r-n}x^{n}\left[ \sum_{k=0}^{k\leq \left( r-\frac{n}{2}\right) }%
\frac{(2r-1)(-1)^{k}}{(2r-2k-1)}\frac{r!}{k!(r-k)!}\frac{\left( 2r-2k\right)
!}{\left( n\right) !\left( 2r-2k-n\right) !}\right] .
\end{eqnarray}
It will be easier now to consider even and odd $n$ separately. For even powers, $n=2m$, we get
\begin{eqnarray}
\frac{g_n}{n!}&=&  (-b)^{2r-2m}x^{2m}\left[ \sum_{k=0}^{k\leq \left( r-\frac{2m}{2}%
\right) }\frac{(2r-1)(-1)^{k}}{(2r-2k-1)}\frac{r!}{k!(r-k)!}\left( 
\begin{array}{c}
2r-2k \\ 
2m%
\end{array}%
\right) \right] \nonumber \\
&=&(-b)^{2r-2m}x^{2m}\left[ \sum_{k=0}^{k\leq \left( r-m\right) }%
\frac{(2r-1)(-1)^{k}}{(2r-2k-1)}\frac{r!}{k!(r-k)!}\frac{\left( 2r-2k\right)
!}{\left( 2m\right) !\left( 2r-2k-2m\right) !}\right]\nonumber \\
&=&(-b)^{2r-2m}x^{2m}\left[ \sum_{k=0}^{k\leq \left( r-m\right) }%
\frac{(2r-1)(-1)^{k}}{(2r-2k-1)}\left( 
\begin{array}{c}
r-m \\ 
k%
\end{array}%
\right)  \frac{(r-m-k)!}{(r-k)!}   \frac{r!}{(r-m)!}  \frac{\left( 2r-2k\right)
!}{\left( 2m\right) !\left( 2r-2k-2m\right) !}\right] \nonumber \\
&=&  x^{2m}  \frac{b ^{2r-2m}(2r-1)r!}{(r-m)!\left( 2m\right) !} \left[ \sum_{k=0}^{k\leq \left( r-m\right) }%
 (-1)^{k}\left( 
\begin{array}{c}
r-m \\ 
k%
\end{array}%
\right)  \frac{(r-m-k)!}{(r-k)!}  \frac{ }{(2r-2k-1)}  \frac{\left( 2r-2k\right)
!}{\left( 2r-2k-2m\right) !}\right] \nonumber \\
&=&  x^{2m}  \frac{2b ^{2r-2m}(2r-1)r!}{(r-m)!\left( 2m\right) !} \left[ \sum_{k=0}^{k\leq \left( r-m\right) }%
 (-1)^{k}\left( 
\begin{array}{c}
r-m \\ 
k%
\end{array}%
\right)  \frac{(r-m-k)!}{(r-k-1)!  }     \frac{\left( 2r-2k-2\right)
!}{\left( 2r-2k-2m\right) !}\right] \nonumber \\
&=&  x^{2m}  \frac{2^m b ^{2r-2m}(2r-1)r!}{(r-m)!\left( 2m\right) !} \left[ \sum_{k=0}^{k\leq \left( r-m\right) }%
 (-1)^{k}\left( 
\begin{array}{c}
r-m \\ 
k%
\end{array}%
\right)     \frac{\left( 2r-2k-3\right)
!!}{\left( 2r-2k-2m-1\right) !!}\right] \nonumber \\
&=&  x^{2m}  \frac{2^m b ^{2r-2m}(2r-1)r!}{(r-m)!\left( 2m\right) !} \left[ \sum_{k=0}^{k\leq \left( r-m\right) }%
 (-1)^{k}\left( 
\begin{array}{c}
r-m \\ 
k%
\end{array}%
\right)    G_{m-1}(k)\right] \nonumber \\ &=&0, \quad m-1 < r-m, \label{evencoef} 
\end{eqnarray}
which vanishes as expected. Now consider odd powers, i.e  $n=2m+1$. For this we get
\begin{eqnarray}
\frac{g_n}{n!}&=& (-b)^{2r-2m-1}x^{2m+1}\left[ \sum_{k=0}^{k\leq \left( r-\frac{2m+1}{2}%
\right) }\frac{(2r-1)(-1)^{k}}{(2r-2k-1)}\frac{r!}{k!(r-k)!}\left( 
\begin{array}{c}
2r-2k \\ 
2m+1%
\end{array}%
\right) \right]   \nonumber \\
&=&(-b)^{2r-2m-1}x^{2m+1}\left[ \sum_{k=0}^{k\leq \left( r-m-1\right) }%
\frac{(2r-1)(-1)^{k}}{(2r-2k-1)}\frac{r!}{k!(r-k)!}\frac{\left( 2r-2k\right)
!}{\left( 2m+1\right) !\left( 2r-2k-2m-1\right) !}\right]  \nonumber \\
&=&x^{2m+1} \frac{-2^{m+1}b^{2r-2m-1}(2r-1) r!}{(r-m-1)! \left( 2m+1\right) ! }\left[ \sum_{k=0}^{k\leq \left( r-m-1\right) }%
{(-1)^{k}}  \left( 
\begin{array}{c}
r-m -1\\ 
k%
\end{array}%
\right)      \frac{    \left( 2r-2k-3\right)
!!}{    \left( 2r-2k-2m-1\right)  !!}\right] \nonumber \\
&=&x^{2m+1} \frac{-2^{m+1}b^{2r-2m-1}(2r-1) r!}{(r-m-1)! \left( 2m+1\right) ! }\left[ \sum_{k=0}^{k\leq \left( r-m-1\right) }%
{(-1)^{k}}  \left( 
\begin{array}{c}
r-m -1\\ 
k%
\end{array}%
\right)      G_{m-1}(k)\right]\nonumber \\  &=&0, \quad m-1 < r-m-1, \label{oddcoef}
\end{eqnarray}
which vanishes as expected. Note that
 in last lines of \eqref{evencoef} and \eqref{oddcoef}, we used the identity \eqref{lowpower}.

This completes the proof that $2r-1$ points in the moduli space given by \eqref{bvalueformAD} and \eqref{svalueformAD} indeed bring the SW curve of pure $SO(2r+1)$ \eqref{SOoddcurve} to maximal cusp form \eqref{CmaxCusp}. In the sense that the SW curve takes a correct form, we call these moduli points candidates for maximal Argyres-Douglas points. In order to claim that they are indeed maximal Argyres-Douglas points, we need further study of scaling dimensions of various operators \cite{Seo}.


\section{Conclusion and open questions \label{conclusion}}

In this paper we studied various methods of computing singularity loci of pure SW theories. 
At discriminant loci $\Delta_x f=0$ of the SW curve, we have vanishing 1-cycles. In certain regions of moduli space, we identified BPS dyon charges of all the 
$2r+1$ and $2(r+1)$ vanishing cycles respectively for pure $Sp(2r)$ and $SU(r+1)$ SW curves.  With the wall-crossing formula, one can in principle write down vanishing 
cycles everywhere in the moduli space. 

When its exterior derivative vanishes i.e when $d \Delta_x f=0$ as well, we know that multiple massless dyons coexist. 
Alternatively one may interpret this as the scenario where the discriminant loci self-intersect at $d\Delta_x f=\Delta_x f=0$, and here the double discriminant also vanishes $\Delta_u \Delta_x f=0$. Note however that the converse does not hold as we provided examples in earlier sections.
If order of vanishing i.e degeneracy of roots to the double discriminant is higher than $3$, then we are at the Argyres-Douglas loci. 
We also discussed the subtlety related to the choice of the moduli variable that appear in the defination of the discriminant operator, 
singling out the modulus that gives a constant term. The latter choice of modulus creates problems because of certain phase ambiguity on the $x$ plane.
  
We also computed the $2r-1$ candidates of maximal Argyres-Douglas points for $SO(2r+1)$, and observe that due to $\mathbb{Z}_{h}$ ($h$ being the dual coxeter number of the gauge group) symmetry on the $x$-plane, $h$ determines the number of maximal AD points for $SO(2r+1)$ and  $Sp(2r)$ \cite{DSW} cases. 
For $SU(r+1)$ and $SO(2r)$ groups, they are given by just 2 points, each one preserving the underlying $\mathbb{Z}_h$ symmetry.  

\paragraph{Open questions }

\begin{itemize}
\item Here we only considered SW curves, and all the analysis was purely based on geometry. However, we need to consider various physical conditions as in \cite{ACSW,AW,GST}. As a first step we can consider the scaling dimensions of Seiberg-Witten one-form and various operators. 
It will be interesting to study the detailed behaviour in the Argyres-Douglas neighbourhood extending the work of \cite{GST, EHIY}
It may also be possible to classify rank $r$ curves, in the same spirit as of \cite{ACSW,AW} done for rank 2 curves.

\item In \cite{WittenM, Gaiotto}, a class of ${\cal N}=2$ gauge theories were interpreted in terms of type IIA string theory and M-theory. It would be interesting to
 understand the theories studied here in terms of M5 branes in M-theory. For example, a question might be what the (maximal) Argyres-Douglas points correspond to in M-theory.  

\item In the same vein, we might want to construct F-theory picture for the pure $SU(r+1)$ and $Sp(2r)$ curves. The Seiberg-Witten curve for the $Sp(2r)$ gauge groups 
with anti-symmetric traceless hypermultiplet and fundamental matters \cite{AMP} have been studied from the F-theory picture. However 
the precise 1-1 mapping between the details of the curve and the corresponding F-theory configuration were never fully spelled out. A recent attempt was done 
 in \cite{DSW}. It will be interesting to complete the dictionary.

\item Another interesting topic is the wall-crossing phenomena that we dealt here only from the massless sector. We need to find some connection to wall-crossing case with the full BPS spectra that includes massive states also. Finding such a connection will shed some light on our observation of the spectra jumps at AD loci. For example in 
\cite{cordovafa} a new {\it mutation} method was developed to obtain the BPS spectra. Again it will be interesting to compare their results on the BPS spectra with ours,
especially near the AD loci.

\item Our work on the $SO$ cases was just the tip of an iceberg. Many questions still remain. For example, we can study massless dyon charges for $SO$ cases, just as we did for the $SU(r+1)$ and $Sp(2r)$ cases here. Because of various subtlety with $\mathbb{Z}_2$ symmetry on the $x$-plane and doubling of the power of hyperelliptic formula, we postpone a more detailed study of the $SO$ case to future works.  
 We can also make comparisons between the $SU(4)$ result here with the the $SO(6)$ result.
 Similarly, moduli space and monodromy of $SO(5)$ can be compared with the $Sp(4)$ case studied here. 

\item We want to check our findings against solutions to Picard-Fuch (PF) equations that give us exact period integrals for many of the pure gauge theories. 
Numerous earlier 
works (see for example \cite{ItoSasakura, KLTeqPF, MSSrank2PF, ItoYang, IMNS}) 
have addressed the solutions of the Picard-Fuch equations to obtain exact expressions for $a$ and $a_D$ and from there one might be 
able to evaluate the vanishing cycles. It'll be interesting to compare our way of getting the vanishing cycles with these techniques.  

	\item Here we only focussed on pure SW theories, but we can also consider SW curves with matter added. In the F-theoretic language this corresponds to adding seven-branes. By controlling the number of the seven-branes we can make the beta function vanish. These super-conformal theories have been extensively studied in recent years following
the work of \cite{Gaiotto}. An F-theory realization of a class of Gaiotto theories were recently addressed in \cite{DSW}. It will be interesting to extend the current techniques developed here to those theories. 
  	
\item The two groups $SU(r+1)$ and $Sp(2r)$ have identical dual coxeter number. Recently in \cite{arbKuchiev} many other connections between SW theories with these gauge 
groups has been pointed out. It will be interesting to investigate this from our monodromy and singularity analysis.
	\end{itemize} 

\acknowledgments
It is a great pleasure to thank David Berman, Heng-Yu Chen, Borun Chowdhury, Jacques Distler, Sungjay Lee, Wei Li, Andy Neitzke, Yuji Tachikawa, Sachin Vaidya, and
Johannes Walcher for helpful discussions. Discussions with Philip Argyres and Alfred Shapere were indispensable. Wolfgang Lerche's review article \cite{LercheReview} and Mathematica notebook file on his website were particularly inspiring.  
Part of our work was completed during
Chicago Great Lakes
string conference 2011, the String-Math 2011 conference at University of Pennsylvania, and Branes, Strings, and M-theory (BSM) workshop at the Newton Institute. We thank the
organizers for
providing a stimulating atmosphere for fruitful discussions. 
 Some of the result presented here already appeared in \cite{SeoUPenn}.
 The work of the authors is supported in
part by NSERC grants.

\newpage

\appendix

\section{How to obtain Seiberg-Witten curve and 1-form for pure $Sp(2r)$ theories \label{PUREfromAS}} 

In this appendix we will derive the curve and the 1-form for pure $Sp(2r)$ theories which will be used in section \ref{sprootstructure}. This appendix has 
also appeared in \cite{DSW}, but we keep it here for completeness. 

Seiberg-Witten curve and Seiberg-Witten one-form for $Sp(2r)$ with $N_f=2r+2$ fundamental hypermultiplets is given in \cite{ArgyresShapere} as following:
\begin{eqnarray}
x y^{2}   & =&\left(  x\prod_{a=1}^{r}\left(  x-\phi_{a}^{2}\right)
+g \prod_{j=1}^{2r+2} m_j \right)  ^{2}-g^2 \prod_{j=1}^{2r+2} \left( x-m_j^2\right) , \label{AS12a}\\
\lambda & =&  \frac{\sqrt{x}}{2\pi i} d \log \left( \frac{   x\prod \left(  x-\phi_{a}^{2}\right) + g \prod  m_j -\sqrt{x} y   }{  x\prod \left(  x-\phi_{a}^{2}\right) + g \prod  m_j +\sqrt{x} y  } \right), \nonumber \\
 & =& a \frac{dx}{2\sqrt{x}}  \log \left( \frac{   x\prod \left(  x-\phi_{a}^{2}\right) + g \prod  m_j +\sqrt{x} y   }{  x\prod \left(  x-\phi_{a}^{2}\right) + g \prod  m_j -\sqrt{x} y  } \right), \label{AS12oneform}\\
\emph{g(}\tau)&=&\frac{\vartheta_{2}^{4}}{\vartheta_{3}^{4}+\vartheta_{4}^{4}} ,%
\label{AS12b}
\end{eqnarray}
in terms of the Jacobi theta functions given below:
\begin{equation}
\vartheta_{2}^{4}=16 q+ {\cal O}(q^3), \qquad \vartheta_{3}^{4}=1+8q+{\cal O}(q^2), \qquad \vartheta_{4}^{4}=1-8q+{\cal O}(q^2).
\label{AS15}
\end{equation}
Combining \eqref{AS12b} and \eqref{AS15} gives us the coupling constant $g(\tau)$ in terms of $q\equiv e^{i \pi \tau}$ as below:
\begin{equation}
g(\tau)=\frac{\vartheta_{2}^{4}}{\vartheta_{3}^{4}+\vartheta_{4}^{4}}%
=\frac{16q+{\cal O}(q^{3})}{1+8q+{\cal O}(q^{2})+1-8q+{\cal O}(q^{2})}=8q+{\cal O}(q^{3}).
\end{equation}
We want to obtain Seiberg-Witten curve for pure (no flavor, $N_{f}=0$) case, which is achieved by taking all the flavors to be infinitely massive $m_{j}\sim M\rightarrow\infty$, while keeping \begin{equation} 8\Lambda^{2r+2}=8qM^{2r+2}=gM^{2r+2} \label{noflavor} \end{equation} finite, as argued in \cite{ArgyresShapere}.
  Inside \eqref{noflavor} allows, $g\prod_{j=1}^{2r+2}m_{j}=gM^{2r+2}$ to be replaced with
$8\Lambda^{2r+2}$ and $g^{2}\prod_{j=1}^{2r+2}(x-m_{j}^{2})=\left(  gM^{2r+2}\right)
^{2}$ with $\left(  8\Lambda^{2r+2}\right)^{2}$ in \eqref{AS12a} and \eqref{AS12oneform}. The former, the Seiberg-Witten curve now becomes
\begin{align}
x y^{2}  &  =\left(  x\prod_{a=1}^{r}\left(  x-\phi_{a}^{2}\right)
+8\Lambda^{2r+2}\right)  ^{2}-\left(  8\Lambda^{2r+2}\right)  ^{2} \nonumber \\
&  =\left(  A+B\right)  ^{2}-\left(  B\right)  ^{2}=A(A+2B)\nonumber \\
&  =\left(  x\prod_{a=1}^{r}\left(  x-\phi_{a}^{2}\right)  \right)  \left(
x\prod_{a=1}^{r}\left(  x-\phi_{a}^{2}\right)  +16\Lambda^{2r+2}\right). \label{ASnoFlavorCurve1}
\end{align}
Dividing with $x$ on both sides of \eqref{ASnoFlavorCurve1}, we finally get:%
\begin{equation}
y^{2}=\left(  \ \prod_{a=1}^{r}\left(  x-\phi_{a}^{2}\right)  \right)
\left(  x\prod_{a=1}^{r}\left(  x-\phi_{a}^{2}\right)  +16\Lambda
^{2r+2}\right). \label{ASnoFlavorCurve}
\end{equation} The Seiberg-Witten one-form becomes
\begin{eqnarray}
\lambda & =& a \frac{dx}{2\sqrt{x}}  \log \left( \frac{   x\prod \left(  x-\phi_{a}^{2}\right) + g \prod  m_j +\sqrt{x} y   }{  x\prod \left(  x-\phi_{a}^{2}\right) + g \prod  m_j -\sqrt{x} y  } \right),\nonumber \\
& =& a \frac{dx}{2\sqrt{x}}  \log \left( \frac{   x\prod \left(  x-\phi_{a}^{2}\right) + 8\Lambda^{2r+2} +\sqrt{x} y   }{  x\prod \left(  x-\phi_{a}^{2}\right) + 8\Lambda^{2r+2} -\sqrt{x} y  } \right). \label{ASnoFlavor1Form}
\end{eqnarray}
The final result in \eqref{ASnoFlavorCurve} and \eqref{ASnoFlavor1Form} is used as the Seiberg-Witten curve and one-form for pure $Sp(2r)$ gauge theory throughout this paper, for example in \eqref{sp1curve1form} and \eqref{OneForm}.

\newpage

\section{Useful binomial identities \label{binoid}}

In this appendix we will derive some binimial identities that will be very useful in extracting certain relations in subsection  \ref{SOoddADmax}. Our 
starting point would be the standard binomial identity
\begin{eqnarray}
\sum_{k}\left( 
\begin{array}{c}
n \\ 
k%
\end{array}%
\right) x^{k}\ & =&(1+x)^{n} ,
\end{eqnarray}
we operate with differential operator $x\frac{d}{dx}$ on both sides to obtain
\begin{eqnarray}
\sum_{k}\left( 
\begin{array}{c}
n \\ 
k%
\end{array}%
\right) x^{k}\ k^i &=& \left( x \frac{d }{dx } \right)^i \left[ (1+x)^{n}\right] , \qquad i \in \mathbb{Z}.
\end{eqnarray}
If $ i<n $, then RHS will have surviving factors of $(1+x)$. 
When we plug in $x=-1$, then RHS will vanish only for $ i<n $. Therefore we obtain
\begin{equation} 
\sum_{k}\left( 
\begin{array}{c}
n \\ 
k%
\end{array}%
\right) (-1)^{k}\ k^i = 0 , \qquad i<n .
\end{equation}
Consider an abitrary polynomial in $k$ with degree $i<n$, 
\begin{equation}G_i(k)=\sum_{j=0}^{i} c_j k^j .\end{equation}
Then we obtain
\begin{equation}
\sum_{k}\left( 
\begin{array}{c}
n \\ 
k%
\end{array}%
\right) (-1)^{k} G_i(k)= \sum_{k}\left( 
\begin{array}{c}
n \\ 
k%
\end{array}%
\right) (-1)^{k} \sum_{j=0}^{i} c_j k^j = \sum_{j=0}^{i} c_j
\sum_{k}\left( 
\begin{array}{c}
n \\ 
k%
\end{array}%
\right) (-1)^{k}  k^j 
=0 , \qquad i<n. \label{lowpower}
\end{equation}
Another trick involves elliptic integrals. 
For example, in order to evaluate $$\sum_{k=0}^{k\leq \left( r-1\right) }\frac{1}{%
(2k+1)}\frac{\left( r-1\right) !}{\left( r-k-1\right) !(k)!}(-1)^{k+1}, $$ 
we evaluate its slight modification as below, 
\begin{eqnarray}
&&\sum_{k=0}^{k\leq \left( r-1\right) }\frac{x^{2k+1}}{(2k+1)}\frac{\left(
r-1\right) !}{\left( r-k-1\right) !(k)!}(-1)^{k+1} \nonumber  \\
&=&\sum_{k=0}^{k\leq \left( r-1\right) }\left( \int dx\ x^{2k}\right) \frac{%
\left( r-1\right) !}{\left( r-k-1\right) !(k)!}(-1)^{k+1}\nonumber \\
&=&-\int dx\left[ \sum_{k=0}^{k\leq \left( r-1\right) }\left(
-x^{2}\right) ^{k}\frac{\left( r-1\right) !}{\left( r-k-1\right) !(k)!}%
\right] \nonumber \\
&=&-\int dx\left[ \ \left( 1-x^{2}\right) ^{r-1}\ \right]  \nonumber \\
&=&-x \ _{2}F_{1}\left( \frac{1}{2},-(r-1),\frac{3}{2},x^{2}\right),\label{hyper1}
\end{eqnarray}
with $x=1$ to be plugged in at the end.
For $x=1$, the hypergeometric function $\ _{2}F_{1}$ can be evaluated as below
\begin{eqnarray}
_{2}F_{1}\left( \frac{1}{2},-(r-1),\frac{3}{2},1\right) &=&\frac{\sqrt{\pi }%
\Gamma \left( r\right) }{2\Gamma \left( \frac{1}{2}+r\right) } \nonumber \\
&=&\frac{\sqrt{\pi }\Gamma \left( r\right) }{2\Gamma \left( \frac{1}{2}%
+r\right) }\frac{\Gamma \left( r\right) \Gamma \left( \frac{1}{2}+r\right) }{%
2^{1-2r}\sqrt{\pi }\Gamma \left( 2r\right) }\nonumber \\
&=&\frac{\ \Gamma \left( r\right) }{\ \ 2^{2-2r}}\frac{\Gamma \left(
r\right) \ }{\ \Gamma \left( 2r\right) }\nonumber \\
&=&\frac{2^{2r-2}(r-1)!(r-1)!}{(2r-1)!} \nonumber \\
&=&\frac{\ (2r-2)!!(2r-2)!!}{(2r-1)!}\nonumber \\
&=&\frac{\ (2r-2)!!}{(2r-1)!!},  \label{hyper1eval}
\end{eqnarray}%
 using the well known Gamma-function identity:
\begin{equation}
\Gamma \left( r\right) \Gamma \left( \frac{1}{2}+r\right) =2^{1-2r}\sqrt{%
\pi }\Gamma \left( 2r\right). \end{equation}
Similar technique will be used to evaluate 
 $\sum_{k=0}^{k\leq r}\frac{(-1)^{k}}{(2k-1)}\frac{r!}{k!(r-k)!}$. We again take a slightly more complicated term $\sum_{k=0}^{k\leq r}\frac{x^{2k-1}}{(2k-1)}\frac{r!}{k!(r-k)!}(-1)^{k} $, and then we will plug in $x=1$ at the end. 
The latter simplifies into an integral as below
\begin{eqnarray}
&&\sum_{k=0}^{k\leq r}\frac{x^{2k-1}}{(2k-1)}\frac{r!}{k!(r-k)!}(-1)^{k} \nonumber   \\
&=&\sum_{k=0}^{k\leq r}\left( \int dx\ x^{2k-2}\right) \frac{r!}{k!(r-k)!}%
(-1)^{k}\nonumber \\
&=&\int dx\left( \sum_{k=0}^{k\leq r}\ \frac{r!}{k!(r-k)!}\frac{%
(-1)^{k}x^{2k}}{x^{2}}\right)\nonumber \\
&=&\int dx\left( \frac{(1-x^{2})^{r}}{x^{2}}\right) =_{2}F_{1}\left( -\frac{1%
}{2},-r,\frac{1}{2},x^{2}\right).\label{hyper0}
\end{eqnarray}%
Its value at $x=1$ is evaluated in the following way
\begin{eqnarray}
_{2}F_{1}\left( -\frac{1}{2},-r,\frac{1}{2},1\right) &=&-\frac{\sqrt{\pi }%
\Gamma \left( 1+r\right) }{\Gamma \left( \frac{1}{2}+r\right) } \nonumber \\
&=&-\frac{2^{2r-1}r!(r-1)!}{(2r-1)!}\nonumber \\
&=&-\frac{(2r)!!(2r-2)!!}{(2r-1)!}\nonumber \\
&=&-\frac{(2r)!!}{(2r-1)!!}.\label{hyper0eval}
\end{eqnarray}

 \newpage

\bibliographystyle{JHEP}
\bibliography{AScurve}

\providecommand{\href}[2]{#2}\begingroup\raggedright\begin{thebibliography}{10}

\bibitem{SeibergWittenNoMatter}
N.~Seiberg and E.~Witten, {\it {Monopole Condensation, And Confinement In N=2
  Supersymmetric Yang-Mills Theory}},  {\em Nucl. Phys.} {\bf B426} (1994)
  19--52, [\href{http://xxx.lanl.gov/abs/hep-th/9407087}{{\tt
  hep-th/9407087}}].

\bibitem{SeibergWittenWithMatter}
N.~Seiberg and E.~Witten, {\it {Monopoles, duality and chiral symmetry breaking
  in N=2 supersymmetric QCD}},  {\em Nucl. Phys.} {\bf B431} (1994) 484--550,
  [\href{http://xxx.lanl.gov/abs/hep-th/9408099}{{\tt hep-th/9408099}}].

\bibitem{Gaiotto}
D.~Gaiotto, {\it {N=2 dualities}},
  \href{http://xxx.lanl.gov/abs/0904.2715}{{\tt arXiv:0904.2715}}.

\bibitem{AGT}
L.~F. Alday, D.~Gaiotto, and Y.~Tachikawa, {\it {Liouville Correlation
  Functions from Four-dimensional Gauge Theories}},  {\em Lett.Math.Phys.} {\bf
  91} (2010) 167--197, [\href{http://xxx.lanl.gov/abs/0906.3219}{{\tt
  arXiv:0906.3219}}].

\bibitem{ArgyresDouglas}
P.~C. Argyres and M.~R. Douglas, {\it {New phenomena in SU(3) supersymmetric
  gauge theory}},  {\em Nucl.Phys.} {\bf B448} (1995) 93--126,
  [\href{http://xxx.lanl.gov/abs/hep-th/9505062}{{\tt hep-th/9505062}}].

\bibitem{GMNwall}
D.~Gaiotto, G.~W. Moore, and A.~Neitzke, {\it {Wall-crossing, Hitchin Systems,
  and the WKB Approximation}},  \href{http://xxx.lanl.gov/abs/0907.3987}{{\tt
  arXiv:0907.3987}}.

\bibitem{AF}
N.~Akerblom and M.~Flohr, {\it {Explicit formulas for the scalar modes in
  Seiberg-Witten theory with an application to the Argyres-Douglas point}},
  {\em JHEP} {\bf 02} (2005) 057,
  [\href{http://xxx.lanl.gov/abs/hep-th/0409253}{{\tt hep-th/0409253}}].

\bibitem{GiveonRocek}
A.~Giveon and M.~Rocek, {\it {Effective actions and gauge field stability}},
  {\em Phys. Lett.} {\bf B363} (1995) 173--179,
  [\href{http://xxx.lanl.gov/abs/hep-th/9508043}{{\tt hep-th/9508043}}].

\bibitem{CH}
M.~Cederwall and M.~Holm, {\it {Monopole and Dyon Spectra in N=2 SYM with
  Higher Rank Gauge Groups}},
  \href{http://xxx.lanl.gov/abs/hep-th/9603134}{{\tt hep-th/9603134}}.

\bibitem{Marino:Moore}
M.~Marino and G.~W. Moore, {\it {The Donaldson-Witten function for gauge groups
  of rank larger than one}},  {\em Commun. Math. Phys.} {\bf 199} (1998)
  25--69, [\href{http://xxx.lanl.gov/abs/hep-th/9802185}{{\tt
  hep-th/9802185}}].

\bibitem{RSVV}
A.~Ritz, M.~A. Shifman, A.~I. Vainshtein, and M.~B. Voloshin, {\it {Marginal
  stability and the metamorphosis of BPS states}},  {\em Phys. Rev.} {\bf D63}
  (2001) 065018, [\href{http://xxx.lanl.gov/abs/hep-th/0006028}{{\tt
  hep-th/0006028}}].

\bibitem{ShapereVafa}
A.~D. Shapere and C.~Vafa, {\it {BPS structure of Argyres-Douglas
  superconformal theories}},
  \href{http://xxx.lanl.gov/abs/hep-th/9910182}{{\tt hep-th/9910182}}.

\bibitem{LercheReview}
W.~Lerche, {\it {Introduction to Seiberg-Witten theory and its stringy
  origin}},  {\em Nucl. Phys. Proc. Suppl.} {\bf 55B} (1997) 83--117,
  [\href{http://xxx.lanl.gov/abs/hep-th/9611190}{{\tt hep-th/9611190}}].

\bibitem{GST}
D.~Gaiotto, N.~Seiberg, and Y.~Tachikawa, {\it {Comments on scaling limits of
  4d N=2 theories}},  {\em JHEP} {\bf 01} (2011) 078,
  [\href{http://xxx.lanl.gov/abs/1011.4568}{{\tt arXiv:1011.4568}}].

\bibitem{EHIY}
T.~Eguchi, K.~Hori, K.~Ito, and S.-K. Yang, {\it {Study of $N=2$ Superconformal
  Field Theories in $4$ Dimensions}},  {\em Nucl. Phys.} {\bf B471} (1996)
  430--444, [\href{http://xxx.lanl.gov/abs/hep-th/9603002}{{\tt
  hep-th/9603002}}].

\bibitem{EH}
T.~Eguchi and K.~Hori, {\it {N=2 superconformal field theories in
  four-dimensions and A-D-E classification}},
  \href{http://xxx.lanl.gov/abs/hep-th/9607125}{{\tt hep-th/9607125}}.

\bibitem{Seo}
J.~Seo, {\it {Scaling behaviour at maximal Argyres-Douglas points of pure
  $SO(2r+1)$ and $Sp(2r)$ Seiberg-Witten theory (work in progress)}}, .

\bibitem{KLYTsimpleADE}
A.~Klemm, W.~Lerche, S.~Yankielowicz, and S.~Theisen, {\it {Simple
  singularities and N=2 supersymmetric Yang-Mills theory}},  {\em Phys. Lett.}
  {\bf B344} (1995) 169--175,
  [\href{http://xxx.lanl.gov/abs/hep-th/9411048}{{\tt hep-th/9411048}}].

\bibitem{ArgyresFaraggiSU}
P.~C. Argyres and A.~E. Faraggi, {\it {The vacuum structure and spectrum of N=2
  supersymmetric SU(n) gauge theory}},  {\em Phys. Rev. Lett.} {\bf 74} (1995)
  3931--3934, [\href{http://xxx.lanl.gov/abs/hep-th/9411057}{{\tt
  hep-th/9411057}}].

\bibitem{HananyOzSU}
A.~Hanany and Y.~Oz, {\it {On the Quantum Moduli Space of Vacua of $N=2$
  Supersymmetric $SU(N_c)$ Gauge Theories}},  {\em Nucl. Phys.} {\bf B452}
  (1995) 283--312, [\href{http://xxx.lanl.gov/abs/hep-th/9505075}{{\tt
  hep-th/9505075}}].

\bibitem{BLSo}
A.~Brandhuber and K.~Landsteiner, {\it {On the monodromies of N=2
  supersymmetric Yang-Mills theory with gauge group SO(2n)}},  {\em Phys.Lett.}
  {\bf B358} (1995) 73--80, [\href{http://xxx.lanl.gov/abs/hep-th/9507008}{{\tt
  hep-th/9507008}}].

\bibitem{SO5DS}
U.~H. Danielsson and B.~Sundborg, {\it {The Moduli space and monodromies of N=2
  supersymmetric SO(2r+1) Yang-Mills theory}},  {\em Phys. Lett.} {\bf B358}
  (1995) 273--280, [\href{http://xxx.lanl.gov/abs/hep-th/9504102}{{\tt
  hep-th/9504102}}].

\bibitem{HananySO}
A.~Hanany, {\it {On the Quantum Moduli Space of N=2 Supersymmetric Gauge
  Theories}},  {\em Nucl. Phys.} {\bf B466} (1996) 85--100,
  [\href{http://xxx.lanl.gov/abs/hep-th/9509176}{{\tt hep-th/9509176}}].

\bibitem{DSW}
K.~Dasgupta, J.~Seo, and A.~Wissanji, {\it {F-Theory, Seiberg-Witten Curves and
  N = 2 Dualities}},  {\em JHEP} {\bf 1202} (2012) 146,
  [\href{http://xxx.lanl.gov/abs/1107.3566}{{\tt arXiv:1107.3566}}].

\bibitem{Sen1O7split}
A.~Sen, {\it {F-theory and Orientifolds}},  {\em Nucl. Phys.} {\bf B475} (1996)
  562--578, [\href{http://xxx.lanl.gov/abs/hep-th/9605150}{{\tt
  hep-th/9605150}}].

\bibitem{BDS1D3}
T.~Banks, M.~R. Douglas, and N.~Seiberg, {\it {Probing F-theory with branes}},
  {\em Phys. Lett.} {\bf B387} (1996) 278--281,
  [\href{http://xxx.lanl.gov/abs/hep-th/9605199}{{\tt hep-th/9605199}}].

\bibitem{DLSmultiD3}
M.~R. Douglas, D.~A. Lowe, and J.~H. Schwarz, {\it {Probing F-theory with
  multiple branes}},  {\em Phys. Lett.} {\bf B394} (1997) 297--301,
  [\href{http://xxx.lanl.gov/abs/hep-th/9612062}{{\tt hep-th/9612062}}].

\bibitem{ShapereTachikawa}
A.~D. Shapere and Y.~Tachikawa, {\it {Central charges of N=2 superconformal
  field theories in four dimensions}},  {\em JHEP} {\bf 0809} (2008) 109,
  [\href{http://xxx.lanl.gov/abs/0804.1957}{{\tt arXiv:0804.1957}}].

\bibitem{AMP}
P.~C. Argyres, R.~Maimon, and S.~Pelland, {\it {The M theory lift of two 06-
  planes and four D6-branes}},  {\em JHEP} {\bf 0205} (2002) 008,
  [\href{http://xxx.lanl.gov/abs/hep-th/0204127}{{\tt hep-th/0204127}}].

\bibitem{ArgyresShapere}
P.~C. Argyres and A.~D. Shapere, {\it {The Vacuum Structure of N=2 SuperQCD
  with Classical Gauge Groups}},  {\em Nucl. Phys.} {\bf B461} (1996) 437--459,
  [\href{http://xxx.lanl.gov/abs/hep-th/9509175}{{\tt hep-th/9509175}}].

\bibitem{KLYTmndrmSU}
A.~Klemm, W.~Lerche, S.~Yankielowicz, and S.~Theisen, {\it {On the monodromies
  of N=2 supersymmetric Yang-Mills theory}},
  \href{http://xxx.lanl.gov/abs/hep-th/9412158}{{\tt hep-th/9412158}}.

\bibitem{LPGg2}
K.~Landsteiner, J.~M. Pierre, and S.~B. Giddings, {\it {On the moduli space of
  N = 2 supersymmetric G(2) gauge theory}},  {\em Phys. Rev.} {\bf D55} (1997)
  2367--2372, [\href{http://xxx.lanl.gov/abs/hep-th/9609059}{{\tt
  hep-th/9609059}}].

\bibitem{APSW}
P.~C. Argyres, M.~Ronen~Plesser, N.~Seiberg, and E.~Witten, {\it {New N=2
  Superconformal Field Theories in Four Dimensions}},  {\em Nucl. Phys.} {\bf
  B461} (1996) 71--84, [\href{http://xxx.lanl.gov/abs/hep-th/9511154}{{\tt
  hep-th/9511154}}].

\bibitem{ACSW}
P.~C. Argyres, M.~Crescimanno, A.~D. Shapere, and J.~R. Wittig, {\it
  {Classification of N = 2 superconformal field theories with two-dimensional
  Coulomb branches}},  \href{http://xxx.lanl.gov/abs/hep-th/0504070}{{\tt
  hep-th/0504070}}.

\bibitem{AW}
P.~C. Argyres and J.~R. Wittig, {\it {Classification of N = 2 superconformal
  field theories with two-dimensional Coulomb branches. II}},
  \href{http://xxx.lanl.gov/abs/hep-th/0510226}{{\tt hep-th/0510226}}.

\bibitem{WittenM}
E.~Witten, {\it {Solutions of four-dimensional field theories via M- theory}},
  {\em Nucl. Phys.} {\bf B500} (1997) 3--42,
  [\href{http://xxx.lanl.gov/abs/hep-th/9703166}{{\tt hep-th/9703166}}].

\bibitem{cordovafa}
M.~Alim, S.~Cecotti, C.~Cordova, S.~Espahbodi, A.~Rastogi, {\em et.~al.}, {\it
  {N=2 Quantum Field Theories and Their BPS Quivers}},
  \href{http://xxx.lanl.gov/abs/1112.3984}{{\tt arXiv:1112.3984}}.

\bibitem{ItoSasakura}
K.~Ito and N.~Sasakura, {\it {Exact and microscopic one instanton calculations
  in N=2 supersymmetric Yang-Mills theories}},  {\em Nucl.Phys.} {\bf B484}
  (1997) 141--166, [\href{http://xxx.lanl.gov/abs/hep-th/9608054}{{\tt
  hep-th/9608054}}].

\bibitem{KLTeqPF}
A.~Klemm, W.~Lerche, and S.~Theisen, {\it {Nonperturbative effective actions of
  N=2 supersymmetric gauge theories}},  {\em Int.J.Mod.Phys.} {\bf A11} (1996)
  1929--1974, [\href{http://xxx.lanl.gov/abs/hep-th/9505150}{{\tt
  hep-th/9505150}}].

\bibitem{MSSrank2PF}
T.~Masuda, T.~Sasaki, and H.~Suzuki, {\it {Seiberg-Witten theory of rank two
  gauge groups and hypergeometric series}},  {\em Int. J. Mod. Phys.} {\bf A13}
  (1998) 3121--3144, [\href{http://xxx.lanl.gov/abs/hep-th/9705166}{{\tt
  hep-th/9705166}}].

\bibitem{ItoYang}
K.~Ito and S.-K. Yang, {\it {Picard-Fuchs equations and prepotentials in N = 2
  supersymmetric {QCD}}},  \href{http://xxx.lanl.gov/abs/hep-th/9603073}{{\tt
  hep-th/9603073}}.

\bibitem{IMNS}
J.~M. Isidro, A.~Mukherjee, J.~P. Nunes, and H.~J. Schnitzer, {\it {A note on
  the Picard-Fuchs equations for N = 2 Seiberg- Witten theories}},  {\em Int.
  J. Mod. Phys.} {\bf A13} (1998) 233--250,
  [\href{http://xxx.lanl.gov/abs/hep-th/9703176}{{\tt hep-th/9703176}}].

\bibitem{arbKuchiev}
M.~Y. Kuchiev, {\it {Supersymmetric N=2 gauge theory with arbitrary gauge
  group}},  {\em Nucl.Phys.} {\bf B838} (2010) 331--357,
  [\href{http://xxx.lanl.gov/abs/0907.2010}{{\tt arXiv:0907.2010}}].

\bibitem{SeoUPenn}
J.~Seo, {\it {Singularity structure and massless dyons of pure Seiberg-Witten
  theories with $SU$ and $Sp$ gauge groups (Strings-Math 2011 proceedings to
  appear)}}, .

\end{thebibliography}\endgroup

\end{document}